\documentclass[aps,reprint,prx]{revtex4-1}
\usepackage{graphicx,amsmath,subcaption}

\usepackage{amsmath}
\usepackage{epsf}
\usepackage{graphicx}

\usepackage{dcolumn}
\usepackage{bm}

\usepackage{amssymb}
\usepackage{array}
\usepackage{varioref}
\usepackage{wrapfig}
\usepackage{etoolbox}
\usepackage{color}
\usepackage[colorlinks, linkcolor=blue, citecolor=red]{hyperref}

\providecommand{\nn}{\nonumber}

\providecommand{\pd}{\partial}

\providecommand{\bv}[1]{\bm{\mathrm{#1}}}

\providecommand{\w}{\omega}
\providecommand{\W}{\Omega}
\providecommand{\q}{\bv{q}}
\providecommand{\Q}{\bv{Q}}
\providecommand{\p}{\bv{p}}

\renewcommand{\k}{\bv{k}}

\providecommand{\ef}{\varepsilon_F}
\providecommand{\vf}{v_F}

\providecommand{\kf}{k_F}

\providecommand{\kf}{k_F}

\renewcommand{\q}{\bv{q}}

\providecommand{\gb}{\bar{g}}

\providecommand{\tp}{2\pi}
\providecommand{\tpp}{\left(2\pi\right)}
\providecommand{\kE}[1][ ]{\epsilon_{#1}}
\providecommand{\ek}{\kE[\k]}
\providecommand{\Sg}{\Sigma}

\providecommand{\sgn}{\sigma}

\newcommand{\inm}{{\mbox{\tiny INM}}}
\newcommand{\sfm}{{\mbox{\tiny SFM}}}
\newcommand{\ET}{{\mbox{\tiny ET}}}
\newcommand{\LT}{{\mbox{\tiny LT}}}

\renewcommand{\ef}{E_F}
\begin{document}

\title{Normal state properties of quantum critical metals at finite temperature}

\author{Avraham Klein}
\affiliation{School of Physics and Astronomy, University of Minnesota, Minneapolis. MN, USA}
\author{Yoni Schattner}
\affiliation{Department of Physics, Stanford University, CA, USA}
\affiliation{Stanford Institute for Materials and Energy Sciences, SLAC National Accelerator Laboratory and Stanford University, Menlo Park, CA, USA}

\author{Erez Berg}
\affiliation{Department of Condensed Matter Physics, Weizmann Institute of Sciences, Rehovot, Israel}
\author{Andrey V. Chubukov}
\affiliation{School of Physics and Astronomy, University of Minnesota, Minneapolis. MN, USA}

\begin{abstract}
  We study the effects of finite temperature on normal state properties of a metal near a quantum critical point to an antiferromagnetic or Ising-nematic state.
  At $T=0$ bosonic and fermionic self-energies are traditionally
   computed within Eliashberg theory and obey
    scaling relations
     with  characteristic power-laws.
  Quantum Monte Carlo (QMC) simulations have shown strong systematic deviations from these predictions, casting doubt on the validity of the theoretical analysis.
  We extend Eliashberg theory to finite $T$ and
 argue that for the $T$ range accessible in the QMC simulations, the scaling forms for both fermionic and bosonic self energies are quite different from those at $T=0$.
  We compare
    finite $T$ results with
     QMC data and find good agreement
      for both systems. This, we argue, resolves
       the key apparent contradiction between the theory and
       the QMC
       simulations.

\end{abstract}
\maketitle

\section{Introduction}
\label{sec:introduction}

Electron-boson models \cite{Hertz1976,Millis1993,Altshuler1994,Abanov2003}
have long been used to study
the behavior of interacting fermions near a metallic quantum critical point (QCP)
In these models, a specific channel of the electron-electron interaction is
assumed to become critical at a QCP and is
represented by a
soft collective boson, while all other channels are assumed to be irrelevant to the low energy dynamics near the QCP. Despite their simplicity, such models predict nontrivial correlation effects such as superconductivity and non Fermi-liquid (NFL)
behavior in the normal state~\cite{Altshuler1994,Abanov2001,Abanov2003,Metlitski2010a,Metlitski2010,Metlitski2010b,Lederer2017,Lee2018,Maslov2010,Fradkin2010,Abanov2001,Wang2016,Raghu2015,Metlitski2015,Lederer2015}.
Correlation effects are stronger in two dimensions (2D) than in 3D,
consistent with the fact that most
systems in which NFL behavior and high-temperature superconductivity have been observed,
e.g. Cu- and Fe- based superconductors, are quasi-2D systems~\cite{Loehneysen2007,Monthoux2007,Abanov2003,Scalapino2012,Sachdev2012,Cyr-Choiniere2018,Fernandes2014,Wang2015,Wang2014}.
Two of the most
studied models of metallic quantum criticality  are the spin-fermion model (SFM)
where the boson
describes fluctuations of an antiferromagnetic order parameter, and the Ising-nematic model (INM),
where the boson represents an
order parameter, which breaks lattice rotational symmetry.

Both the SFM and the INM have been analyzed within
the low-energy
theory which is termed
Eliashberg
theory (ET) due to its similarity with the Eliashberg theory of the
electron-phonon interaction.
The theory assumes
that
near a QCP, a  soft
collective boson is a slow mode compared to
a
dressed fermion.  This
effectively decouples the fermionic and bosonic degrees of freedom \cite{Millis1992,Abanov2003,Rech2006} and makes the problem analytically tractable.
In technical terms, in ET the fermionic self-energy $\Sigma (k, \omega)$  is large, but it can be approximated by
its value at the Fermi energy $\Sigma (k_F, \omega)$ and computed perturbatively at one-loop order. Higher-order
corrections (commonly termed as vertex corrections) are neglected by the argument that    
in the processes giving rise to vertex corrections 
fermions oscillate at 
frequencies near the bosonic mass shell,
which are far away from their own mass shell. 
In addition,
the
integration over internal momenta in the one-loop
diagram for the fermionic self-energy
factorizes into 
the one transverse to the Fermi surface (FS), 
which involves 
only fermionic degrees of freedom, and 
the one parallel to the FS, 
involving only bosonic degrees of freedom.  
In this situation, characteristic momentum deviations from the FS are small, and the integration can be carried out by linearizing the fermionic dispersion near $k_F$.    
By the same argument, the bosonic self-energy is also computed perturbatively,
at one-loop order.

The decoupling of the fermions and bosons leads
to fermionic self-energy which depends
much more strongly on the frequency $\omega$
than  on deviation of the momentum from $k_F$  transverse to the FS.
The magnitude of the $\omega$-dependent self-energy
in turn depends on the location of $k_F$ on the FS.
In the INM, the  Eliashberg self-energy  scales as $\Sigma (\omega, {\bf k}_F) \sim  \omega^{1/3}_\inm\omega^{2/3}$  over the whole FS, except at special points (cold spots), where FL behavior survives. 
In the SFM, the Eliashberg self-energy  scales as $\Sigma (\omega, {\bf k}_F) \sim \w_\sfm^{1/2}\omega^{1/2}$  at special FS points (hot spots), connected by
a momentum vector corresponding to
antiferromagnetic order,
while everywhere else on the FS the self-energy has a FL form at the lowest frequencies and crosses over to $\omega^{1/2}$ behavior at a characteristic frequency proportional to the deviation from a hot spot.
The $\omega_\sfm$ and $\omega_\inm$ are characteristic
frequencies, 
which  we discuss below. Both remain finite at a QCP.  

The validity of
the 
ET
in the case when a soft boson is a collective mode of fermions
is a more tricky issue than for the  original Eliashberg
theory of
superconductivity,
where the boson
is an independent degree of freedom (a phonon).
For that theory, the applicability condition is the smallness of
the ratio 
$\omega_0/\ef \sim v_0/v_F $,  where 
$\omega_0$ is the  characteristic phonon frequency and $v_0$ is the corresponding boson velocity
(the dressed Debye frequency and sound velocity for an acoustic phonon)
and $\ef$ is the Fermi energy 
(this allows one to factorize the momentum integration), 
and
the smallness of the ratio $\alpha^2/(
\omega_0 \ef)$, where $\alpha$ is the effective fermion-boson coupling (this allows one to neglect vertex corrections).
The last condition is not satisfied when 
$\omega_0$ vanishes, but for large enough $\ef$
it holds
in a wide range of 
$\omega_0$ (Ref. \cite{Kivelson2020}).  When
the boson is a collective mode in the
spin or charge channel, its bare velocity is of order of $\vf$,
so at the bare level the ET is
inapplicable. However, in both the SFM and INM a dressed collective boson is Landau overdamped due to decay into particle-hole pairs. This opens up a possibility that at low energies a dressed boson becomes slow compared to a dressed fermion, i.e., ET becomes applicable as an effective theory, which describes dressed bosons and fermions at low energies.
For the INM, 
the Landau-overdamped boson is 
slow compared to dressed fermions by
$(\omega/\omega_{\inm})^{1/3}$
(Refs.  \onlinecite{Altshuler1994,Rech2006,Lee2009}), which justifies factorization of momentum integration leading to 
$\omega^{2/3}$  scaling at a QCP.
Vertex corrections diverge at a QCP, when calculated with free fermions,
but remain finite within the effective ET. The lowest-order vertex correction
is of order one, but can be
made parametrically small if one extends the theory to $N \gg 1$ fermionic flavors~\cite{Altshuler1994,Rech2006}.
two loops, however, there are unavoidable logarithmical  singularities for both $\Sigma ({\bf k}_F, \omega)$ and $\Sigma ({\bf k}, 0)$ (Refs. \onlinecite{Metlitski2010a,LKCunp}).
These logarithms come from special ``planar diagrams'', which describe hidden 1D processes with momentum transverse either $0$ or $2k_F$ (Ref. \onlinecite{Lee2009}).
Logarithmical corrections were also reported for a bosonic propagator in 5-loop calculations~\cite{Holder2015,Holder2015a}.
These logarithms are not accounted for in the effective ET.
\footnote {Whether these logarithmic corrections give rise to the appearance of   an anomalous fermionic residue
  but preserve the
  $\omega^{2/3}$ scaling
  for the self-energy is not known.}
For
the
SFM, the
velocities of dressed fermions and bosons are comparable, i.e., corrections to $\Sigma_{SFM} (\omega) =\omega^{1/2} \omega^{1/2}_{SFM}$ are of order one. 
This can be
cured by extending the theory to
$N \gg 1$ fermionic flavors,
in which case  the corrections to factorization  are small in $1/N$. 
The $k-$dependent self-energy and vertex are also  small in $1/N$.  However, just like in the INM, there are
logarithmical corrections to the effective ET. Moreover, in the SFM, logarithms appear already in one-loop $\Sigma (k)$ and vertex corrections~\cite{Abanov2003,
  Metlitski2010,Lee2018}.
\footnote{ It was
  argued~\cite{Lee2018}
  that because of these logarithms, the system eventually flows towards the new fixed point with the 
  dynamical exponent $z=1$.}

This 
analysis shows that for both models
the effective  ET
becomes invalid 
below some characteristic  frequency, at which logarithmic corrections become of order one.
However, for the breakdown of ET to occur, this frequency
must be larger than 
superconducting $T_c$, otherwise the logarithmic singularities will be cut off by the opening of a gap due to superconductivity.
In the
SFM, $T_c$ 
generally of order
$\w_\sfm$, and in the INM $T_c \sim \w_\inm$ (Refs. \onlinecite{Abanov2001,Abanov2003,Metlitski2015,Wang2013,Bonesteel1996}).
At such frequencies, some calculations show~\cite{Monthoux1997} that  
corrections to ET may be small numerically, in which case 
the effective ET should remain valid, 
at 
least qualitatively.

The validity of the effective ET at a QCP has
been recently
tested
in
a series of sign-free quantum Monte Carlo (QMC) simulations of both the SFM and INM. \cite{Schattner2016a,Lederer2017,Gerlach2017,Schattner2016,Wang2017,Liu2018,Xu2017,Xu2017a}.
Such simulations are numerical experiments that test effective models of quantum-critical metals \cite{Xu2017a,Xu2019,Berg2019}.
QMC data were taken at temperatures above $T_c$, where ET is expected to work.
Analysis of the QMC data revealed
that some properties, most strikingly the superconducting $T_c$ of the SFM, agreed well with predictions of ET\cite{Wang2017}.
However, other properties
showed
systematic deviations from ET. In particular, 
for  both SFM and INM,
fermionic self-energies in the normal state, extracted from QMC,
do not show the
power-law forms,
expected from the theory, and appear to saturate at a finite value even at
the smallest fermionic Matsubara frequency $\omega= \pi T$.  In addition, the bosonic self-energy in the INM does not show the expected $\Omega/q$ scaling
of a Landau-overdamped boson.
The
apparent contradiction with the numerical experiments has
cast into doubt the validity of ET.

In this work we argue that the discrepancies of the QMC data with the
ET can be reconciled by properly accounting for finite temperature effects
within
appropriately modified ET (MET), which, we argue, differs qualitatively  from the ET at $T=0$ (we will keep the notation ET for the $T=0$ Eliashberg theory)
Several previous works have studied finite temperature
effects within perturbation theory \cite{Abanov2003,Yamase2012,Punk2016}.
We argue that at finite temperature 
one has to go beyond perturbation theory and compute fermionic $\Sigma (\omega)$ and bosonic $\Pi (\Omega)$
self-consistently and without factorization of momentum integration. 
Specifically, we argue
that the fermionic self-energy
on the Matsubara axis, $\Sigma (\omega_m)$
(the one which can be directly compared with QMC results)
is the sum of
thermal and quantum parts,
\begin{equation}
  \label{eq:sigma-main-1}
  \Sigma(\omega_m) = \Sigma_T(\omega_m) + \Sigma_Q(\omega_m),
\end{equation}
where $\Sigma_T(\omega_m)$ is the thermal contribution, coming from the static bosonic propagator, 
and $\Sigma_Q(\omega_m)$ comes from the dynamic propagator.  
The thermal piece $\Sg_T$ needs to be calculated self-consistently without factorizing the momentum integration. The dynamical $\Sg_Q$
does not need to be computed self-consistently, but 
at $\omega_m \sim T$ one cannot factorize the momentum integration for this term as well.

We show that at finite $T$ there are
two
characteristic scales, a larger one and a smaller one.  The larger scale, $\omega_T$, is the same for SFM and INM and up to a logarithmic factor
is
\begin{equation}
  \label{eq:omega-T-intro}
  \omega_T \sim \sqrt{{\bar g} T},
\end{equation}
where ${\bar g}$ is  the effective fermion-boson coupling (defined below).
The smaller scale is, again up to a  logarithmic factor, 
\begin{equation}
  \label{eq:omega-T-intro_1}
  \omega'_T \sim \omega_T   \left\{
    \begin{array}{ll}
      \left(\frac{T}{{\bar g}}\right)^{1/2}& \text{SFM} \\
      \frac{T}{E_F} & \text{INM}
    \end{array}
  \right.
\end{equation}
We assume $T \ll {\bar g}, E_F$, such that in both models $\omega'_T \ll \omega_T$.

At the smallest Matsubara frequencies, $\omega_m \ll  \omega'_T$,
the two components of the self energy have the form
\begin{equation}
  \label{eq:Sg-regime-I}
  \Sigma_T(\w_m) \sim \w_T
  ,~\Sg_Q(\w_m) \sim \frac{\w_m}{\w_T}.
\end{equation}
We call this regime \emph{strongly thermal}.
At high Matsubara frequencies, $\omega_m \gg \omega_T$,
the self energy components have the form
\begin{equation}
  \label{eq:se-intro-approx}
  \Sigma_T(\omega_m) \sim  \frac{\omega_T^2}{\omega_m},  ~\Sg_Q(\w_m) \sim
  \left\{
    \begin{array}{ll}
      \omega_m^{1/2}& \text{SFM} \\
      \omega_m^{2/3} & \text{INM}
    \end{array}
  \right.
\end{equation}
We call this regime \emph{almost critical}.
In between these two regimes,
i.e., at  $\omega'_T < \omega_m < \omega_T$, 
the system behavior is
rather complex and 
there is no particular scaling behavior for both $\Sigma_T(\omega_m)$ and $\Sigma_Q(\omega_m)$.  
We argue that most of QMC data in Refs. ~\onlinecite{Schattner2016a,Gerlach2017,Lederer2017} fall into this intermediate frequency region.

We compute $\Sigma_T(\omega_m)$ and $\Sg_Q(\w_m)$
within MET
and compare the result with QMC data.
For both SFM and INM
we show  that
$\Sigma (\omega_m) = \Sigma_T(\omega_m) + \Sigma_Q(\omega_m)$
agrees with  QMC results.
The agreement holds for both  the magnitude of $\Sigma (\omega_m)$ and 
its dependence on frequency. 
We show that in the temperature range of the QMC simulations, 
thermal effects are essential, and  $\Sigma_{\sfm} (\omega_m)$ and  $\Sigma_{\inm} (\omega_m)$ , obtained in
the MET,
are quite flat
functions of frequency.  The bosonic propagator, $D (\Omega_m,q)$, obtained within MET, also agrees with QMC result. This is particularly significant for the INM, where QMC shows that the frequency dependence of $D^{-1} (\Omega_m, q)$ is proportional to $|\Omega_m|$ rather than $|\Omega_m|/q$, expected 
for 
Landau damping.  The absence of $|\Omega_m|/q$ scaling at the smallest $q$ is due to the fact that the Ising-nematic order parameter is not a conserved quantity, but the near-absence of the $q$ dependence over a wide range of $q$ is 
chiefly  the consequence of the flatness of $\Sigma (\omega_m)$ in the $T$ range probed by QMC.

For comparison,
we also compute  $\Sigma_T(\omega_m)$ and $\Sg_Q(\w_m)$  using the same equations as in the MET, but integrate over the internal fermionic momenta
in
the full Brillouin zone
(i.e., compute the self-energy without linearizing the fermionic dispersion near the FS). We call this the lattice theory (LT).  We show that the forms of the self-energies in MET and LT are qualitatively similar, but
with differences in the details.  We note in this regard that 
while in both MET and LT  one neglects vertex corrections and extracts $\Sigma$ from self-consistent
analysis, 
$\Sigma$ in LT contains an additional piece coming from high-energy fermions,
with energies of the order of the bandwidth.
Because vertex corrections also
predominantly come from high-energy fermions, by comparing the self-energies in MET and LT to the one extracted from QMC, one can verify whether there is at least a partial cancellation  between the 
vertex corrections and contributions from high-energy fermions to 
the  self-energy.  
We show that QMC data agree somewhat better with MET than with LT, particularly for the INM. This
suggests  that there may be 
some cancellation between different contributions from high-energy fermions.

Our results demonstrate that (a) current QMC data are
consistent with the MET (i.e., ET, properly extended to finite $T$) and that (b) the
comparison between MET, LT, and QMC
provides a
framework
to identify
the strength of the
correlation effects that are not captured by the low-energy MET.
We hope that our results will provide useful input both to further analytical work and for analysis of upcoming numerical results.

The rest of the manuscript is organized as follows. In Sec. \ref{sec:spin-fermion-ising}
we introduce the SFM and INM and review the low temperature predictions of ET.
In Sec. \ref{sec:self-energies-at} we discuss the modified  ET at a
finite temperature, which takes into account thermal fluctuations.
In Sec. \ref{sec:lattice-theory} we briefly review the lattice models which we use in the LT calculation of the self-energy.  Finally, in Sec. \ref{sec:comp-monte-carlo} we compare our
results to the QMC data. We present our conclusions in Sec. \ref{sec:summary}.

\section{Eliashberg theory for the spin-fermion and Ising-nematic  models}
\label{sec:spin-fermion-ising}

\subsection{The models}

\begin{figure}
  \centering
  \begin{subfigure}{0.45\hsize}
    \includegraphics[width=\hsize]{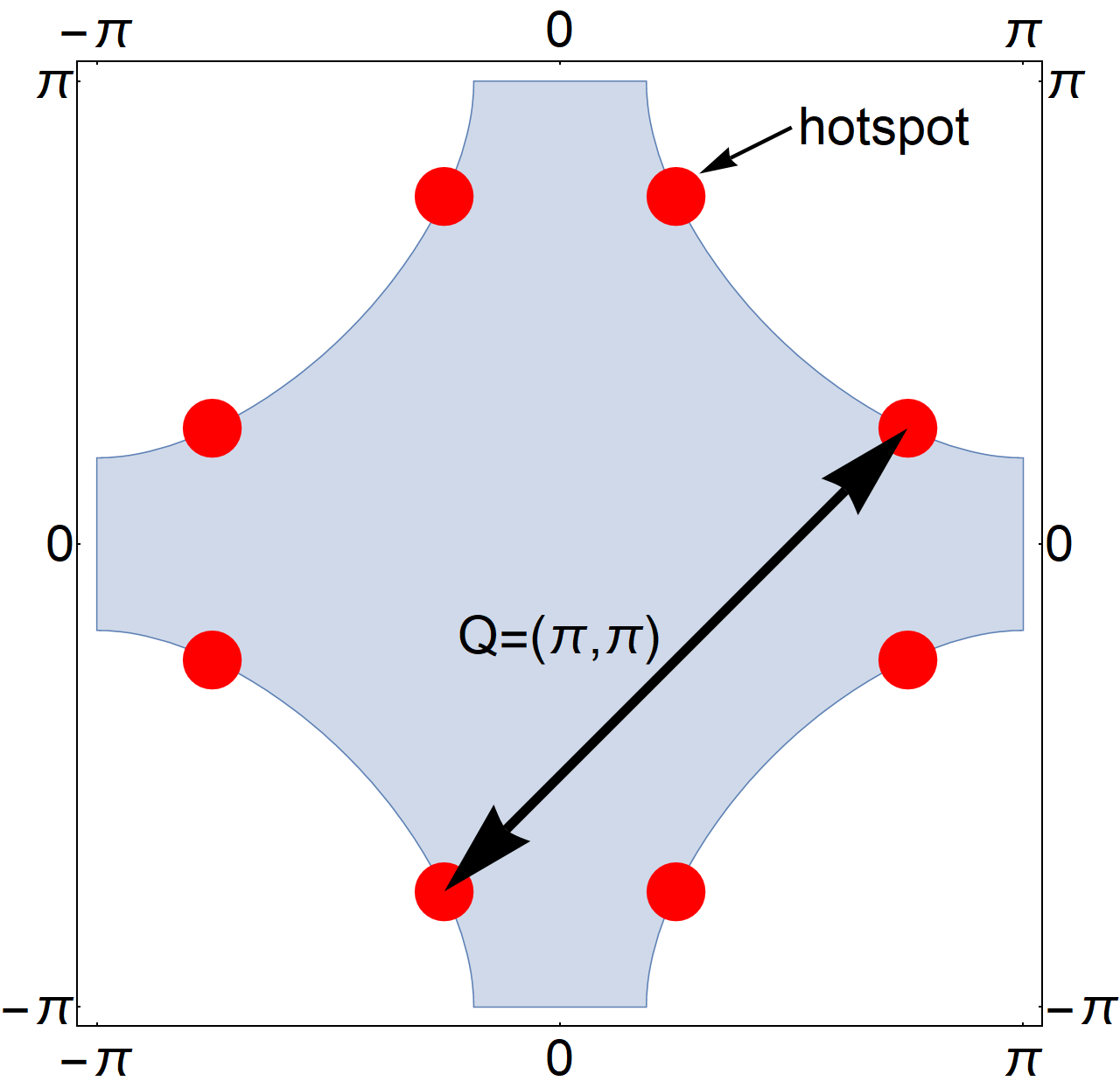}
    \caption{\label{fig:sf-band}}
  \end{subfigure}\hfill
  \begin{subfigure}{0.45\hsize}
    \includegraphics[width=\hsize]{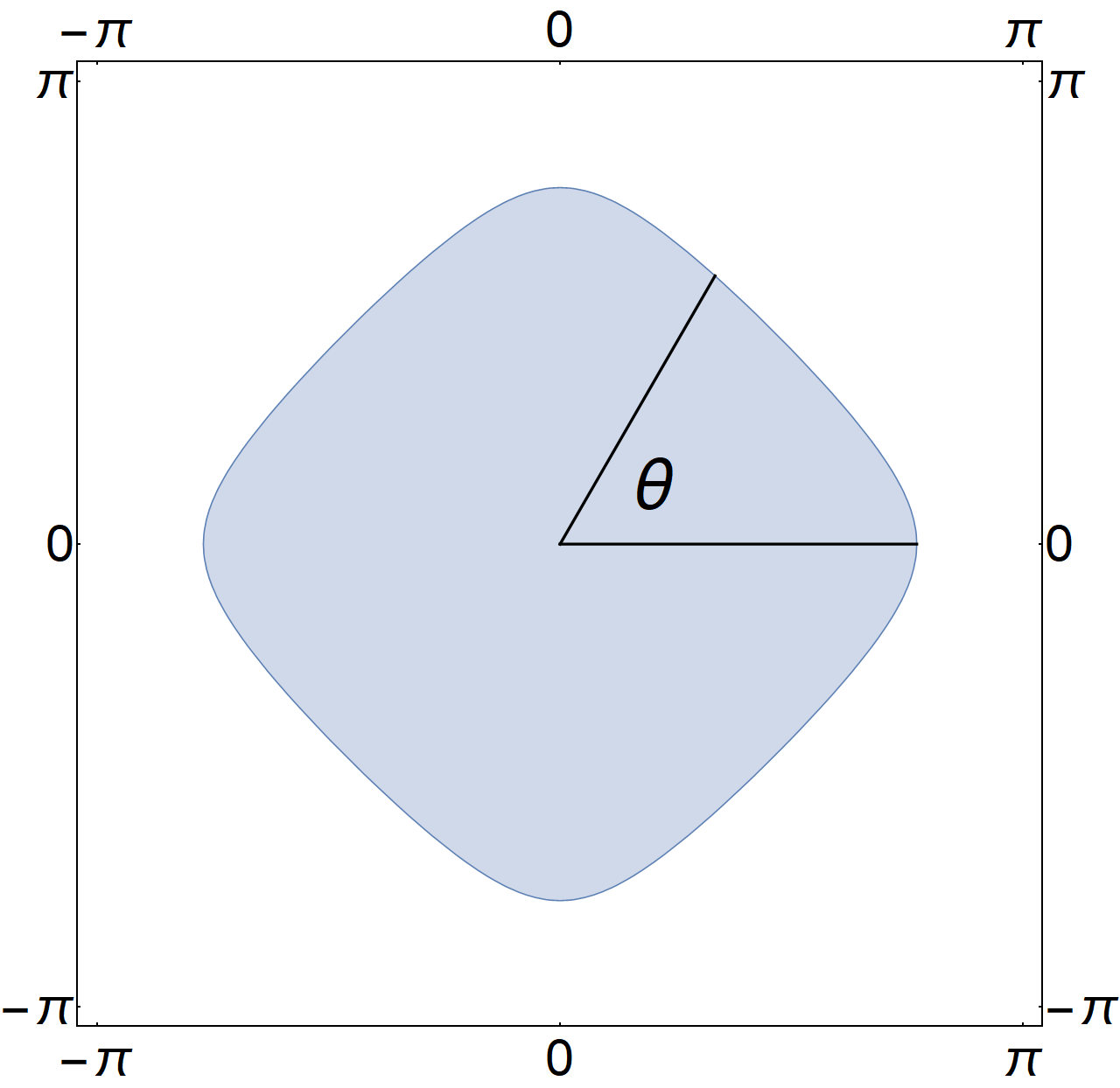}
    \caption{\label{fig:in-band}}
  \end{subfigure}
  \caption{\label{fig:bands} FS structure of the spin-fermion  and Ising-nematic metals. (\ref{fig:sf-band}) The FS structure of the SFM is 4 hole pockets, with ``hotspots'' connected by the AF $(\pi,\pi)$ vector.  (\ref{fig:in-band}) The electron FS in the nematic problem, with the FS angle $\theta$ denoted in the figure.}
  \label{fig:band-figs}
\end{figure}

Both the SFM and the INM are described within a single framework. We assume a 2D system of spinful fermions coupled to a single bosonic field. The
effective field theory is
\begin{equation}
  \label{eq:S-tot}
  \mathcal{S} = \mathcal{S}_f + \mathcal{S}_b + \mathcal{S}_I.
\end{equation}
It is the sum of  three terms -- a fermionic action $\mathcal{S}_f$, a bosonic action $\mathcal{S}_b$, and an interaction $\mathcal{S}_I$.
The fermionic action is
\begin{equation}
  \label{eq:S-f}
  \mathcal{S}_f = \int d\tau \sum_{\p,\sigma} \psi^\dagger_\sigma(\p,\tau)
  (\pd_\tau-\ek)\psi_\sigma(\p,\tau),
\end{equation}
where $\sigma$ sums over fermion spins and $\ek$ is the band dispersion. Fig. \ref{fig:bands} depicts schematically the FS
for the SFM and INM. The bosonic action is
\begin{equation}
  \label{eq:S-b}
  \mathcal{S}_b = \int d\tau \sum_{\q,j}\phi_j(\q,\tau)\frac{M^2_0 + |\q-\Q|^2}{D_0}\phi_j(\q,\tau),
\end{equation}
where $j=1\ldots N_b$ sums over $N_b$
boson components, $D_0$ is a constant with units of (area$\times$energy)$^{-1}$, and
measures the distance of the bosons from the QCP. We treat $M_0=M_0 (T)$ as a parameter, and our results don't explicitly depend on the temperature dependence of $M_0$, which can have e.g. Curie-Weiss form,
$M^2_0 \propto T-T_{0}$.
$\Q$ is the momentum at which a collective boson softens at a QCP: $\Q_{\sfm} = (\pi/a,\pi/a)$
and $\Q_{\inm} = 0$
($a$ is the interatomic spacing). The boson is assumed to have no bare dynamics, and acquires it solely from the interaction with fermions. (More accurately, the assumption is that the bare boson dynamics exists, but is irrelevant compared to the acquired one.)
The interaction is described by
\begin{align}
  \label{eq:S-I}
  \mathcal{S}_I &= g \int d\tau \sum_{\begin{array}{c} \q\p\\j\sigma\sigma'\end{array}}\phi_j(\q)t^j_{\sigma\sigma'}f(\p) \nn\\
                &\qquad\qquad\qquad\psi_\sigma^\dagger\left(\p-\frac{\q}{2}\right)\psi_{\sigma'}\left(\p+\Q+\frac{\q}{2}\right).
\end{align}
where $g$ is the coupling  constant, and $t^j_{\sigma\sigma'}, f(\p)$ express the spin and momentum form-factors of the interaction.

\paragraph*{Spin-fermion model--}
\label{sec:spin-fermion-model}

In the SFM the bosons represent spin fluctuations, so $\phi$ has up to three components, $t^j=\tau^j$ are Pauli matrices, and $f(\p) \approx 1$.
The momentum $\Q = (\pi/a,\pi/a)$
couples
$N=8$ discrete points on the FS, in sets of two (see Fig. \ref{fig:in-band}). Consequently
the most
relevant fermionic degrees of freedom are the ones near these ``hot spots''.
The interaction between  hot fermions and a $(\pi/a,\pi/a)$ boson has the form
\begin{align}
  \label{eq:sf-int}
  \mathcal{S}^{\sfm}_I &= \int d\tau ~g\sum_{\q\p l\sigma\sigma'}\vec{\phi}(\Q+\q)\cdot\nn\\
                     &\qquad\psi_{l \sigma}^\dagger \left(\p-\frac{\q}{2}\right)\vec{\tau}_{\sigma\sigma'}\psi_{l\sigma'}\left(\p+\Q+\frac{\q}{2}\right),
\end{align}
where $l=1\ldots N$ label the hot spots.

\paragraph*{Ising-nematic model--}
\label{sec:ising-nematic-model}

In the INM, the
boson  represents a nematic deformation along one of two symmetry axes. Accordingly, $\phi$ is a scalar.
Because $\Q = 0$, particles on the entire FS
participate  in the critical  dynamics, and the interaction has the form
\begin{equation}
  \label{eq:inm-int}
  \mathcal{S}^{\inm}_I = \int d\tau ~g\sum_{\q,\p,\sigma}\phi(\q)\psi_{\sigma}^\dagger (\p-\q/2)f(\p)\psi_{\sigma}(\p+\q/2),
\end{equation}
where
$f(\p) = (\bar{k}_F)^{-2}(p_x^2-p_y^2)$ encodes the nematic form-factor and $\bar{k}_F = \tpp^{-1}\int d\theta \kf(\theta)$ is the FS averaged Fermi vector, where $\theta$ traces out
a direction
on the FS (see Fig. \ref{fig:in-band}). Near the FS, $f(\p)$ can be approximated  as a function of
$\theta$:
$f(\theta) = (\kf(\theta)/\bar{k}_F)^2\cos2\theta$.

\subsection{Review of the diagrammatic theory}
\label{sec:review-t=0-theory}

We now briefly review diagrammatic perturbation theory for the two models.
The dynamics of fermions and bosons is encoded in their self-energies $\Sigma(\omega,\p)$ and $\Pi(\Omega,\q)$ (the latter is also called a polarization bubble).
We consider the self-energies on the Matsubara axis, where  $\omega = \omega_m = 2\pi(m+1/2)$ and $\Omega =\Omega_m = 2\pi m$.
As we noted in the Introduction, we compare the results on the Matsubara axis
with the QMC data.
We use latin letters to denote frequency-momentum 3-vectors, $q=(\Omega_m,\q), k=(\omega_m,\k)$. The self-energies are  related to bosonic and fermionic propagators as
\begin{equation}
  \label{eq:G-def}
  G(k) = \left(i\w_m + i\Sigma(k) - \epsilon(\k)\right)^{-1}
\end{equation}
and
\begin{equation}
  \label{eq:D-def}
  D(q) = D_0\left(M^2_0 + (\q-\Q)^2 + \Pi(q)\right)^{-1}
\end{equation}
Both self-energies
are represented by
diagrammatic series in  the effective coupling ${\bar g} = g^2 D_0$. We assume that
${\bar g}$ is small compared to the Fermi energy $\ef$. Then we can safely restrict to the lowest-order expansion in $\q-\Q$ in Eq. (\ref{eq:D-def}) and restrict $\k$ in Eq. (\ref{eq:G-def}) to be near the FS.

\begin{figure}
  \centering
  \includegraphics[width=\hsize]{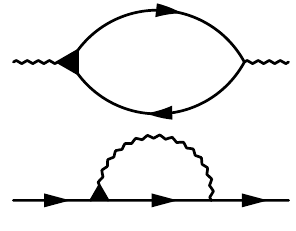}
  \caption{Diagrammatic expressions for $\Pi(q)$ (top) and $\Sg(k)$ (bottom). Solid lines are fully dressed fermion propagators $G(k)$, wavy lines are fully dressed boson propagators $D(q)$, and solid triangles are fully dressed vertex functions.}
  \label{fig:diags}
\end{figure}
The diagrammatic expressions for  $ G(k)$ and $\Pi (q)$ are presented in Fig. \ref{fig:diags}. Thick lines in this figure are fully dressed $G(k)$ and
$D(q)$
and the solid triangles are fully renormalized vertices.
In the ET, as well as in
several other computational methods,
e.g.
fluctuation-exchange
(FLEX) and dynamical mean-field approximations,
vertex corrections are neglected.
As stated in the Introduction, for the cases when  bosons are collective modes of fermions, there is no small parameter to justify this step, unless one extends the theory to, e.g.,
a large number of fermionic flavors. Furthermore, vertex corrections coming from low-energy fermions are logarithmically singular.  However,  in both SFM and INM these corrections  are rather small numerically at frequencies above
superconducting $T_c$, and we proceed by just neglecting them without further discussion.  There are also vertex corrections coming from high-energy fermions (the ones with energies of order bandwidth). These  corrections are regular and, to first approximation, can be absorbed into the renormalization of the coupling ${\bar g}$.

Without vertex corrections, $\Sigma (\k)$ and $\Pi (\q)$ are self-consistently expressed as
\begin{align}
  i\Sigma(k) &= N_b \gb T \sum_n \int \frac{d^2p}{\tpp^2}f^2(\k+\p/2)G\left(p+k\right)D\left(p\right),\label{eq:1loop-sig} \\
  \Pi(q) &= 2\gb T \sum_{n}\int \frac{d^2p}{\tpp^2}f^2(\p+\q/2) G\left(p+q\right)G\left(p\right). \label{eq:1loop-Pi}
\end{align}
Here the factor $b$ in $\Sg$ comes from summation over boson components and the factor 2 in $\Pi$ comes from spin summation.

The r.h.s.'s of Eqs. (\ref{eq:1loop-sig}) and (\ref{eq:1loop-Pi}) contain  integrals over fermionic momenta.
In principle, these integrals are over the whole Brillouin zone.
In the ET for fermion-boson models it is further assumed
that the contributions from high-energy fermions to  the r.h.s. of Eqs. (\ref{eq:1loop-sig}) and (\ref{eq:1loop-Pi}) can be absorbed into
${\bar g}$. This renormalization of the coupling is in addition to the ones discussed above due to high energy static vertex corrections.
Therefore, we restrict to
contributions from only low-energy fermions treating ${\bar g}$ in
(\ref{eq:1loop-sig}) and (\ref{eq:1loop-Pi}) as the effective, dressed couplings, which include both vertex corrections and the contributions from
high-energy fermions to the r.h.s.'s of  (\ref{eq:1loop-sig}) and (\ref{eq:1loop-Pi}). The dressed couplings in
these two equations are generally not equal, but the non-equivalence only affects the numerical factors in the formulas below, and for simplicity we keep the same ${\bar g}$ in the equations for $\Sigma (k)$ and $\Pi (q)$.

The static contribution to the polarization $\Pi(q)$ serves to renormalize the parameters $D_0$ and $M_0$ in Eq.~\eqref{eq:D-def}.
In particular, it shifts the bosonic mass to
\begin{equation}
  \label{eq:m-def}
  M^2 = M_0^2 -  \Pi(\W_m=0,\q=\Q).
\end{equation}
The dynamical contribution, which gives rise to Landau damping of a boson, can be obtained most straightforwardly
by first integrating in  Eq. \eqref{eq:1loop-Pi} over momentum
and then summing up over Matsubara frequencies \cite{Abrikosov1975}. We shift the incoming bosonic momentum to the vicinity of the ordering vector $\q \to \Q+\q$, and
symmetrize Eq. \eqref{eq:1loop-Pi} by shifting $\p \to \p - \q/2$. We
linearize the fermionic dispersion near the FS as $\epsilon(\p+\Q+\q) \approx \epsilon + \mathbf{v}_F \cdot \q$, where $\mathbf{v}_F$ is the Fermi velocity at $\p_F+\Q$, and $\p_F(\theta)=k_F(\theta)(\cos\theta,\sin\theta)$ is the Fermi vector with FS angle $\theta$. We then replace the momentum integration
by an
integral over the dispersion, as  $d^2p \approx d\theta \nu_F(\theta) d\epsilon$, where
$\nu_F(\theta)=k_F(\theta)/\vf(\theta)$ is the
density of states (DOS). We integrate over $\epsilon$ and obtain
\begin{widetext}
  \begin{align}
    \label{eq:Pi-low}
    &\Pi(\Q+\q,\W_m) = 2 i \gb T\sum_n\int \frac{d\theta}{2\pi} \nu_F(\theta)f^2(\theta) \frac{\Theta(\omega_n+\Omega_m)-\Theta(\omega_n)}{i(\W_m +\Sg(\w_n+\W_m)-\Sg(\w_n)) - (\epsilon(\p_F(\theta)+\Q+\q/2)-\epsilon(\p_F(\theta)-\q/2))},
  \end{align}
\end{widetext}
In Eq. \eqref{eq:Pi-low} $\epsilon(\p_F(\theta) + \cdots)$ should be understood as the
dispersion linearized near the FS.

The fermionic self energy  for a fermion on the FS,
i.e. for $|\k| = \kf(\theta_k)$, is obtained by
linearizing the dispersion of
an internal fermion
near the FS as
$\epsilon(\k+\Q+\p) \approx \epsilon(\k) +  v_F(\theta_{k+Q})|\p|\cos(\theta-\theta_{k+Q})$, where $\theta_{k+Q}$ is the FS angle of $\k+\Q$ and $\theta$ is the angle of $\p$.
Evaluating the angular integral,
we obtain
\begin{widetext}
\begin{flalign}
  \label{eq:sig-low}
  \Sigma(k) \approx
  N_b
  \gb T
  \sum_n \int \frac{|\p| dp}{\tp}\frac{\sigma(\w_n)}{\sqrt{(\w_n+\Sigma(\w_n))^2+[v_F(\theta_{k+Q})]^2 |\p|^2}}
  \frac{f^2(\theta_{k+Q/2})}{M^2+|\p|^2+\Pi(|\p|\hat n(\theta_{k+Q}),\w_m-\w_n)},
\end{flalign}
\end{widetext}
where
$\hat n(\theta_{k+Q}) = \left.\frac{(-{\mathbf{v}_F}_y,{\mathbf{v}_F}_x)}{v_F}\right|_{\theta=\theta_{k+Q}}$ is a unit vector pointing parallel to the FS at the angle $\theta_{k+Q}$.
$\sigma(x)$ is the sign function.

Eqs. \eqref{eq:Pi-low} and \eqref{eq:sig-low} form a self-consistent set of equations for the low energy dynamics of fermions near the FS. We next review how they are rendered analytically tractable in the ET.

\subsection{Eliashberg theory at $T=0$}
\label{sec:t=0-theory}
The main technical simplification step in ET
\cite{Altshuler1994,Chubukov2005,Abanov2003,Rech2006}
is
the
additional assumption that at $T=0$ the
\emph{typical} momentum transfer of a fermion near its mass shell $v_F(|\k|-\kf) \sim \omega + \Sigma(\omega)$ is much smaller than the typical momentum transfer of a boson, i.e.
\begin{equation}
  \label{eq:eliash-assump}
  |\w + \Sg(\w)| \ll \vf |\q|,
\end{equation}
where $\w$ is a typical fermionic frequency and $|\q|$ is a typical bosonic momentum.
As noted above, this allows one to factorize the momentum interaction between directions along and transverse to the FS.  In addition,  the ET assumes that relevant frequencies are much smaller than $E_F$  and integrates over momenta in infinite limits.

We will first show the results within the ET and then
discuss its validity
for the SFM and the INM.

We begin with the bosonic self-energy.
In Eq. \eqref{eq:Pi-low}, the dominant contributions come from regions where $\delta\epsilon = \epsilon(\p+\Q+\q/2) - \epsilon(\p-\q/2)$ vanishes. This is the source of the different bosonic dynamics for the SFM and INM. In the SFM, $\delta\epsilon$ vanishes near the hotspots
as $\delta\epsilon \approx v_F(\theta_{hs})Q_{hs}\sin(\theta-\theta_{hs})$, where $\theta_{hs}$ is the FS angle of a hotspot and
$Q_{hs}=k_F(\theta_{hs})|{\mathbf{v}_F}^2_y-{\mathbf{v}_F}^2_x|/\vf^2$.
Expanding near the hotspots, replacing the Matsubara sum by an integral and integrating,
we obtain
\begin{equation}
  \label{eq:Pi-SFM}
  \Pi^{\sfm} (\Omega_m) \approx {\bar g}
  \frac{\nu_F(\theta_{hs})N}{2\pi}\frac{|\W_m|}{v_F(\theta_{hs})Q_{hs}}.
\end{equation}
In the INM, $\delta\epsilon \approx v_F(\theta)q\cos(\theta-\theta_q)$ depends strongly on $|\q|,\theta_q$, and is dominated by $\theta \approx \theta_q \pm \pi/2$, yielding
\begin{equation}
  \label{eq:Pi-INM}
  \Pi^{\inm} (\Omega_m, \q) \approx {\bar g} \frac{\nu_F(\theta_q)}{\pi}f^2(\theta_q)\frac{|\W_m|}{v_F(\theta_q)|\q|}.
\end{equation}
We emphasize that in both cases the result does not depend on fermionic $\Sigma (\omega_m)$. This is because in ET the polarization bubble is the convolution of two DOS's, and each DOS $\nu(\omega) = -(\nu_F /\pi) {\text Im} \int d\k G(\omega,\k) = \nu_F \sigma (\omega)$ is independent of $\Sigma (\omega)$.

We now turn to the fermionic self-energy.
In Eq. \eqref{eq:sig-low}, we can approximate $(\sqrt{(\w_m+\Sg(\w_m))^2+\vf^2|\p|^2})^{-1} \approx (\vf |\p|)^{-1}$,
because of the assumption of ET, Eq. \eqref{eq:eliash-assump}. This term then  just contributes a factor $\propto |\p|^{-1}$ to the integral, which reduces the effective dimensionality of the $\p$ integral and renders it one dimensional. Physically this means that the boson momentum $\p$ is confined to be parallel to the FS. Then the momentum and frequency integrals are straightforward.
For the SFM at the QCP $M = 0$
we obtain
\begin{equation}
  \label{eq:sig-SFM}
  \Sigma^{\sfm}(\w_m) \approx \w_{\sfm}^{1/2} |\w_m|^{1/2}\sigma(\w_m)
\end{equation}
where
$\w_{\sfm} = \frac{
N_b^2 \gb Q_{hs}}{2\pi N \kf(\theta_{hs})}$, while for the INM we get,
\begin{equation}
  \label{eq:sig-INM}
  \Sigma^{\inm}(\w_m) \approx \w_{\inm}^{1/3} |\w_m|^{2/3}\sigma(\w_m).
\end{equation}
where
$\w_{\inm} = \frac{\gb^2f^4(\theta_q)}{8\pi^23^{3/2}\nu_F(\theta_q)\vf^2(\theta_q)}$.

Now we check the justification of the assumption we made, Eq. \eqref{eq:eliash-assump}. Consider first the SFM. In both Eqs. \eqref{eq:Pi-low} and \eqref{eq:sig-low} the Heaviside and sign functions limit internal Matsubara frequencies to be on order of the external frequency. As we discussed in the introduction, both the NFL behavior and superconductivity in the SFM, emerge at typical external frequencies on order of $\w \sim \w_\sfm \sim \gb/N$, see Eq. \eqref{eq:sig-SFM}. In the bosonic self energy, the typical bosonic momentum is $\vf Q_{hs} \sim \ef$ so the assumption is well justified as long as $\gb \ll \ef$. On the other hand, in the fermionic self energy, the typical boson momentum is $\vf p\sim \sqrt{\Pi(\w_\sfm)} \sim \gb $, so the bosonic and fermionic momenta are of the same order unless $N \gg 1$. In practice, $ N = 8$ of the SFM is sufficiently large that there is just a reduction of the scale $\w_\sfm$ by a factor of order one. In the INM, the typical frequencies are of order $\w \sim \w_\inm \sim \gb^2/\ef \ll \ef$, while the typical bosonic momenta are of order $\vf p \sim \sqrt{\Pi(\w_\inm)} \sim \gb$. Thus, the Eliashberg condition is well obeyed for the INM.

We see that at low enough temperatures, all fermionic self energies scale as power-laws in $\omega_m$, with no additional temperature dependence. Therefore, the fermionic self energies at different temperatures should collapse onto one another, and should vanish at the lowest frequencies. Similarly, the bosonic self energies in the SFM should collapse onto one another, while in the INM they should scale with $1/|\q|$. In the next section, we show that finite temperature effects destroy these scaling properties.

\section{Self-energies at finite temperatures}
\label{sec:self-energies-at}

We now turn to study the bosonic and fermionic self energies at finite temperatures. We do not assume that the typical fermionic momentum is smaller than bosonic momentum. In addition, we do not replace the Matsubara sum by a frequency integral, which means that we must account for the thermal piece which we discussed in the Introduction.

We start with the fermionic self-energy. We explicitly split the Matsubara sum into a thermal and nonthermal part and rewrite $\Sigma$ as,
\begin{equation}
  \label{eq:sigma-split}
  \Sigma(\w_m) \approx \Sigma_T(\w_m) + \Sg_Q(\w_m),
\end{equation}
where $\Sigma_T(\omega_m)$ contains only the $\omega_m = \omega_n$ term in the sum, and $\Sigma_Q(\w_m)$ accounts for the nonthermal dynamical contributions. $\Sigma_T$ should be understood as the thermal scattering rate of the fermions, akin to scattering due to static disorder.
As a guide for developing a useful approximation, we notice that the Eliashberg $T=0$ results, Eqs.~(\ref{eq:Pi-SFM}--\ref{eq:sig-INM}), are actually the same as what would be obtained without self-consistency, i.e. the Eliashberg approximation reduces to the one-loop result. Therefore, we will assume that the thermal $\Sigma_T$ must be evaluated self-consistently, but $\Sg_Q$ does not and only requires the self-consistent $\Sigma_T$ as an input. We first present the results and then discuss the validity of the approximation.

The self-energy $\Sigma_T$ is the solution of the self-consistent equation
\begin{align}
  \label{eq:gamma-eq}
  \Sigma_T(\w_m,\theta_k) &\approx
                            N_b\gb T \nn\\
                          &\quad\int \frac{|\p| dp}{\tp}\frac{\sigma(\w_m)}{\sqrt{(\w_m+\Sigma_T(\w_m))^2+\tilde{v}_F^2|\p|^2}}
                            \frac{f^2(\theta_k)}{m^2+|\p|^2} \nn \\
                          &= \frac{
                            \gb T f^2(\theta_k)\sigma(\w_m)}{\tp |\w_m+\Sigma_T(\w_m)|}\mathcal{S}\left(\frac{\vf M}{|\w_m+\Sigma_T(\w_m)|}\right),
\end{align}
where
\begin{equation}
  \label{eq:scaleF-1}
  \mathcal{S}(x) = \frac{\cosh^{-1}(1/x)}{\sqrt{1-x^2}}\approx\left\{
  \begin{array}{ll}
    \log(2/x) & x \ll 1 \\
    \pi/(2x)  & x \gg 1
  \end{array}\right. .
\end{equation}
and $\tilde{v}_F = \vf(\theta_{k+Q})$, i.e. $\tilde{v}_F = |\vf(\theta_{hs})|$ for the SFM and $\tilde{v}_F = \vf(\theta_k)$ for the INM.
Recall that we set $f(\theta) =1$ for the SFM. For the INM, the typical momentum transfers along the FS are small, and this explains the presence of $f^2 (\theta_k)$ in the r.h.s. of (\ref{eq:gamma-eq}).

\begin{figure}
  \centering
  \includegraphics[width=0.85\hsize]{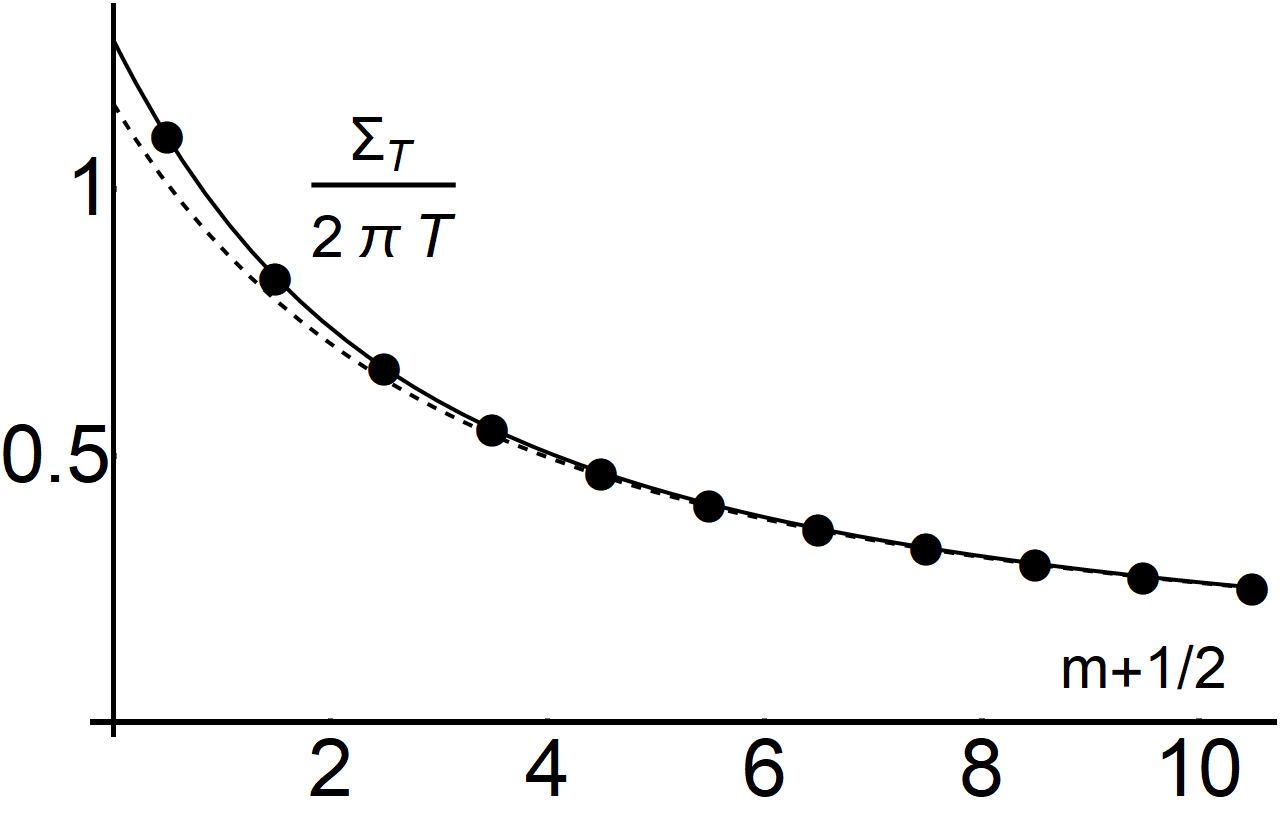}
  \caption{The thermal contribution $\Sigma_T(\w_m)$ to the fermionic self-energy. The black dots are a numerical solution of Eq. \eqref{eq:gamma-eq}. The dashed line is the analytic approximation Eq. \eqref{eq:gamma-sol} and the solid black line is a solution with higher accuracy, up to $\log\log\log(\cdots)$ terms. The numerical solution was obtained for $ \frac{\tp T}{\vf M }= \frac{\gb}{\tp T} = 10$.}
  \label{fig:GammaT}
\end{figure}

We assume that close enough to the QCP, $\vf M \ll |\w_m|+|\Sigma_T(\w_m)|$ for all Matsubara numbers
 (this will allow us to obtain analytic expressions in what follows).
Using the form of $S(x)$ in this limit,
we can solve Eq. \eqref{eq:gamma-eq} to obtain an analytic expression for
 $\Sg_T(\w_m)$. For $\w_m > 0$,
\begin{align}
  \label{eq:gamma-sol}
  &\Sigma_T(\w_m,\theta_k) \nn\\
  &\approx \sqrt{\omega_T(\theta_k)^2 \log \frac{\sqrt{4\omega_T(\theta_k)^2+\w_m^2} + \w_m}{\vf M} + \frac{\w_m^2}{4}}- \frac{\w_m}{2} \nn\\
  &= \left\{
    \begin{array}{ll}
      \omega_T(\theta_k)
      - \frac{\w_m}{2}& \w_m \ll \omega_T(\theta_k)
      \\
      \frac{\omega_T(\theta_k)^2}{\w_m}
                      & \w_m \gg \omega_T(\theta_k)
    \end{array}
\right.
\end{align}
where
\begin{align}
  \label{eq:eta-def}
  \w_T(\theta_k)  &=  \w_T^0 |f(\theta_k)|\sqrt{\log\frac{2\w_T^0|f(\theta_k)|}{\vf M}},
  \w_T^0 = \sqrt{\frac{N_b \gb T}{2 \pi}}
\end{align}
and we neglected $\log(\log(\cdots))$ terms.
We show the result in Fig. \ref{fig:GammaT}.
The crossover between the two asymptotic behaviors of $\Sigma_T$ occurs at $\omega_m \sim \omega_T$, i.e., at a typical Matsubara number
\begin{equation}
  \label{eq:matsu-cross}
   m_T \equiv \frac{\w_T}{T} \sim \sqrt{\frac{\gb}{T}} \sqrt{\log\frac{\sqrt{\gb T}}{v_F M}}.
\end{equation}

Next, we determine the dynamic contribution $\Sg_Q$  -- the sum over non-zero bosonic frequencies in Eq. (\ref{eq:sig-low}).
As we discussed earlier, this contribution does not require self-consistency as it is sufficient to replace $\Sigma (\omega_m)$ by  $\Sigma_T$ in the r.h.s. of (\ref{eq:sig-low}).
For the SFM we find,
\begin{align}
  \label{eq:sig0-SFM}
  & \Sigma_Q^{\sfm}(\w_m,\theta_k
    = \theta_{hs}) \approx \frac{
    N_b \gb T}
    {\tp} \nn \\
  &\qquad\sum_{n\neq m}\frac{\sigma(\w_n)} {|\w_n+\Sigma_T(\w_n)|}\mathcal{S}\left(\frac{ \vf \sqrt{M^2+\Pi^{\sfm}(\w_m-\w_n)}}{|\w_n+\Sigma_T(\w_n)|}\right),
\end{align}
where
$\mathcal{S}(x)$ is the same as in (\ref{eq:scaleF-1}), but the argument now contains the polarization bubble $\Pi^{\sfm}$ next to $M^2$. This implies that the crossover between two limiting forms of $\mathcal{S}(x)$ now holds even if $M=0$, i.e., there are two different behaviors of $\Sigma_Q^{\sfm}(\w_m,\theta_k)$ near a QCP.
To find out at what $m$ the crossover occurs, we need to know $\Pi^{\sfm} (\w_m-\w_n,Q_{hs})$ at a finite temperature.  Examining  Eq. \eqref{eq:Pi-low} for $\Pi (\Omega_m,Q_{hs})$ we find that corrections to
the zero-temperature form, Eq. (\ref{eq:Pi-SFM}), are suppressed by powers of  $T/E_F$  and are therefore irrelevant. Therefore, we simply plug the $T=0$ result into (\ref{eq:sig0-SFM}).
We assume and then verify that the argument of $\mathcal{S}$ in (\ref{eq:sig0-SFM}) is of order one at  $m < m_T$,
where $\omega_m < \Sigma_T (\omega_m)$. Neglecting $\omega_m$ compared to $\Sigma_T (\omega_m)$, using the fact that
typical internal $n$ are comparable to $m$ and using  $\Sigma_T(\w_m)\sim\w_T\sim\sqrt{\gb T\log(\sqrt{{\bar g}T}/\vf M)}$,
$\sqrt{\Pi(\w_m-\w_n,Q_{hs})} \sim \sqrt{\nu_F \gb \omega_m/\vf Q_{hs}}$, and $\nu_F v_F/ Q_{hs} = O(1)$,
we find that
the crossover in $\Sigma_Q^{\sfm}(\w_m,\theta_{hs})$ occurs at $\omega_m \sim \omega'_T \sim \omega_T (T/\gb)^{1/2} \ll \omega_T$,  i.e., at a typical Matsubara number
\begin{equation}
  \label{eq:Sg-Q-cross-m}
  m_T' \sim
  \log\frac{\sqrt{\gb T}}{\vf M}.
\end{equation}
Clearly $m_T' \ll m_T$ at low enough temperatures. This justifies the use $m <m_T$ in the derivation of
(\ref{eq:Sg-Q-cross-m}).
The two limiting forms of  $\Sigma_Q^{\sfm}(\w_m,\theta_{hs})$ are
\begin{equation}
  \label{eq:sig0-SFM-low}
  \Sigma_Q^{\sfm}(\w_m,\theta_{hs})\approx \frac{
    \gb
    \w_m}{2\pi^2\w_T (\theta_{hs})
  }
  L(m,\w_m)
\end{equation}
at $\w_m \ll \w_T'$ , where
\begin{equation}
  \label{eq:F-def}
  L \approx  \log\left(\frac{\w_T
    }{\vf \sqrt{{\bar g}|\omega_m|}}\right),
\end{equation}
and
\begin{equation}
  \label{eq:sig0-SFM-hi}
  \Sigma_Q^{\sfm}(\w_m,
  \theta_{hs})
  \approx \w_{\sfm}^{1/2}\w_m^{1/2}.
\end{equation}
for  $\omega_m  \gg \omega'_T$. This form is the same
as at $T=0$.

We now combine our results for the thermal and quantum parts
of the self-energy
and show
that there are two asymptotic behaviors for $\Sg(\w_m)$ separated by a wide regime.

\paragraph*{Region I: the strongly thermal regime --}
\label{sec:region-i:-strongly}

The strongly thermal regime occurs at $\omega_m \ll \omega'_T \ll \omega_T$, i.e.,  $m \ll m_T'  \ll  m_T$. In this case, adding up the appropriate limits from Eqs. \eqref{eq:gamma-sol} and \eqref{eq:sig0-SFM-low} we find that the self energy has the form
\begin{equation}
  \label{eq:se-form-small-wm}
  \Sigma^{\sfm}(\w_m) \approx \w_T
  \left(1 + A \frac{\omega_m}{\w_T}
  \right)
\end{equation}
where the leading term comes from frequency-independent part of $\Sigma_T$, and the subleading term, with prefactor $A$, is the combination of
 frequency-dependent term in $\Sigma_T$ and from $\Sigma_Q$, Eqs. \eqref{eq:gamma-sol} and \eqref{eq:sig0-SFM-low}.
 At vanishing $M$, $A \approx -1/2$, and the full $ \Sigma(\w_m)$ comes from $\Sigma_T$. However, the $O(\omega_m)$ term from $\Sigma_Q$ is only logarithmically reduced and in practice contributes nearly equally to $A$.

\paragraph*{Region II: the almost critical regime --}
\label{sec:region-ii:-almost}

The almost
critical
regime occurs for $\omega_m \gg \omega_T \gg \omega'_T$, i.e., $m \gg m_T \gg  m_T'$.  Here,  summing up the appropriate expressions in Eqs. \eqref{eq:gamma-sol} and \eqref{eq:sig0-SFM-hi}, we find that the total fermionic self energy has the form
\begin{equation}
  \label{eq:se-large-omega}
  \Sigma^{\sfm} (\omega_m) \approx  \w_{\sfm}^{1/2}\w_m^{1/2} + \frac{\w_T^2}{\w_m}.
\end{equation}
The first term (the same one as at $T=0$) comes from $\Sigma_Q$, the second one comes from $\Sigma_T$.
Because $\w_{\sfm} \sim {\bar g}$, the first term is the dominant one for $m >m_T$, i.e., the self-energy is predominantly determined by dynamical quantum fluctuations.

Note that because $\Sigma_Q (\omega_m)$ changes its behavior at  $\omega_m \sim \omega_T'$ and $\Sigma_T (\omega_m)$ changes its behavior at $\omega_m \sim \omega_T$,
there is no  single crossover from the region where thermal self energy dominates to the one where the dynamical self energy dominates.
In between the two limiting regimes $\omega_m \ll \omega'_T$ and $\omega_m \gg \omega_T$,  there is a wide intermediate region of $\omega'_T \ll \omega \ll \omega_T$, ($m_T' \ll m \ll m_T$),  where
quantum and thermal contributions are comparable.
This is the result we stated in the Introduction, Eq. \eqref{eq:se-intro-approx}.
The implication of Eqs. \eqref{eq:se-large-omega} and \eqref{eq:se-form-small-wm}
is that the
behavior of $\Sigma$ at the critical point is
quite involved.
While at high frequencies the behavior will tend to a power law, at lower frequencies the self-energy saturates and appears to reach a plateau.

Before moving on to the INM, we need to go back and check whether we were justified in neglecting the contribution of
$\Sigma_Q$ in comparison with $\Sigma_T$ in the
self-consistent calculation of $\Sigma_T$ and subsequent calculation of $\Sg_Q$.
From Eqs. \eqref{eq:se-form-small-wm} and \eqref{eq:se-large-omega}, it is evident that in the strongly thermal region $\Sg_Q \ll \Sigma_T$ by a logarithmic factor, while in the almost critical regime
$\Sg_T \ll \w_m$, and we already know from ET (see Sec. \ref{sec:t=0-theory}) that the contribution of $\Sg_Q$ can be neglected when computing $\Sg_Q$.
In the intermediate regime
the approximation is not
rigorously  justified since
both
$\Sigma_T$ and $\Sg_Q$ scale as $\sqrt{\gb T}$, and both are larger than $\omega_m$, but it
serves as a way to interpolate between the two regimes.

For the INM, the behavior is somewhat more complex due to the momentum dependence of the polarization. The thermal contribution $\Sigma_T(\w_m)$ is identical
to that for the SFM and
is given by Eq. \eqref{eq:gamma-sol}. The dynamical contribution is, from Eq. \eqref{eq:sig-INM},
\begin{align}
  \label{eq:sig0-INM}
  &\Sigma_Q^{\inm}(\w_m) \approx \frac{
    \gb T f^2(\theta_k)}{\tp} \nn \\
  &~~~\sum_{n\neq m}\frac{\sigma(\w_n)} {|\w_n+\Sigma_T(\w_n)|}\mathcal{T}\left(\frac{\vf M}{p_n},\frac{(\Pi^{\inm}(\w_n-\w_m, p_n))^{1/3}}{p_n^{2/3}}\right),
\end{align}
where $\vf p_n = |\w_n +\Sigma_T(\w_n)|$, $\Pi^{\inm}(\Omega_m, q)$ is given by Eq. (\ref{eq:Pi-INM}), and
\begin{align}
  \label{eq:T-def}
  \mathcal{T}(x,y) &= \int_0^{\infty} \frac{z^2 dz}{\sqrt{z^2+1}(z^3+ z x^2 + y^3)} \nn \\
                   &=\left\{
                     \begin{array}{lll}
                       \log(2/x) & y=0,x \to 0 \\
                       \log(2/y) & x=0, y \to 0 \\
                       2\pi/(3\sqrt{3}y) & x=0, y \gg 1
                     \end{array}\right. .
\end{align}
A
similar analysis to the one we performed for the SFM shows that at $M =0$, $\Sigma_Q^{\inm}(\w_m)$ undergoes a crossover at
$\omega'_T \sim \omega_T (\omega^2_T/({\bar g} E_F)) \sim \omega_T (T \log{(\sqrt{{\bar g} T}/v_F M)}/E_F)$
between $\Sigma_Q^{\inm}(\w_m) \sim (\omega_m/\omega_T) \log{(\nu_F {\bar g} \w_m v^2_F/\omega^3_T)}$ at $\omega_m < \omega'_T$ and
$\Sigma_Q^{\inm} (\w_m) \sim \omega^{2/3}_m (\omega_{\inm})^{1/3}$ at  $\omega_m > \omega'_T$. The crossover frequency $\omega'_T$ is different from the one for the SFM but like in that case,
$\omega'_T \ll \omega_T$.
Comparing the results for $\Sigma_Q^{\inm} (\w_m)$ and $\Sigma_T (\w_m)$, Eq.
\eqref{eq:scaleF-1}, we obtain the limiting forms of $\Sigma^\inm$ in region $I$, $\omega_m < \omega'_T$ and region II, $\omega_m > \omega_T$:
\begin{equation}
  \label{eq:se-INM-forms}
  \Sg(\w_m) \approx \left\{
    \begin{array}{ll}
      \w_T
      \left(1 + B \frac{\omega_m}{\w_T}\right) & \text{region I}\\
      \w_{\inm}^{1/3}\w_m^{2/3}+ \frac{\w_T^2}{\w_m} & \text{region II}     \end{array}
  \right.,
\end{equation}
where $B$ is a constant.
(Note that since $\w_T'/\w_T \sim \w_T^2/\gb \ef$, for low enough temperatures $\w_T' < T$ and region I is inaccessible.)

We now discuss the bosonic self energy
in more detail. From Eq. \eqref{eq:Pi-low} we see that corrections to the $T=0$ form of $\Pi$ comes from the term $\w_m + \Sg(\w_m)$, which means that the behavior of $\Pi$ depends on whether we are in region I or II. In region II, we may neglect self-energy corrections to $\Pi$ and it will retain its $T=0$ form. In region I, there is a different behavior for the SFM vs. the INM.
In the SFM, the typical momentum transfer is of order $v_FQ_{hs} \sim \ef$, hence $v_FQ_{hs} \gg \w_T$ even at finite $T$, and the $T = 0$ result holds. In the INM, the
 typical momentum transfer is $\vf |\q| \sim (\gb\ef T)^{1/3} \gg \w_T$, so the the $T=0$ result can be used in evaluating the fermionic self energy.
 However, since QMC simulations also measure $\Pi(\W_m,|\q|)$ for a given \emph{external} $\W_m, |\q|$, and since there is a large parameter range where
$|\w_m| \ll v_F|\q| \ll \w_T$, we need to calculate $\Pi$ taking into account the self-energy contribution.

For the
computation $\Pi$ in the INM,
it is important to realize that a nematic order parameter
is not a conserved quantity, such as e.g. a spin order parameter (the total magnetization)  in a ferromagnet. For a conserved order parameter, a Ward identity insures that within ET vertex corrections cancel out exactly, so that the one-loop calculation which is analogous to Eq. \eqref{eq:Pi-INM} is exact\footnote{In actuality, it is only exact for $\W_m/q \to 0$ or $\W_m/q \to \infty$, but because of the constraint imposed at these two limits, the corrections for finite $\W_m/q$ are at most of order one.}. In the INM this cancelation does not occur, hence at low frequencies self-energy corrections become important. This is true even in the $T=0$ limit\cite{Punk2016}.

To gain a qualitative understanding of the impact of the self-energy we evaluate Eq. \eqref{eq:Pi-low}, keeping only the thermal part of the fermionic self-energy. To leading order in $\w$ we may treat the self-energy as a constant
$\Sigma_T(\w_n,\theta_k) \approx \w_T(\theta_k)\sgn(\w_n)$. We obtain,
\begin{widetext}
  \begin{align}
    \label{eq:Pi-INM-T-0}
    \Pi^{in}(\W_m,\q) &\approx 2i\gb T \sum_n\int \frac{d\theta}{\tp} \nu_F(\theta) f^2(\theta)
                        \frac{\Theta(\w_{n+m})-\Theta(\w_n)}{i\w_T(\theta)(\sgn(\w_{n+m})-\sgn(\w_n)) - \vf(\theta)q\cos(\theta-\theta_q)} \\
                      &\approx \frac{\gb}{2\pi^2}|\W_m| \int d\theta \nu_F(\theta)f^2(\theta)\frac{2|\w_T(\theta)|}{4\w_T^2(\theta) + \vf^2(\theta)q^2\cos^2(\theta-\theta_q)} \nn\\
                      &\approx
                        \gb\frac{\tilde \nu_F}{\tp} \frac{|\W_m|}
                        {\w_T(\theta=0)} \label{eq:Pi-INM-T}
  \end{align}
\end{widetext}
where $\tilde \nu_F = \tpp^{-1}\int \nu_F(\theta)|f(\theta)|$,
and the last line is valid for $q \to 0$.
The detailed behavior of $\Pi$ as a function of $\W_m,\Sg(\w_m),q$ and $\theta$ is more complicated and is obtained by a numerical integration of Eq. \eqref{eq:Pi-INM-T-0}.

Eqs. \eqref{eq:sig0-INM} and \eqref{eq:Pi-INM-T} for the INM  and Eqs. \eqref{eq:Pi-SFM} and \eqref{eq:sig0-SFM-low} for the SFM  show remarkably similar behavior. The fermionic response is the same for both
systems, up to some  model-dependent constants. The bosonic response is featureless and linear in $\W_m$,
which implies that the fermionic self-energy will be qualitatively the same as for the SFM in both models.

\section{Lattice theory}
\label{sec:lattice-theory}

Before beginning our analysis of the QMC data, we first briefly describe the lattice version of the self consistent equations \eqref{eq:1loop-sig} and \eqref{eq:1loop-Pi}, which we refer to as the lattice theory (LT). As we discussed there, comparing ET, LT and QMC results
gives us insight into
the role of high energy fermions in contributing to the self energy.

The SFM and INM are defined on a finite space-time lattice of $L\times L \times L_\tau$ sites with periodic boundary conditions in the spatial directions, and periodic (antiperiodic) boundary conditions for the bosons (fermions) in the imaginary time direction.
For both models, the bosonic part of the action is given by the lattice version of Eq.\eqref{eq:S-b}, with an additional dynamical term
\begin{equation}
  \label{eq:lattice-Sb}
  \mathcal{S}_b = D^{-1}_0\int d\tau \sum_{\q}\phi(\q)(M^2_0 - \partial^2_x - \partial^2_y - c^{-2}\partial^2_\tau)\phi(\q).
\end{equation}
Here $\partial_{x,y,\tau}$ is understood as a discretized derivative, and $c$ is the bare velocity of the boson.
We introduce this additional dynamical term to better match the QMC lattice models, where the bosonic fields have their own independent dynamics.

In the Ising-nematic case, the form factor is slightly modified compared to the definition used in Sec.~\ref{sec:ising-nematic-model},
\begin{equation}
  \label{eq:lattice-fk-INM}
  \mathcal{S}_{\mathrm{I}}=\frac{g}{L}\int d\tau\sum_{\mathbf{kq}\sigma}\psi^{\dagger}_{\sigma}(\mathbf{k}-\mathbf{q})\psi_{\sigma}(\mathbf{k})\phi(\mathbf{q})f(\mathbf{k},\mathbf{q}),
\end{equation}
where $f(\k,\q)=\cos\left(q_x/2\right)\cos\left(k_x-q_x/2\right)-\cos\left(q_y/2\right)\cos\left(k_y-q_y/2\right) = \cos k_x -\cos k_y + O(q)$.

The self energies then take the form
\begin{eqnarray}
  i\Sigma(k)=\frac{g^2T}{L^2}\sum_{q}f^{2}(\mathbf{k},\mathbf{q})D(q)G(k+q)\\
  \Pi(q)=\frac{2g^{2}T}{L^2}\sum_{k}f^{2}(\mathbf{k},\mathbf{q})G(k)G(k+q)
\end{eqnarray}
The fermionic part of the action is given by Eq. \eqref{eq:S-f}, where $\epsilon_\mathbf{k}$ is a nearest-neighbor tight-binding dispersion.

\begin{figure}
  \centering
  \includegraphics[width=0.5\hsize]{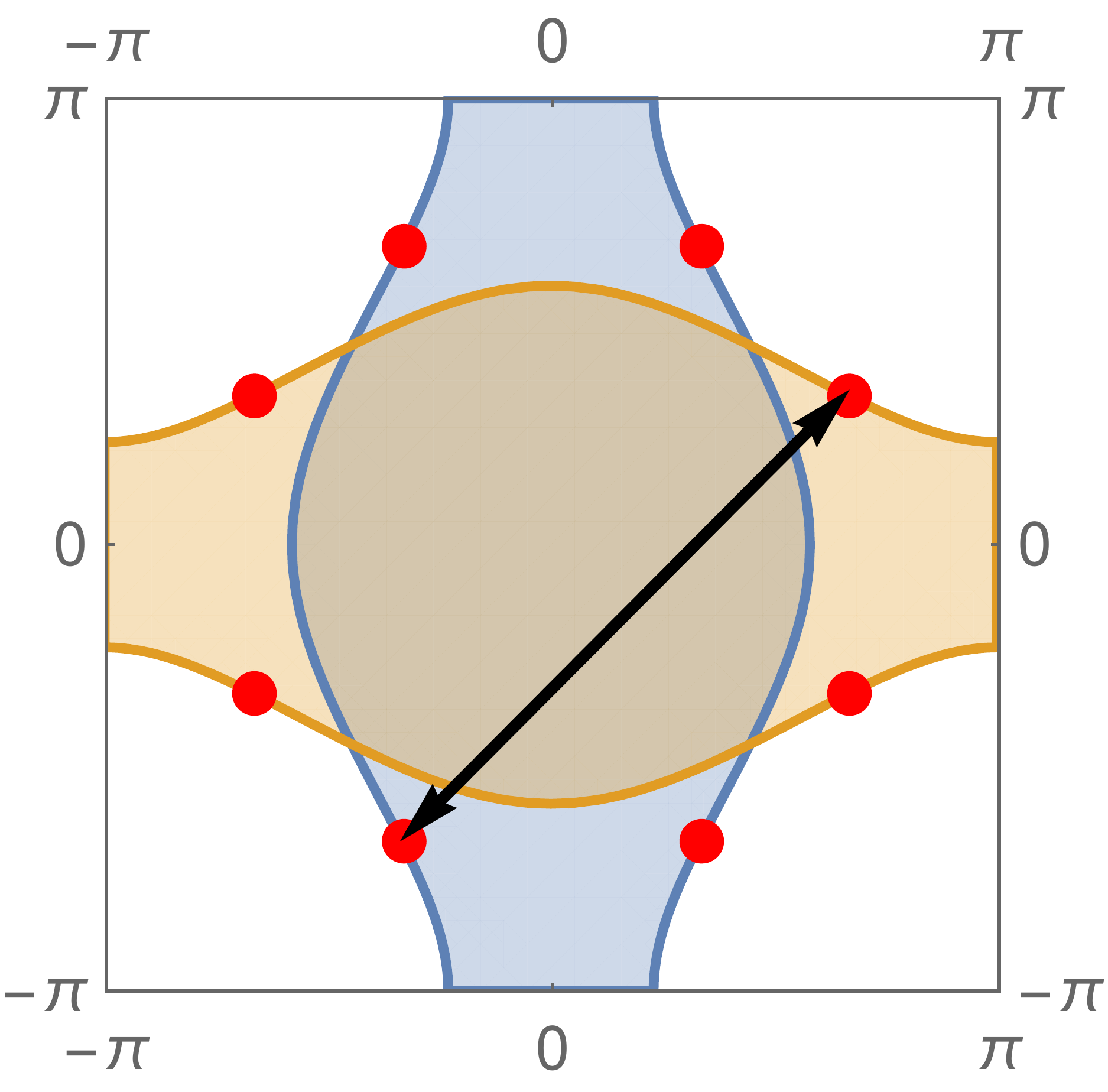}
  \caption{FS structure
    of the two-band spin-fermion model used in the QMC simulations and the LT
    calculations. Hotspots are indicated in red, and are connected by the AF wavector $\mathbf{Q}=(\pi,\pi)$, denoted by an arrow.}
  \label{fig:sfm-fs2}
\end{figure}

In the spin-fermion case, we consider the two-band model
depicted in Fig. \ref{fig:sfm-fs2}. Again, the reason for this is to better match the sign problem-free model that was simulated in QMC. The fermionic part of the action is
\begin{equation}
  \mathcal{S}_{f} = \int d\tau \sum_{\mathbf{k}\eta\sigma}\psi_{\eta\sigma}^{\dagger} (\mathbf{k})(\partial_\tau -\epsilon_{\eta,\mathbf{k}}) \psi_{\eta\sigma}(\mathbf{k}),
\end{equation}
where $\eta=\pm 1$ is the band index.
The interaction part of the action is
\begin{equation}
  \mathcal{S}_{I} = \frac{g}{L}\int d\tau\sum_{\mathbf{kq}\sigma\sigma'\eta} \vec{\phi}(\mathbf{Q}+\mathbf{q})\psi^\dagger_{\eta\sigma}(\mathbf{k}) \vec\tau_{\sigma\sigma'}\psi_{-\eta\sigma'}(\mathbf{k}+\mathbf{Q}+\mathbf{q})
\end{equation}
Finally, the
LT
self energies are given by
\begin{eqnarray}
  i\Sigma_{\eta}(k)=\frac{g^{2}N_b T}{L^2}\sum_{q}D(q)G_{-\eta}(k+q),\\
  \Pi(q)=\frac{2g^{2}T}{L^2}\sum_{k\eta}G_{\eta}(k)G_{-\eta}(k+q).
\end{eqnarray}

We solve the
LT
equations by using the iteration method. The local nature of the boson-fermion coupling allows us to rewrite the Eliashberg equations in real space and imaginary time. We then evaluate the self-energies by performing a fast Fourier transform, with a computational cost of $L^2L_\tau\log(L^2L_\tau)$ per iteration.
Note that a na\"ive Fourier transform of the fermionic Green's function to Matsubara frequencies generates an error which scales as $O(1/T L_\tau)$. For better convergence, we use the `Filon-Trapezoidal' rule\cite{tuck_simple_1967} to reduce the error to $O(1/T L_\tau)^2$.

\section{Comparison to Lattice theory and to Quantum Monte Carlo simulations}
\label{sec:comp-monte-carlo}

\begin{figure*}
  \centering
  \begin{subfigure}{0.32\hsize}
    \includegraphics[width=\hsize]{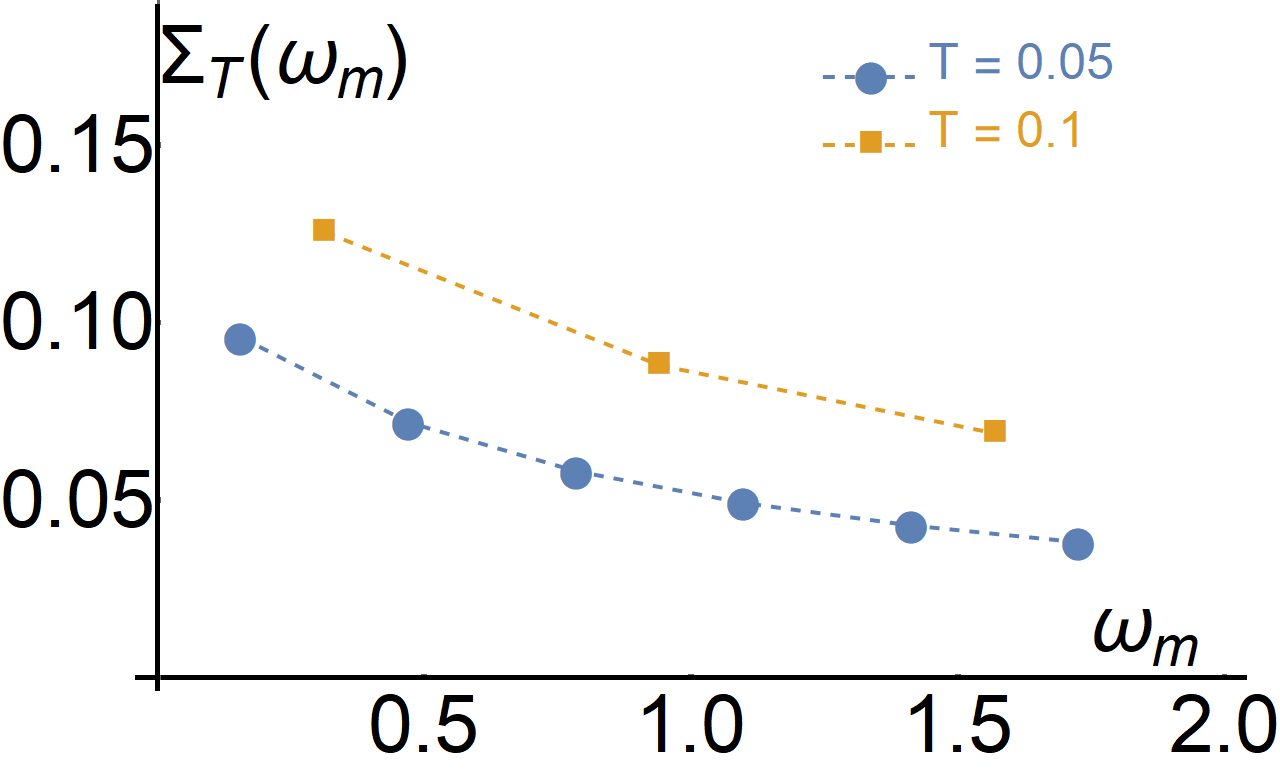}\caption{}
  \end{subfigure}
  \begin{subfigure}{0.32\hsize}
    \includegraphics[width=\hsize]{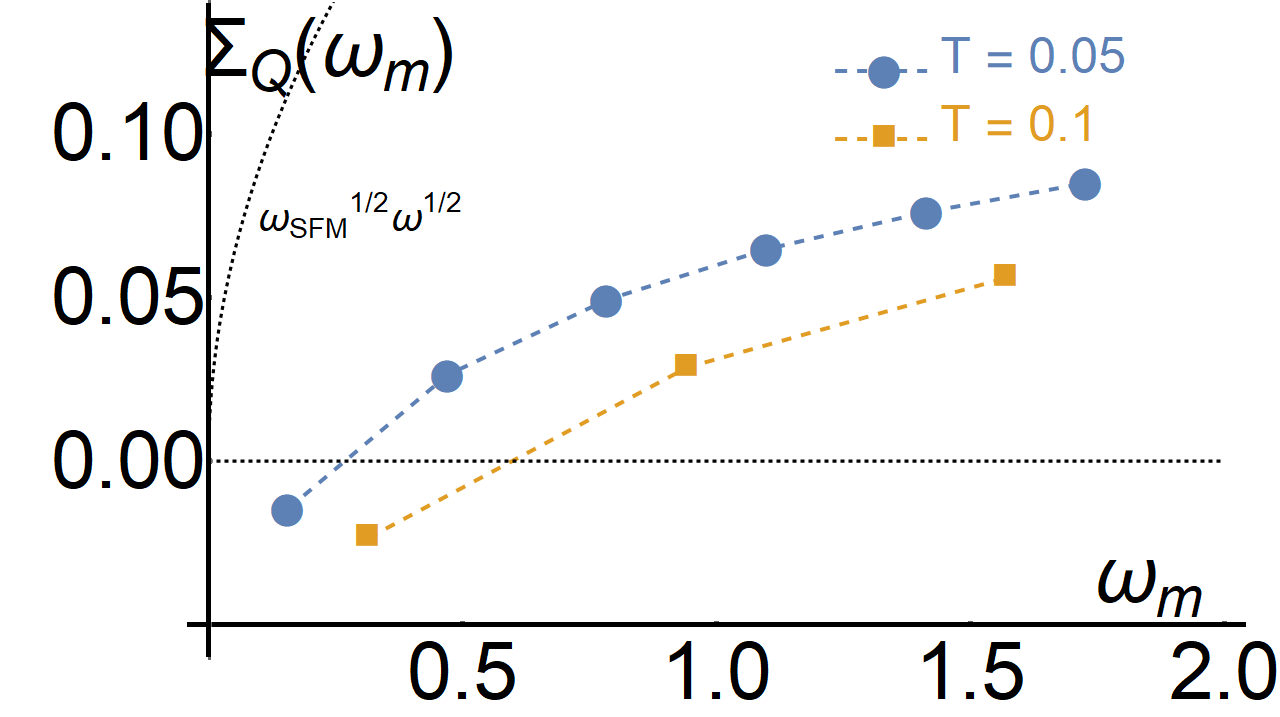}\caption{}
  \end{subfigure}
  \begin{subfigure}{0.32\hsize}
    \includegraphics[width=\hsize]{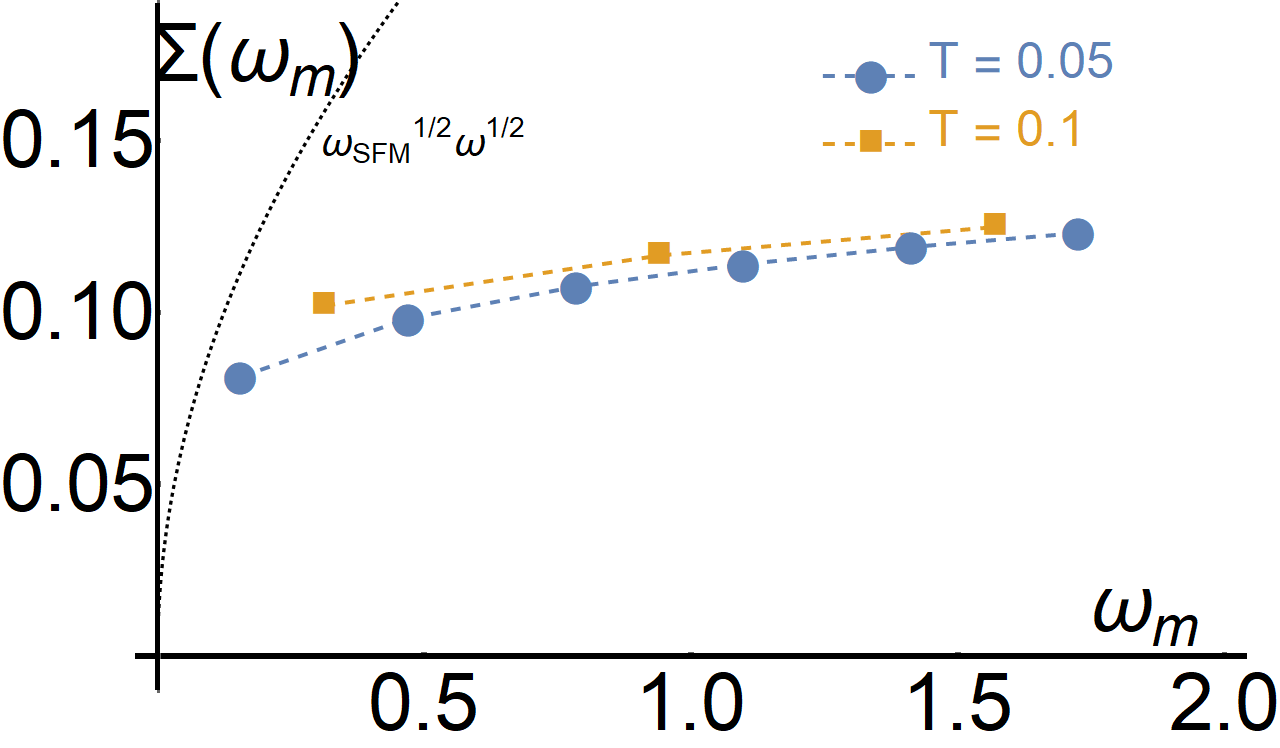}\caption{}
  \end{subfigure}\\
  \begin{subfigure}{0.32\hsize}
    \includegraphics[width=\hsize]{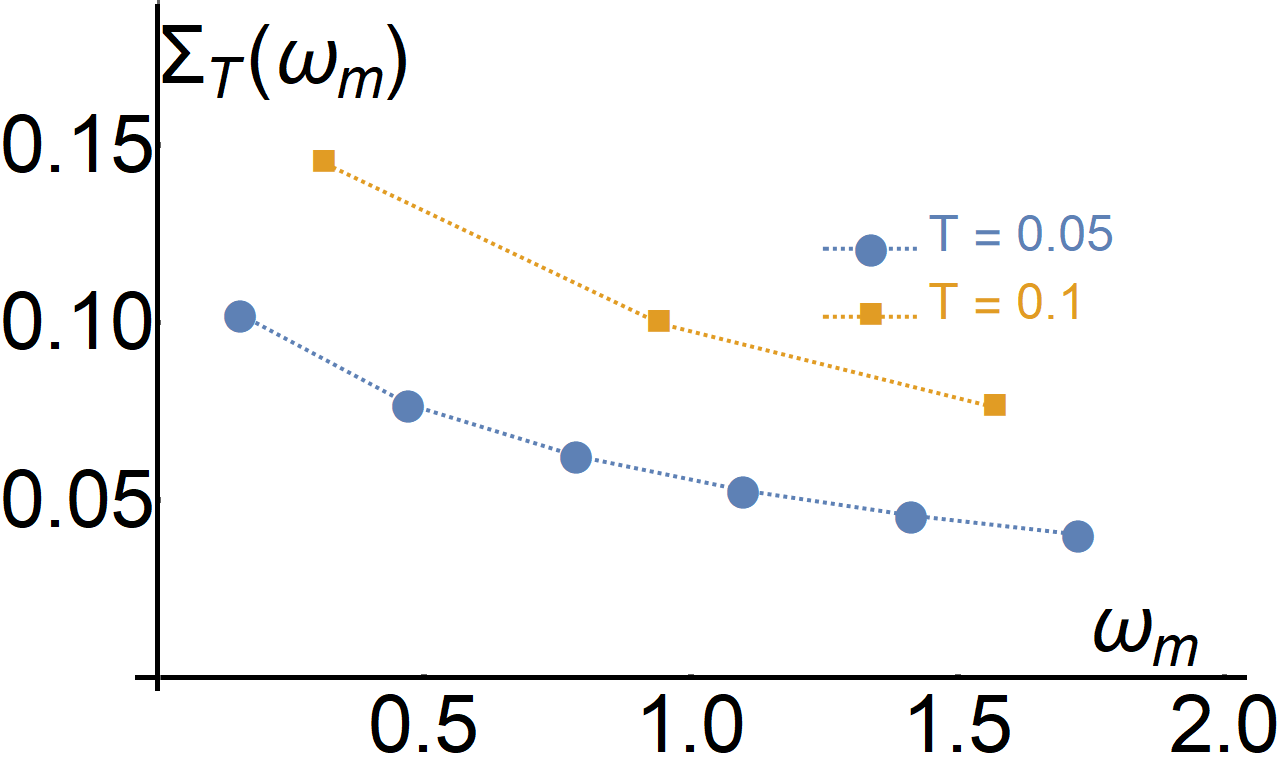}\caption{}
  \end{subfigure}
  \begin{subfigure}{0.32\hsize}
    \includegraphics[width=\hsize]{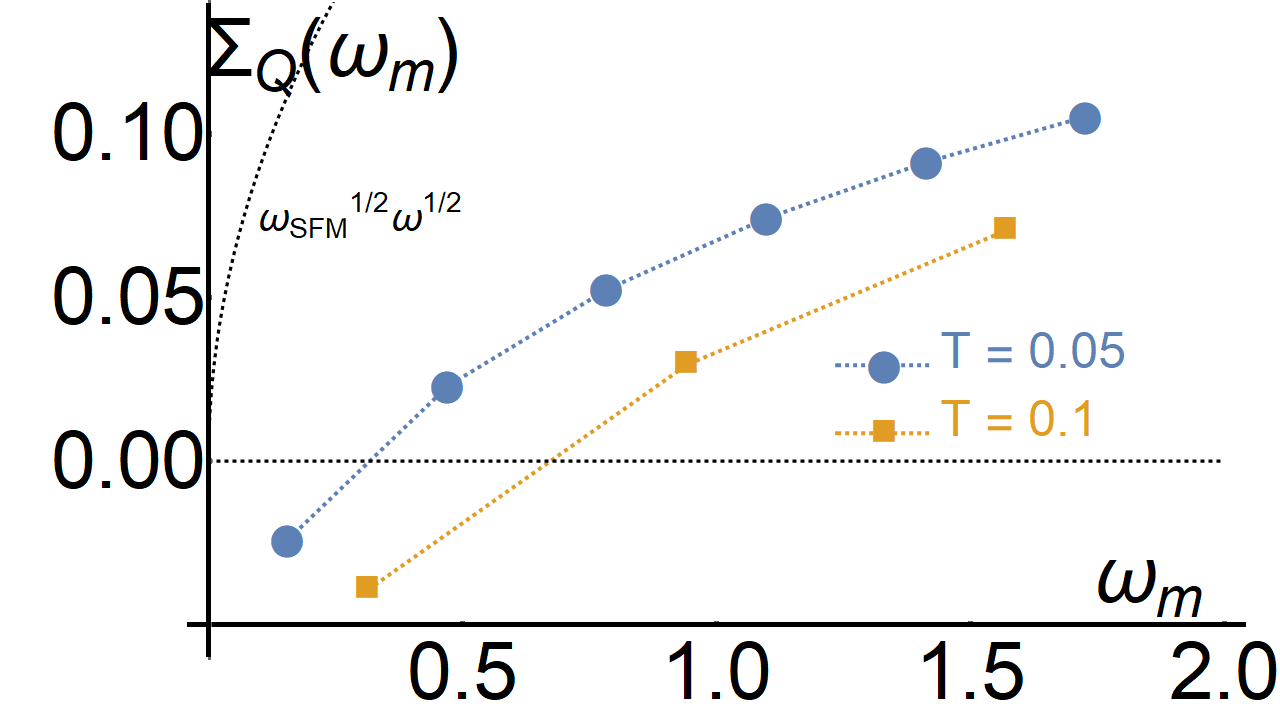}\caption{}
  \end{subfigure}
  \begin{subfigure}{0.32\hsize}
    \includegraphics[width=\hsize]{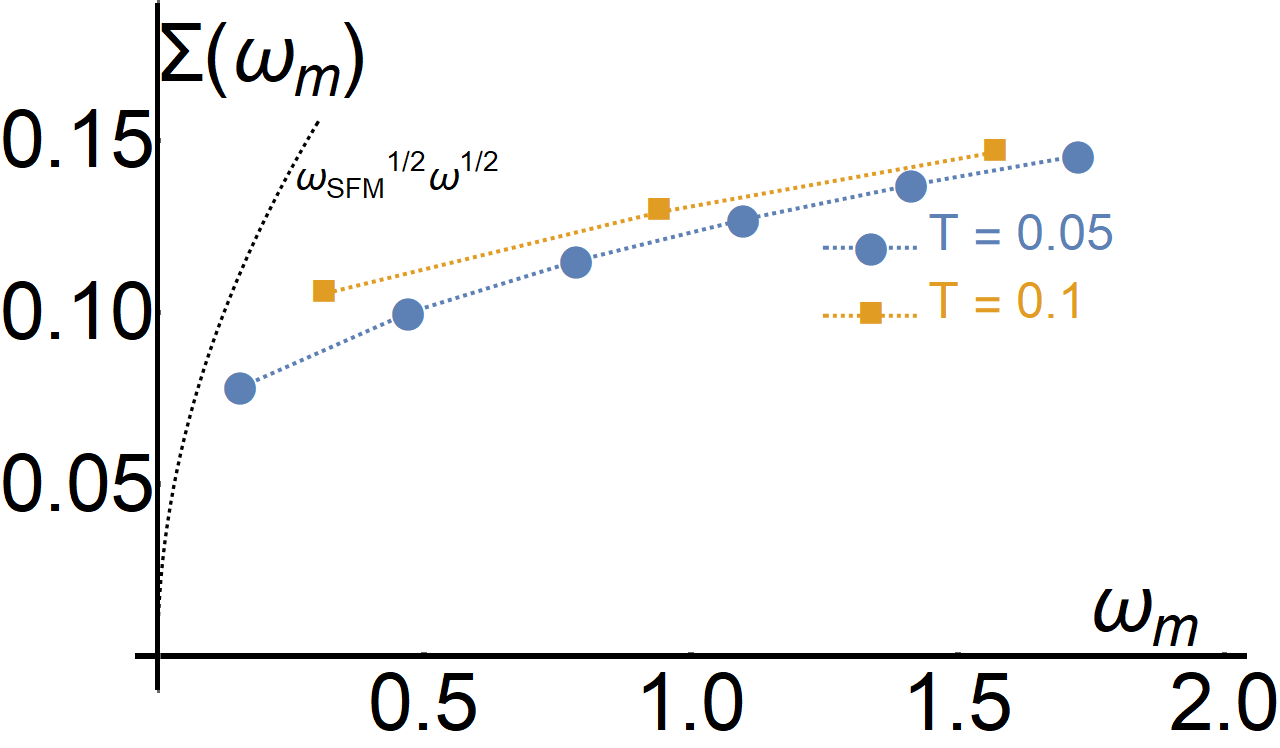}\caption{}
  \end{subfigure}
  \caption{
    The fermionic self-energy for the SFM. (a, b, c -- dashed traces) Fermionic self energy from MET, Eq.
    \eqref{eq:sigma-split}, showing the (a) thermal, (b) quantum  and (c) total self energy. The black dashed line on the right is the asymptotic $T=0$ prediction of ET. (d, e, f -- dotted traces) Fermionic self energy from LT.}
  \label{fig:sfm-an}
\end{figure*}

In this section, we compare our finite-temperature
results
for the self-energy with QMC results for the SFM and the INM.
We compare QMC results to the expressions for $\Sigma (k)$ and $\Pi (q)$, which we obtained within MET as well as within LT.
We will see that the functional forms of $\Sigma (k)$ and $\Pi (q)$ obtained within MET and LT
are similar but with some differences.
By comparing these three forms of $\Sg$ and $\Pi$ we are able to both verify the validity of the MET, and identify the strength of vertex corrections, not included into MET and LT, and lattice effects, which are not included in MET but are present in LT.

Before going to a detailed comparison, we discuss the relationship between MET, LT, and QMC calculations, and how to compare them.

The models used in the QMC studies differ from the ones we introduced in Sec.~\ref{sec:spin-fermion-ising} in several ways. First, in the QMC models the bosonic degrees of freedom have their own dynamics. Nevertheless, at low frequencies, relevant to the physics near the QCP, the leading term in the bosonic dynamics is the Landau damping due to coupling to electronic degrees of freedom. This is since the Landau damping term scales as $|\W|$, while the dynamical term of a boson scales as $\W^2$.
Second, the models used in the QMC studies contain additional interactions, which are not present in the MET or LT. These include boson-boson interactions, as well as (in the INM) additional, non-critical bosonic modes which couple to the fermions.
Finally, the three methods differ in the proper definition of the parameters of $D_0$, $M(T)$ and $g$. The QMC starts from the theory with bare parameters and returns the numerically exact $G(k),D(q)$,
so that $D_0, M(T)$ are outputs of the calculation. 
In the MET and LT,  $D_0, M(T)$ are considered as inputs to the theory.
The bare parameter values
are renormalized by vertex corrections and, in the case of MET, one-loop self-energy corrections from high-energy fermions. These renormalizations should be absorbed into the input to the theory. This implies that the effective $g$ may be different in MET and LT.
We obtain $D_0$ and $M(T)$ by fitting the static $D(q,\W_m = 0)$ from QMC, and use the bare value of $g$, as elaborated below. We find that the bare $g$ reproduces the QMC data well for both the SFM and INM.

We first show our results for the SFM. Refs.~\onlinecite{Schattner2016a,Gerlach2017,Berg2019} presented extensive QMC data on a realization of the SFM. To compare with our work, we use the data from those papers for $g = 1.5, T=0.05\ldots 0.2$,
where all energies are in units of the bare hopping used in Refs.~\onlinecite{Schattner2016a,Gerlach2017,Berg2019}.
We note that for lowest temperature, $T = 0.05$, the  thermal scale,
given by Eq. \eqref{eq:eta-def},
$ \w_T^0 = 0.16$, is almost the same as the lowest Matsubara frequency $\pi T$, and in addition $\vf M(T) \gtrsim \pi T,\w_T^0$.
The implication is that the QMC data
mostly fall into the
the intermediate regime between regions I and II,
and the system is not fully critical.
For this reason, for the quantitative comparison of the low-energy MET
with QMC we performed
the Matsubara summations in
Eq. \eqref{eq:sig0-SFM} numerically.
\begin{figure*}
  \centering
  \begin{subfigure}{0.45\hsize}
    \includegraphics[width=\hsize]{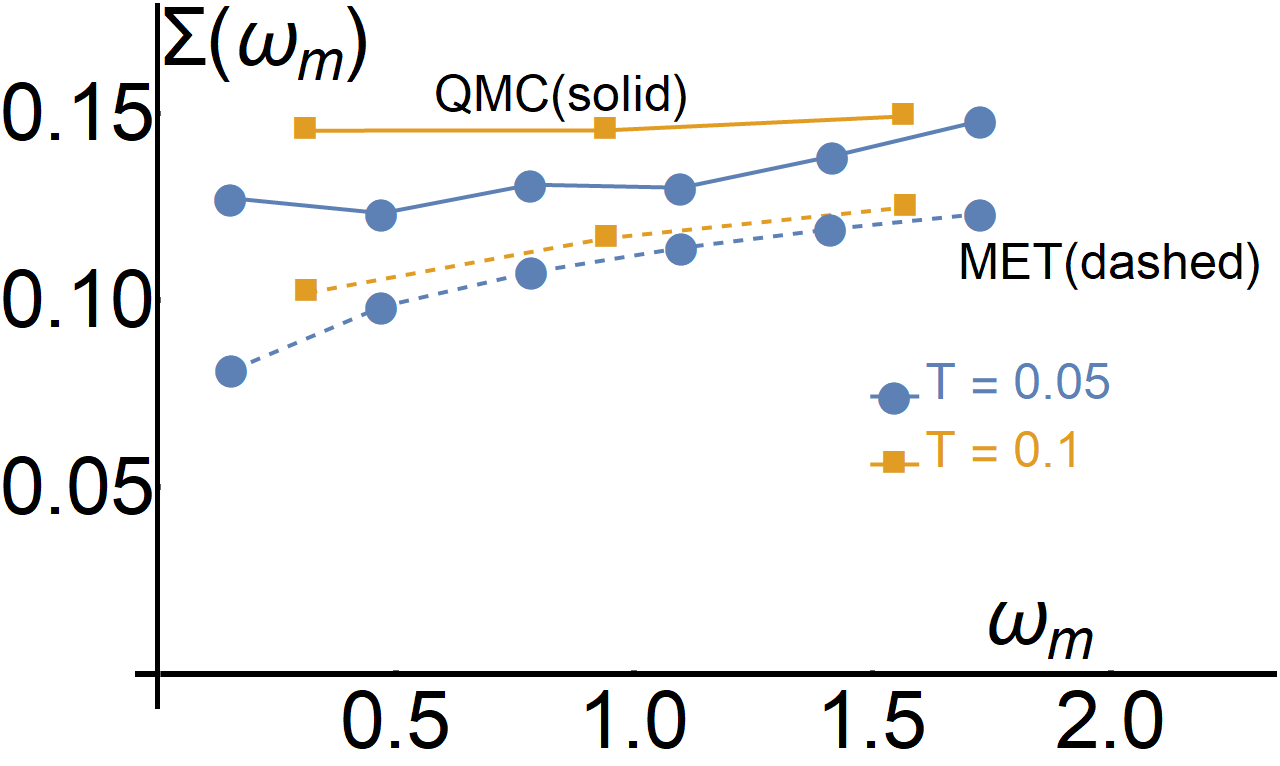}\caption{}
  \end{subfigure}
    \begin{subfigure}{0.45\hsize}
    \includegraphics[width=\hsize]{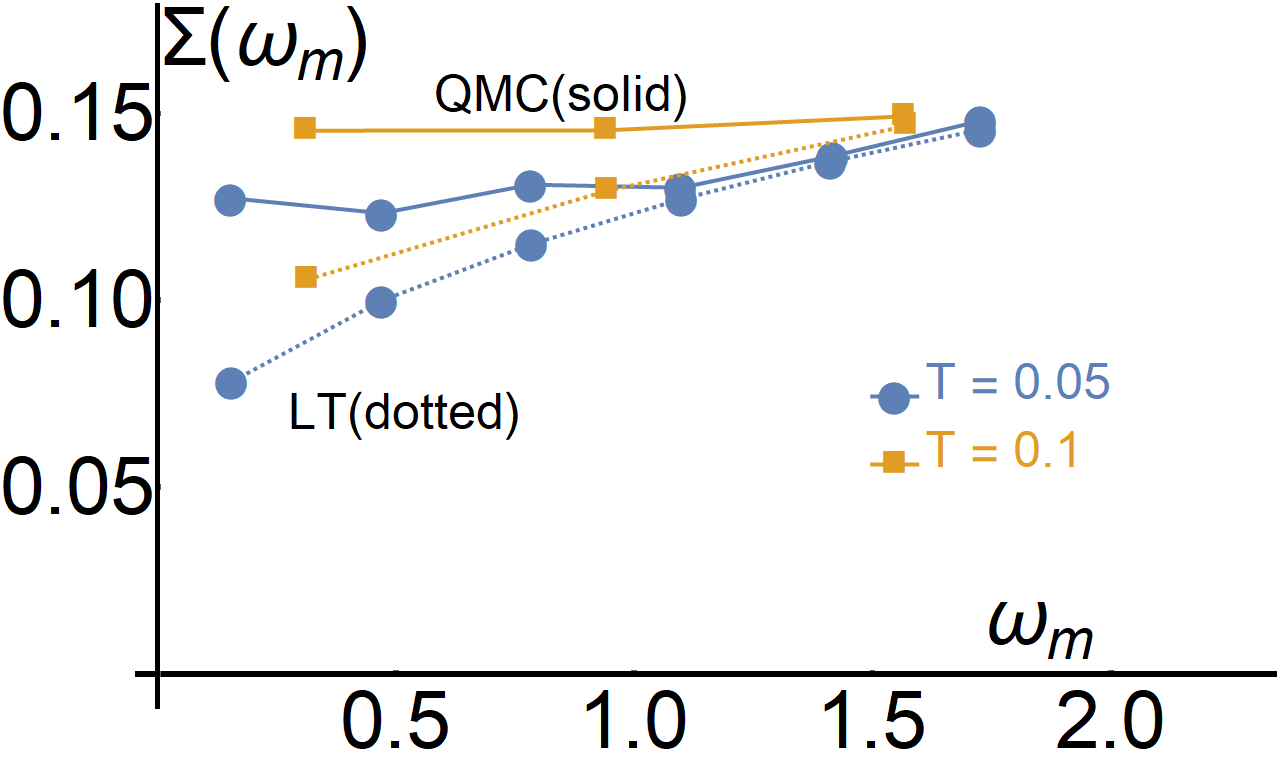}\caption{}
  \end{subfigure}
  \caption{Fermionic self energies in the SFM.
    (a) Comparison of the MET and QMC self energy. (b) Comparison of the LT and QMC self energy.}
  \label{fig:QMC-SFM}
\end{figure*}

\begin{figure}
  \centering
  \includegraphics[width=0.9\hsize]{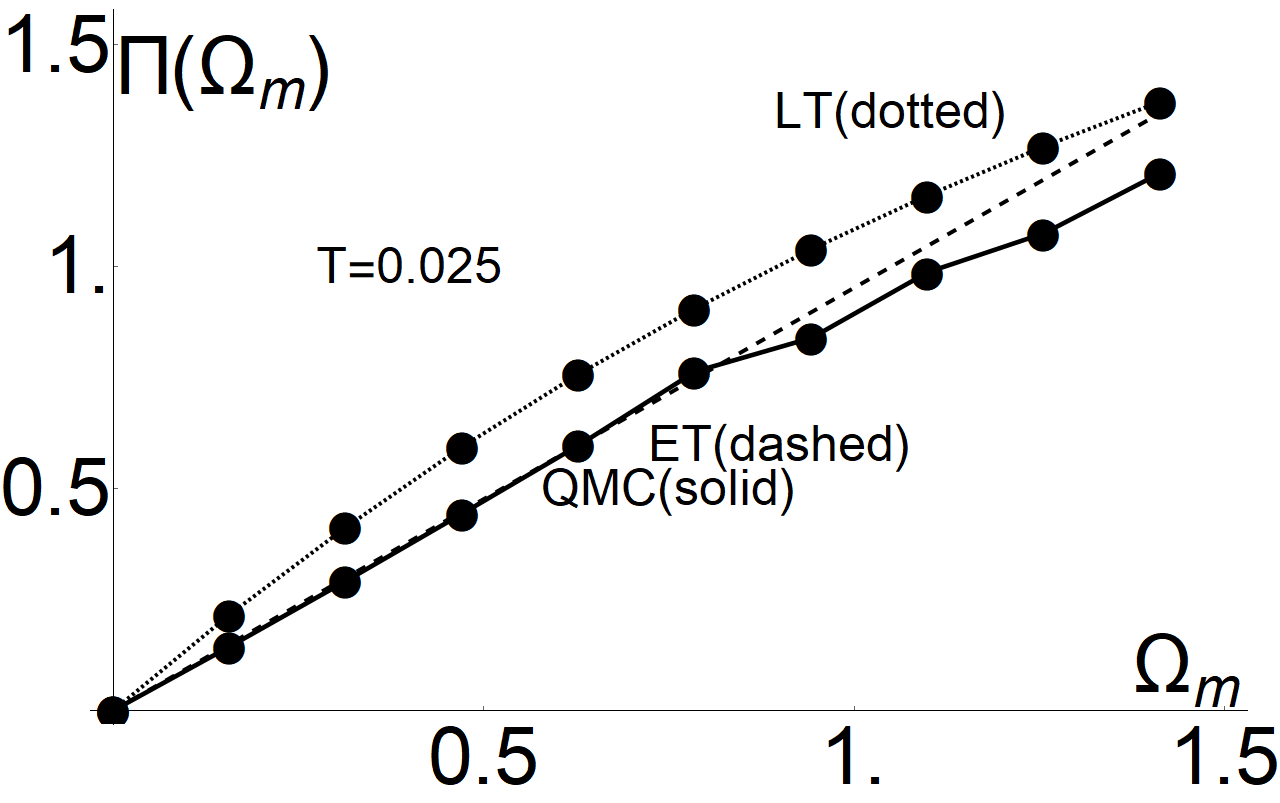}
  \caption{The bosonic self-energy $\Pi(\W_m)$ in the SFM for $q=0$. In this figure, the dashed trace corresponds to the $T=0$ (ET) expression, Eq. \eqref{eq:Pi-SFM}.\label{fig:SFM-Pi}}
\end{figure}

We begin by comparing the self-energies $\Sigma_T (\w_m)$ and $\Sigma_Q (\w_m)$ within MET and LT.
Fig. \ref{fig:sfm-an} presents $\Sg_T,\Sg_Q$ and $\Sg$ at $\theta=\theta_{hs}$ for both methods.
Although there are some deviations, the overall results are similar.
We also note the somewhat curious feature, shown in panels \ref{fig:sfm-an}b and \ref{fig:sfm-an}e, that for both theories the quantum self energy at the first Matsubara frequency $\Sg_Q(\pi T)$ is \emph{negative} (it is also true for the INM, Figs. \ref{fig:INM-se-panels}b, e). We explain this property of $\Sg_Q(\pi T)$ in the Appendix.

In Fig. \ref{fig:QMC-SFM} we compare the QMC data with both calculations. Except for a global discrepancy of about 20\% for the MET, and a slight difference in slope for the LT, there is good agreement with QMC.  The overall magnitude agrees well,
and both the MET and  the LT reproduce the QMC result
that $\Sg$ seems to extrapolate to a non-zero value as $\w_m\to 0$.
We note that the overall good agreement between MET, LT, and QMC calculations holds despite the fact that the coupling constant $\gb \approx 1.67$ is not small compared to $\ef \approx 2.2$, implying e.g. vertex corrections
should generally be $ O(1)$. In Figs. \ref{fig:sfm-an} \ref{fig:QMC-SFM} we show only two of the temperatures so as to keep the images clear, but the same agreement holds for the other temperatures.
For completeness, we also display a comparison of QMC, MET and LT for $\Pi(\W_m)$ in Fig. \ref{fig:SFM-Pi}, showing good agreement.

Next we consider the INM. While in the spin-fermion case, the model used in QMC allows us to directly read off the bare value for the coupling constant $g$, the nematic modes in the QMC studies \onlinecite{Schattner2016,Lederer2017,Berg2019} consist of pseudo-spin $1/2$ degrees of freedom that are located on the bonds of the lattice, and do not directly match the form of Eqs~(\ref{eq:inm-int},\ref{eq:lattice-fk-INM}). Nevertheless, the bare $g$ can be obtained by comparing the interaction term for a uniform ($\q=0$) configuration of the nematic modes. This yields $g=3.3\alpha$ for MET and $g=2 \alpha$ for the LT, where $\alpha$ is the coupling to the pseudo-spin degrees of freedom in the notation of Refs. \onlinecite{Schattner2016,Lederer2017,Berg2019}.

Here, we focus on the data for $\alpha=1$ and $T = 0.1,0.167,0.25$\footnote{For $T < 0.1$ the QMC data shows that the nematic coherence length is already cut off by superconductivity, so we did not use that data}.
Extracting $D_0$ from the QMC data, we find the bare coupling used in MET is $\gb = g^2D_0  = 8.25$. 
We note in passing that QMC simulations show a $T_c = 0.04$, which is in good agreement with the ET prediction of $T_{c,\ET}=0.066$ for the given $\gb$. The Eliashberg $T_c$ in the INM scales as
$ \gb^2/\ef \propto g^4$, so that the difference in critical temperatures can be interpreted as a 15\% renormalization of $g$.
In the LT, we find $T_{c,\LT} = 0.03$.
\ref{fig:INM-se-panels}.
\begin{figure*}
  \centering
  \begin{subfigure}{0.32\hsize}
    \includegraphics[width=\hsize]{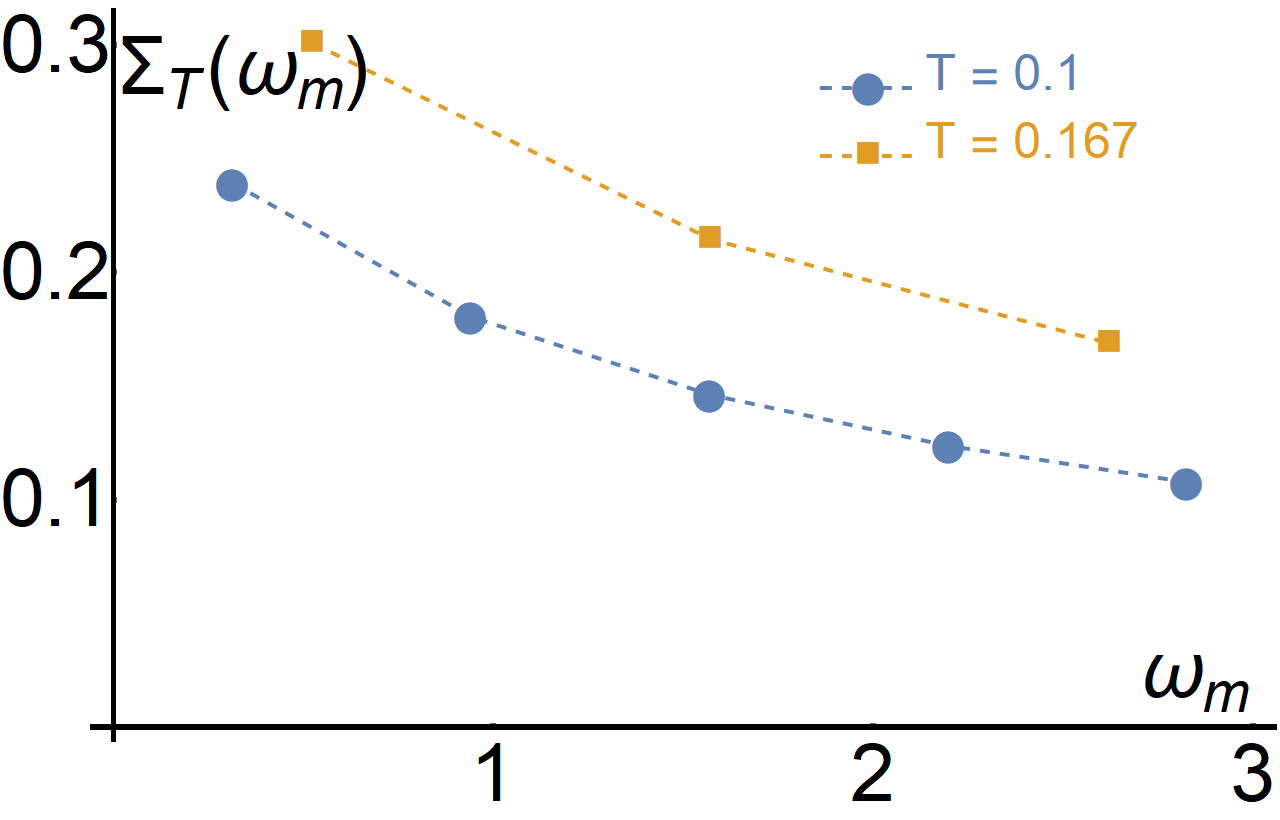}\caption{}
  \end{subfigure}
  \begin{subfigure}{0.32\hsize}
    \includegraphics[width=\hsize]{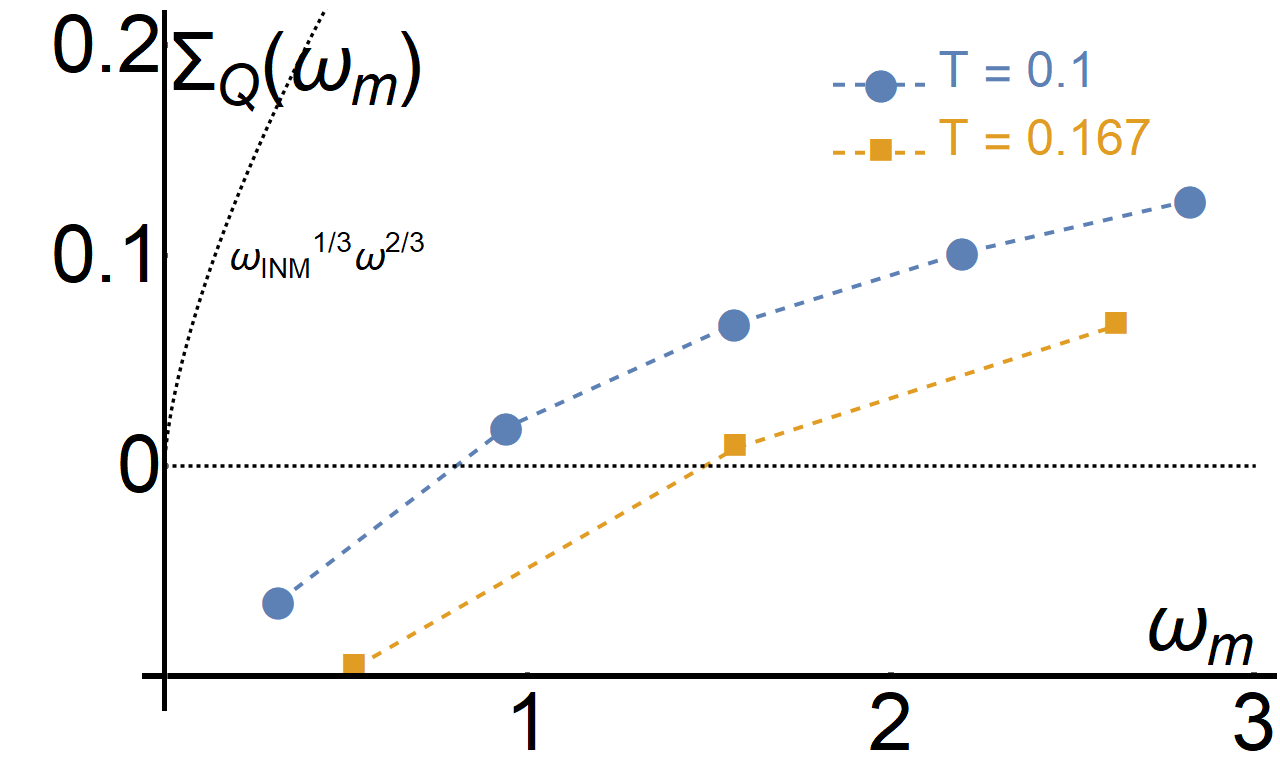}\caption{}
  \end{subfigure}
  \begin{subfigure}{0.32\hsize}
    \includegraphics[width=\hsize]{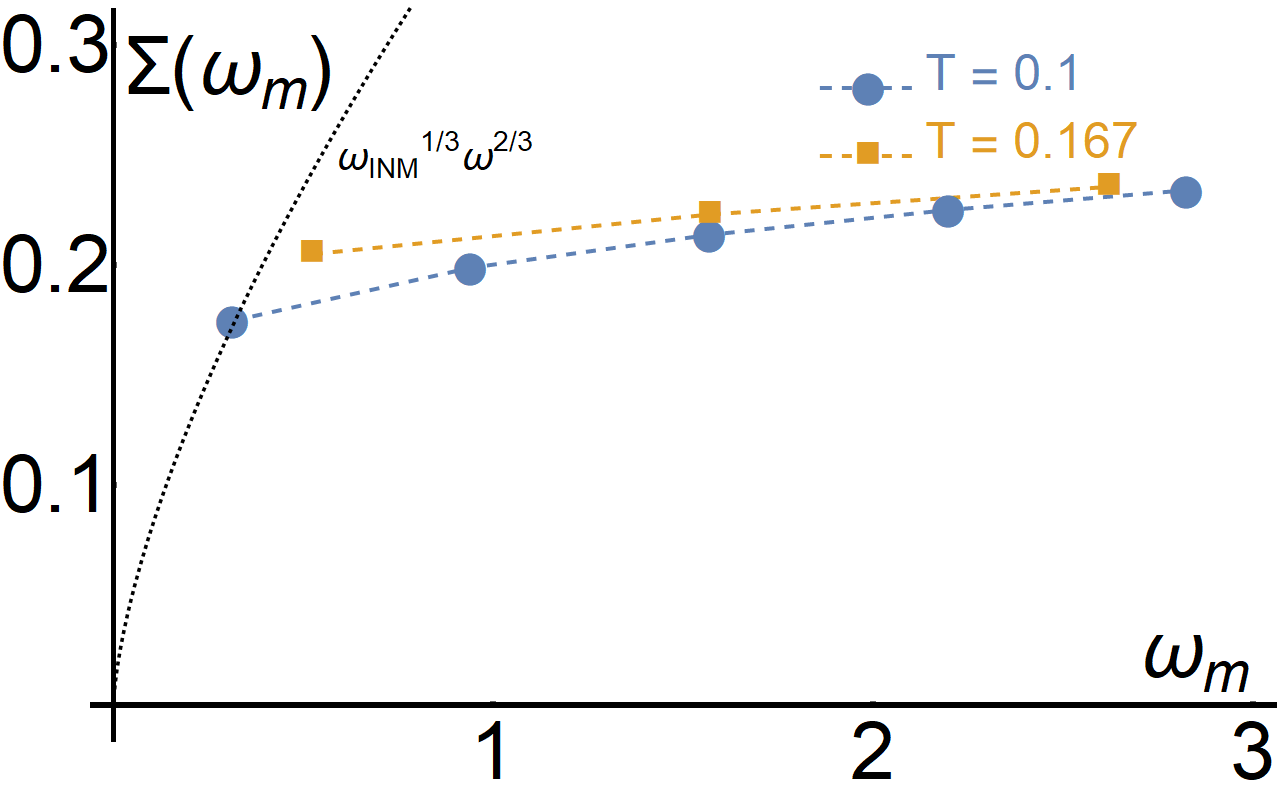}\caption{}
  \end{subfigure}\\
  \begin{subfigure}{0.32\hsize}
    \includegraphics[width=\hsize]{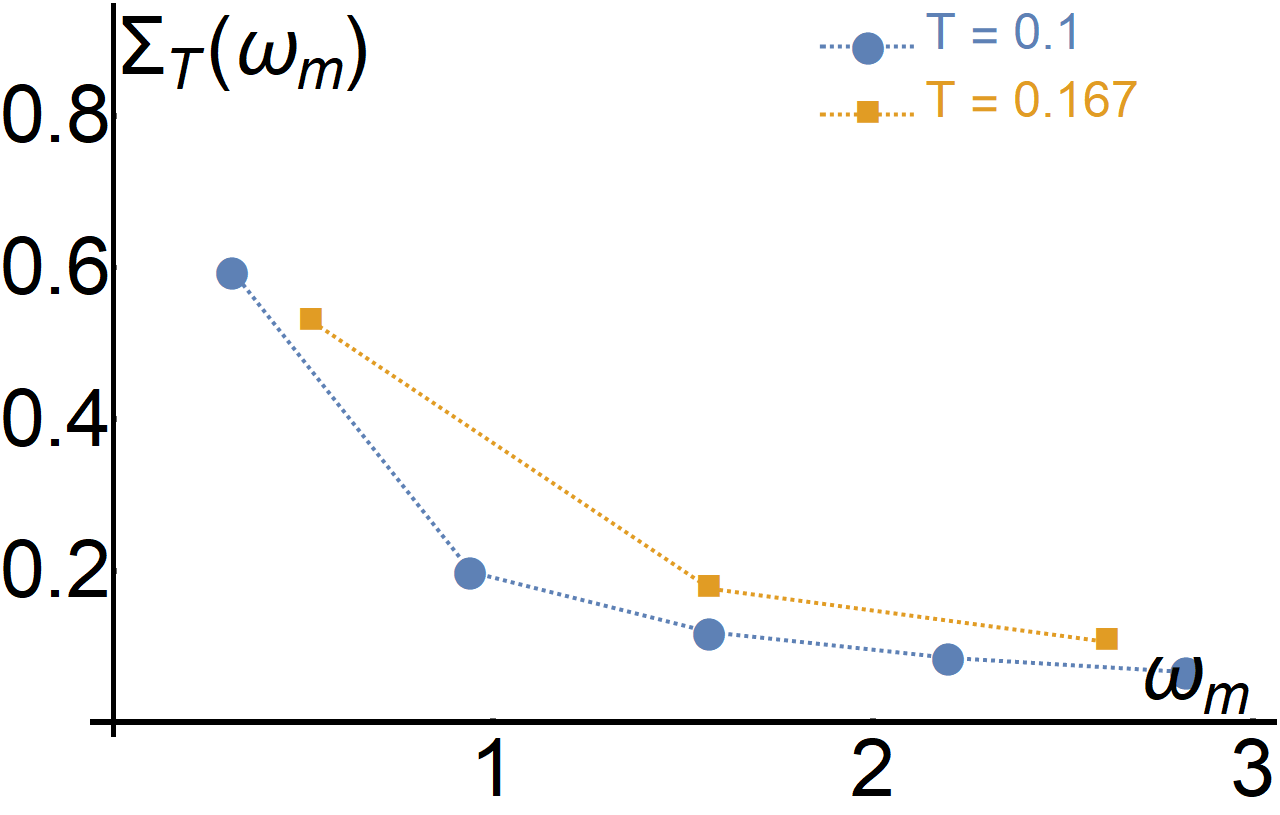}\caption{}
  \end{subfigure}
  \begin{subfigure}{0.32\hsize}
    \includegraphics[width=\hsize]{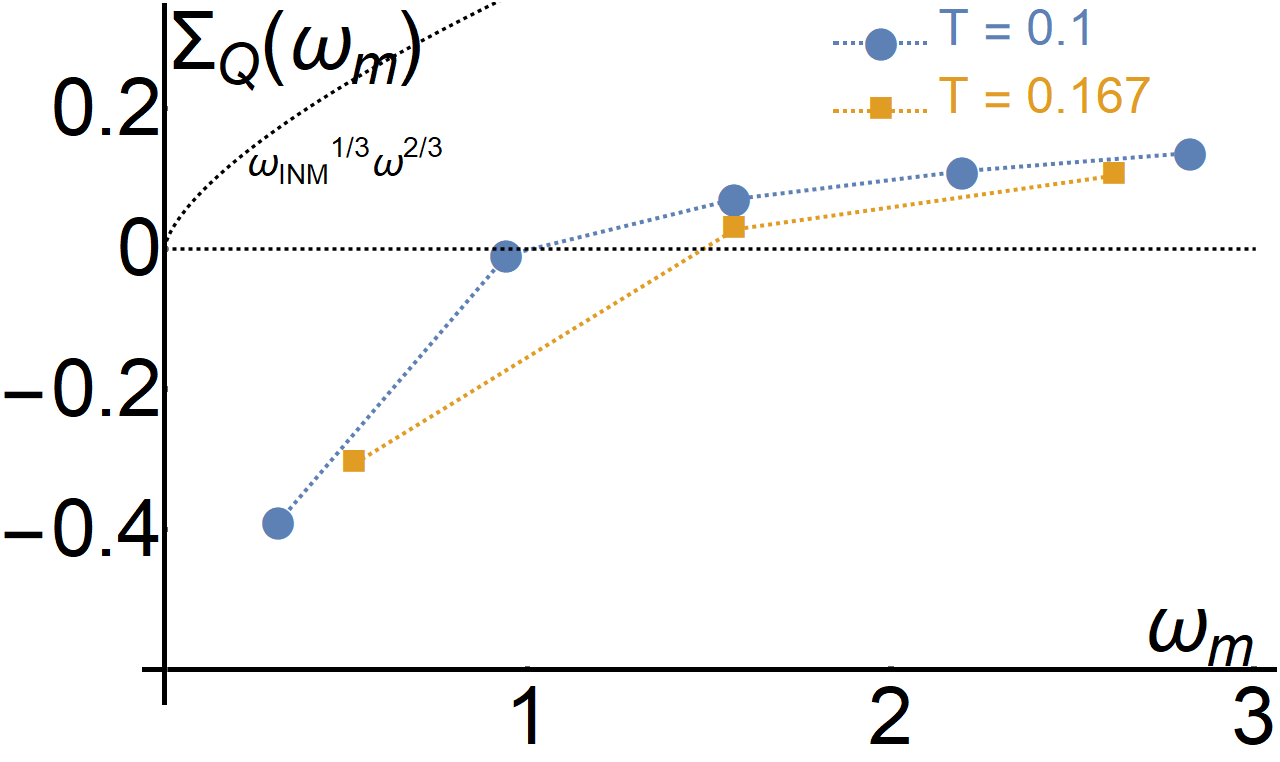}\caption{}
  \end{subfigure}
  \begin{subfigure}{0.32\hsize}
    \includegraphics[width=\hsize]{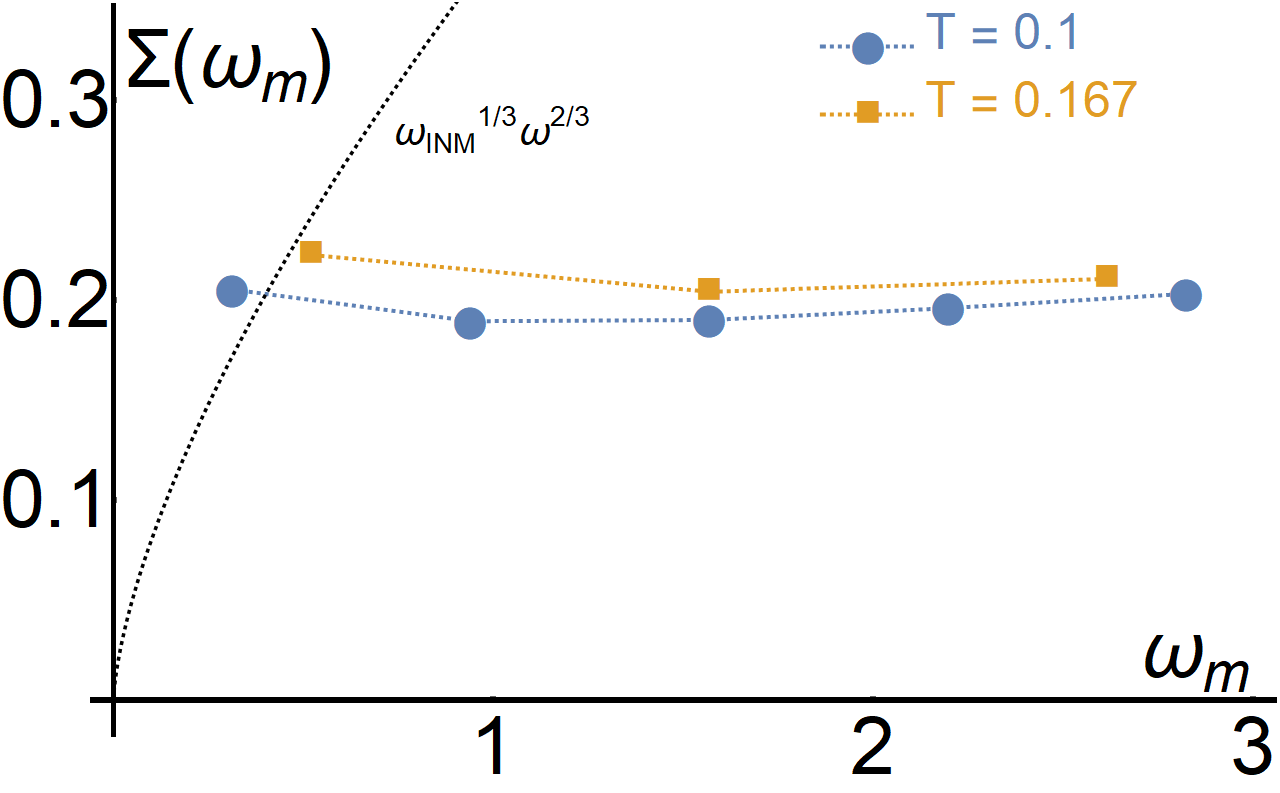}\caption{}
  \end{subfigure}
  \caption{The fermionic self-energy for the INM. (a, b, c -- dashed traces) Fermionic self energy from MET, showing the (a) thermal, (b) quantum and (c) total self energy.  The black dashed lines on the right are the asymptotic $T=0$ prediction of ET. (d, e, f -- dotted traces) Fermionic self energy from LT.}
  \label{fig:INM-se-panels}
\end{figure*}

\begin{figure*}
  \centering
  \begin{subfigure}{0.45\hsize}
    \includegraphics[width=\hsize]{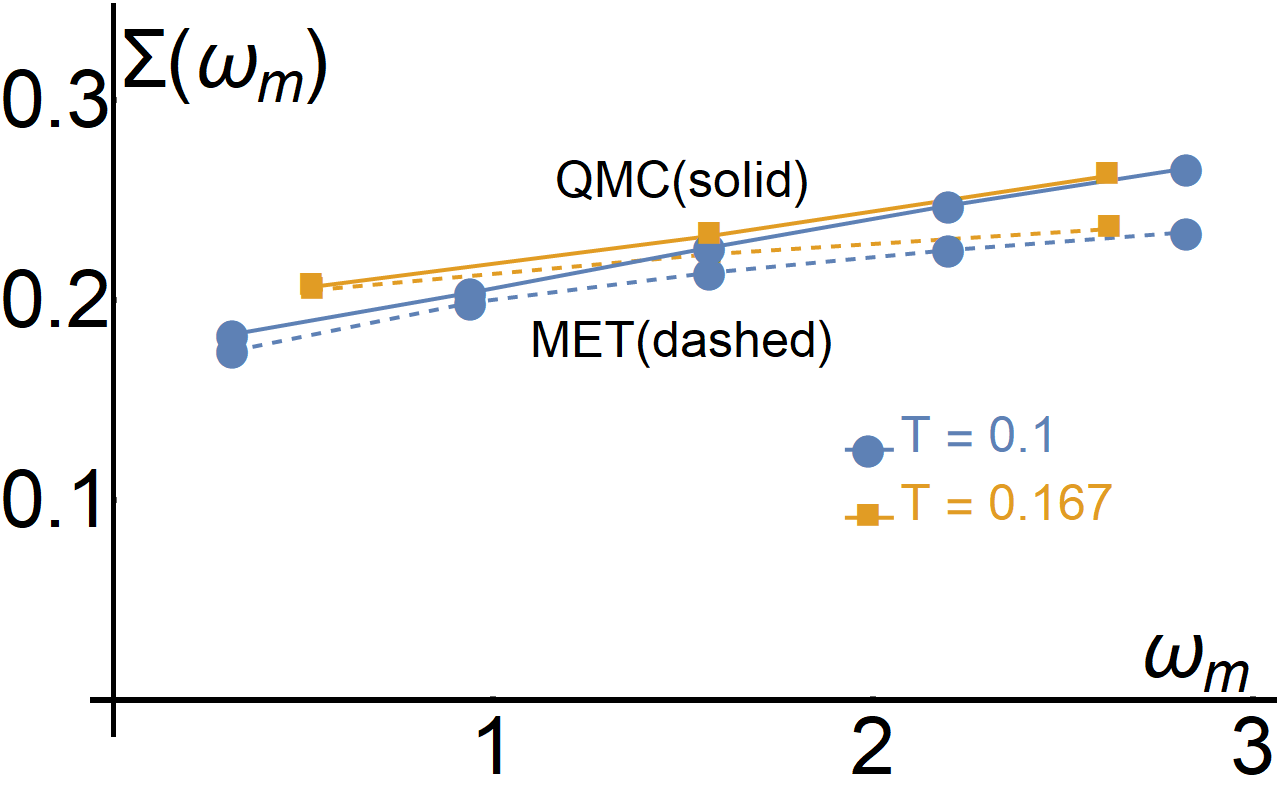}
  \end{subfigure}
    \begin{subfigure}{0.45\hsize}
    \includegraphics[width=\hsize]{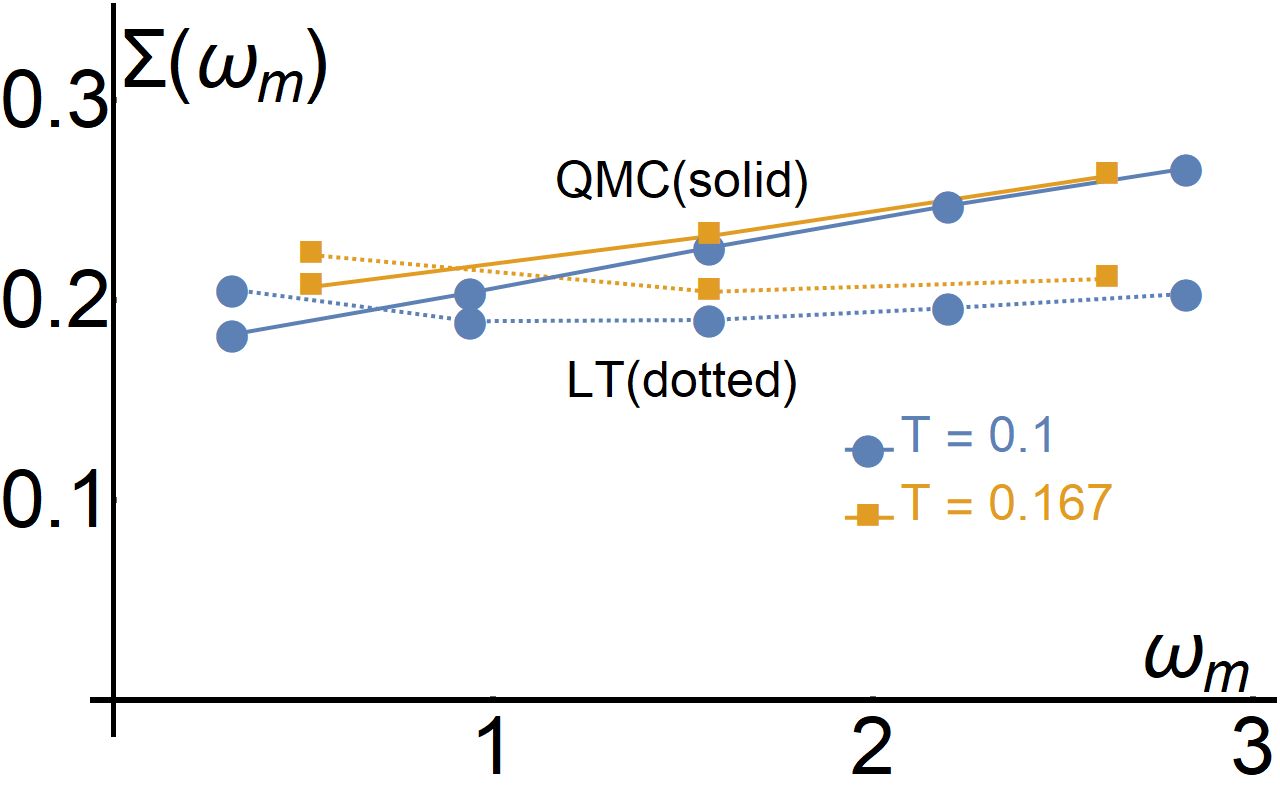}\caption{}
  \end{subfigure}
  \caption{
    Fermionic self-energy of the INM. (a) Comparison of the MET and QMC self energy.
    (b) Comparison of the LT and QMC self energy.
  }
  \label{fig:inm-se}
\end{figure*}
Fig. \ref{fig:inm-se} presents a comparison of the QMC data for $\Sg$ with both the MET and LT calculations.
The MET calculation shows an excellent agreement with the data. The LT calculation has some systematic deviations in the frequency behavior. From all these results, we conclude that similarly to the SFM there are only moderate high-energy vertex corrections and lattice renormalizations in the INM, and that
they are somewhat better accounted for in the low energy MET.

\begin{figure}
  \centering
  \includegraphics[width=0.9\hsize]{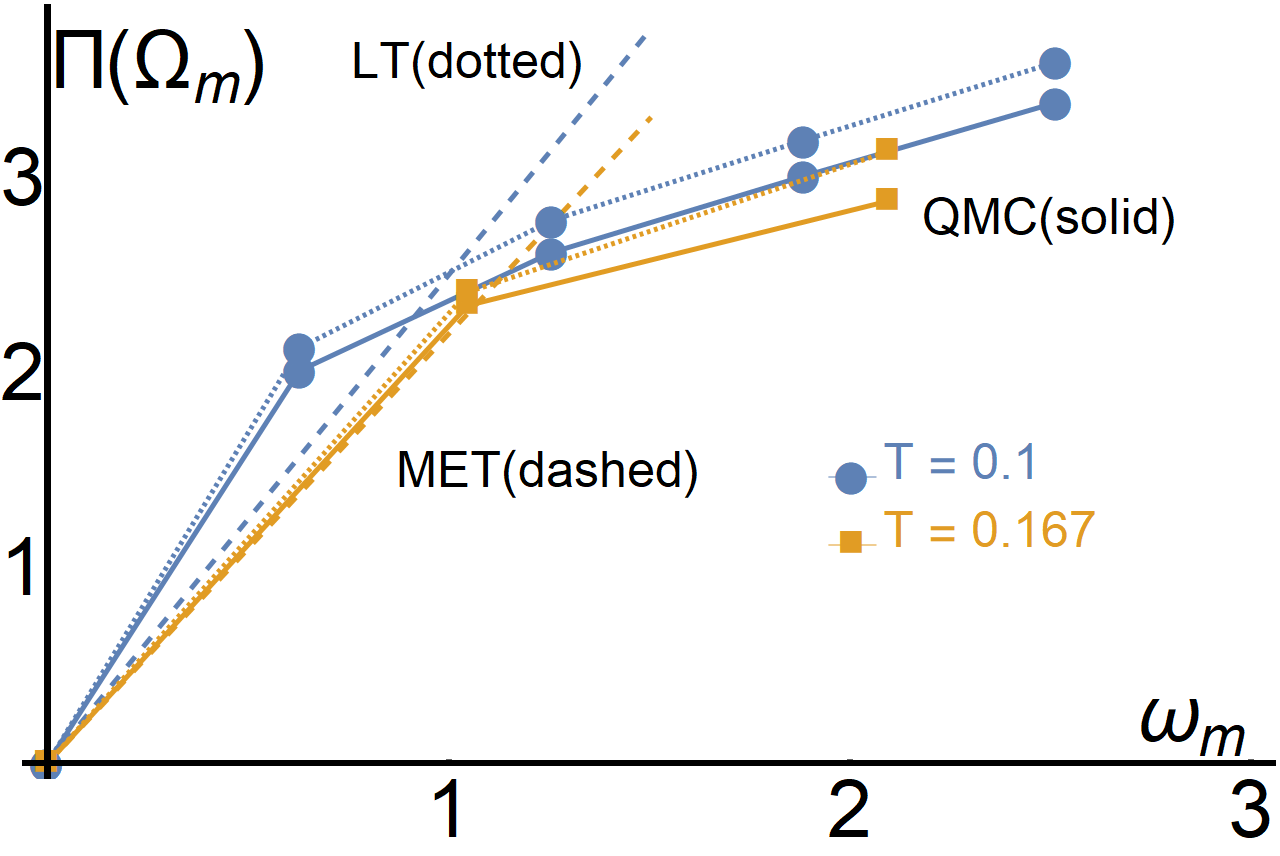}
  \caption{The bosonic self energy $\Pi(\W_m)$ in the INM for the lowest $q=2\pi/L$, where $L$ is the system size. Solid lines denote the QMC data, and dashed and dotted lines denote the MET and LT calculations.}
  \label{fig:inm-pi}
\end{figure}
We also compared MET and lattice calculations to the QMC data for $\Pi(q)$. In Fig. \ref{fig:inm-pi} we present the behavior of $\Pi$ at the lowest $q=2\pi/L$, where $L$ is the QMC system size, along with a lattice calculation of $\Pi$ and the MET prediction of the low frequency behavior of $\Pi$, Eq. \eqref{eq:Pi-INM-T}\footnote{Note that the asymptotic expression in Eq. \eqref{eq:Pi-INM-T} for the MET applies when $\vf q \ll \w_T$, which is not the case for the QMC data. We performed the angular integration in that equation numerically.}.
There is good agreement between MET/LT and the QMC data.

We find it striking that most of the
features of the QMC data are reproduced by
our ET/MET calculations.
We believe that our results imply the QMC data is a validation of, rather than a challenge to, the applicability of ET to quantum-critical models of interacting electrons.

We used the differences between the theories to identify the importance of high energy vertex and lattice renormalizations to the low energy theory.
The fact that the bare $\gb$ reproduces both $T_c$ and the self energy well in both models implies that the contribution to fermionic self-energies from vertex corrections due to high-energy fermions are likely small.
In addition, we show that MET
reproduces the QMC self energies quite well, but the comparison of QMC with LT
shows some systematic deviations.
This suggests that there is at least a partial cancellation between vertex corrections and the contribution from high-energy fermions to one loop self-energy.

\section{Summary}
\label{sec:summary}

In this work, we studied the effect of thermal fluctuations on metals near a QCP, either to a spin density wave state or to an Ising nematic state. We calculated the deviation from the scaling behavior predicted by ET, in the regime where the thermal contributions do not permit a separation of scales between fermionic and bosonic degrees of freedom. We found, that at low temperatures, close to the QCP the thermal contribution to fermionic self energy scales as $\w_T \propto \sqrt{\gb T}$ and dominates the contribution from quantum dynamics. We showed that once this additional physics is taken into account,
by an appropiately modified Eliashberg theory, it
reproduces properties that were found in recent QMC simulations, above the superconducting $T_c$. 

In the absence of pair-breaking, the regime of metallic quantum critical behavior is limited by the large enhancement of the superconducting $T_c$ near the QCP, and the associated strong superconducting fluctuations. QMC studies report no clear separation between the energy scales associated with superconductivity and NFL behavior.
As we discussed in the Introduction, this is in agreement with the predictions of ET, where both superconductivity and NFL appear on a scale of $\w_\sfm,\w_\inm$ for the SFM and INM respectively. Recent work has indicated \cite{chubukov2019interplay} that superconducting fluctuations play a significant role in the normal state.
%
Our analysis of the normal-state self neglects superconducting fluctuations, and so
the
combined effect of thermal
and superconducting fluctuations remains an open question.
We hope our findings will motivate further numerical and analytic work on quantum critical metals.

\begin{acknowledgments}
  We thank S. Lederer, M.H. Christensen, X. Wang, R. M. Fernandes,
  X.-Y. Xu, K. Sun, Z.Y. Meng
  and L. Classen for helpful conversations. This work was supported by the US-Israel Binational Science Foundation (BSF). EB acknowledges support from the European Research Council (ERC) under grant HQMAT (grant no. 817799) and from the Minerva foundation. YS was supported by the Department of Energy, Office of Basic Energy Sciences, under contract no. DE-AC02- 76SF00515 at Stanford, and by the Zuckerman STEM Leadership Program.
\end{acknowledgments}

\bibliography{QCsuperconductivity}

\begin{thebibliography}{56}%
\makeatletter
\providecommand \@ifxundefined [1]{%
 \@ifx{#1\undefined}
}%
\providecommand \@ifnum [1]{%
 \ifnum #1\expandafter \@firstoftwo
 \else \expandafter \@secondoftwo
 \fi
}%
\providecommand \@ifx [1]{%
 \ifx #1\expandafter \@firstoftwo
 \else \expandafter \@secondoftwo
 \fi
}%
\providecommand \natexlab [1]{#1}%
\providecommand \enquote  [1]{``#1''}%
\providecommand \bibnamefont  [1]{#1}%
\providecommand \bibfnamefont [1]{#1}%
\providecommand \citenamefont [1]{#1}%
\providecommand \href@noop [0]{\@secondoftwo}%
\providecommand \href [0]{\begingroup \@sanitize@url \@href}%
\providecommand \@href[1]{\@@startlink{#1}\@@href}%
\providecommand \@@href[1]{\endgroup#1\@@endlink}%
\providecommand \@sanitize@url [0]{\catcode `\\12\catcode `\$12\catcode
  `\&12\catcode `\#12\catcode `\^12\catcode `\_12\catcode `\%12\relax}%
\providecommand \@@startlink[1]{}%
\providecommand \@@endlink[0]{}%
\providecommand \url  [0]{\begingroup\@sanitize@url \@url }%
\providecommand \@url [1]{\endgroup\@href {#1}{\urlprefix }}%
\providecommand \urlprefix  [0]{URL }%
\providecommand \Eprint [0]{\href }%
\providecommand \doibase [0]{https://doi.org/}%
\providecommand \selectlanguage [0]{\@gobble}%
\providecommand \bibinfo  [0]{\@secondoftwo}%
\providecommand \bibfield  [0]{\@secondoftwo}%
\providecommand \translation [1]{[#1]}%
\providecommand \BibitemOpen [0]{}%
\providecommand \bibitemStop [0]{}%
\providecommand \bibitemNoStop [0]{.\EOS\space}%
\providecommand \EOS [0]{\spacefactor3000\relax}%
\providecommand \BibitemShut  [1]{\csname bibitem#1\endcsname}%
\let\auto@bib@innerbib\@empty
\bibitem [{\citenamefont {Hertz}(1976)}]{Hertz1976}%
  \BibitemOpen
  \bibfield  {author} {\bibinfo {author} {\bibfnamefont {J.~A.}\ \bibnamefont
  {Hertz}},\ }\bibfield  {title} {\bibinfo {title} {Quantum critical
  phenomena},\ }\href {https://doi.org/10.1103/PhysRevB.14.1165} {\bibfield
  {journal} {\bibinfo  {journal} {Phys. Rev. B}\ }\textbf {\bibinfo {volume}
  {14}},\ \bibinfo {pages} {1165} (\bibinfo {year} {1976})}\BibitemShut
  {NoStop}%
\bibitem [{\citenamefont {Millis}(1993)}]{Millis1993}%
  \BibitemOpen
  \bibfield  {author} {\bibinfo {author} {\bibfnamefont {A.~J.}\ \bibnamefont
  {Millis}},\ }\bibfield  {title} {\bibinfo {title} {Effect of a nonzero
  temperature on quantum critical points in itinerant fermion systems},\ }\href
  {https://doi.org/10.1103/PhysRevB.48.7183} {\bibfield  {journal} {\bibinfo
  {journal} {Phys. Rev. B}\ }\textbf {\bibinfo {volume} {48}},\ \bibinfo
  {pages} {7183} (\bibinfo {year} {1993})}\BibitemShut {NoStop}%
\bibitem [{\citenamefont {Altshuler}\ \emph {et~al.}(1994)\citenamefont
  {Altshuler}, \citenamefont {Ioffe},\ and\ \citenamefont
  {Millis}}]{Altshuler1994}%
  \BibitemOpen
  \bibfield  {author} {\bibinfo {author} {\bibfnamefont {B.~L.}\ \bibnamefont
  {Altshuler}}, \bibinfo {author} {\bibfnamefont {L.~B.}\ \bibnamefont
  {Ioffe}},\ and\ \bibinfo {author} {\bibfnamefont {A.~J.}\ \bibnamefont
  {Millis}},\ }\bibfield  {title} {\bibinfo {title} {Low-energy properties of
  fermions with singular interactions},\ }\href
  {https://doi.org/10.1103/PhysRevB.50.14048} {\bibfield  {journal} {\bibinfo
  {journal} {Phys. Rev. B}\ }\textbf {\bibinfo {volume} {50}},\ \bibinfo
  {pages} {14048} (\bibinfo {year} {1994})}\BibitemShut {NoStop}%
\bibitem [{\citenamefont {Abanov}\ \emph {et~al.}(2003)\citenamefont {Abanov},
  \citenamefont {Chubukov},\ and\ \citenamefont {Schmalian}}]{Abanov2003}%
  \BibitemOpen
  \bibfield  {author} {\bibinfo {author} {\bibfnamefont {A.}~\bibnamefont
  {Abanov}}, \bibinfo {author} {\bibfnamefont {A.~V.}\ \bibnamefont
  {Chubukov}},\ and\ \bibinfo {author} {\bibfnamefont {J.}~\bibnamefont
  {Schmalian}},\ }\bibfield  {title} {\bibinfo {title} {Quantum-critical theory
  of the spin-fermion model and its application to cuprates: Normal state
  analysis},\ }\bibfield  {booktitle} {\emph {\bibinfo {booktitle} {Advances in
  Physics}},\ }\href {https://doi.org/10.1080/0001873021000057123} {\bibfield
  {journal} {\bibinfo  {journal} {Advances in Physics}\ }\textbf {\bibinfo
  {volume} {52}},\ \bibinfo {pages} {119} (\bibinfo {year} {2003})}\BibitemShut
  {NoStop}%
\bibitem [{\citenamefont {Abanov}\ \emph {et~al.}(2001)\citenamefont {Abanov},
  \citenamefont {Chubukov},\ and\ \citenamefont {Finkel'stein}}]{Abanov2001}%
  \BibitemOpen
  \bibfield  {author} {\bibinfo {author} {\bibfnamefont {A.}~\bibnamefont
  {Abanov}}, \bibinfo {author} {\bibfnamefont {A.~V.}\ \bibnamefont
  {Chubukov}},\ and\ \bibinfo {author} {\bibfnamefont {A.~M.}\ \bibnamefont
  {Finkel'stein}},\ }\bibfield  {title} {\bibinfo {title} {Coherent vs .
  incoherent pairing in 2d systems near magnetic instability},\ }\href
  {http://stacks.iop.org/0295-5075/54/i=4/a=488} {\bibfield  {journal}
  {\bibinfo  {journal} {EPL (Europhysics Letters)}\ }\textbf {\bibinfo {volume}
  {54}},\ \bibinfo {pages} {488} (\bibinfo {year} {2001})}\BibitemShut
  {NoStop}%
\bibitem [{\citenamefont {Metlitski}\ and\ \citenamefont
  {Sachdev}(2010{\natexlab{a}})}]{Metlitski2010a}%
  \BibitemOpen
  \bibfield  {author} {\bibinfo {author} {\bibfnamefont {M.~A.}\ \bibnamefont
  {Metlitski}}\ and\ \bibinfo {author} {\bibfnamefont {S.}~\bibnamefont
  {Sachdev}},\ }\bibfield  {title} {\bibinfo {title} {Quantum phase transitions
  of metals in two spatial dimensions. i. ising-nematic order},\ }\href
  {https://doi.org/10.1103/PhysRevB.82.075127} {\bibfield  {journal} {\bibinfo
  {journal} {Phys. Rev. B}\ }\textbf {\bibinfo {volume} {82}},\ \bibinfo
  {pages} {075127} (\bibinfo {year} {2010}{\natexlab{a}})}\BibitemShut
  {NoStop}%
\bibitem [{\citenamefont {Metlitski}\ and\ \citenamefont
  {Sachdev}(2010{\natexlab{b}})}]{Metlitski2010}%
  \BibitemOpen
  \bibfield  {author} {\bibinfo {author} {\bibfnamefont {M.~A.}\ \bibnamefont
  {Metlitski}}\ and\ \bibinfo {author} {\bibfnamefont {S.}~\bibnamefont
  {Sachdev}},\ }\bibfield  {title} {\bibinfo {title} {Quantum phase transitions
  of metals in two spatial dimensions. ii. spin density wave order},\ }\href
  {https://doi.org/10.1103/PhysRevB.82.075128} {\bibfield  {journal} {\bibinfo
  {journal} {Phys. Rev. B}\ }\textbf {\bibinfo {volume} {82}},\ \bibinfo
  {pages} {075128} (\bibinfo {year} {2010}{\natexlab{b}})}\BibitemShut
  {NoStop}%
\bibitem [{\citenamefont {Metlitski}\ and\ \citenamefont
  {Sachdev}(2010{\natexlab{c}})}]{Metlitski2010b}%
  \BibitemOpen
  \bibfield  {author} {\bibinfo {author} {\bibfnamefont {M.~A.}\ \bibnamefont
  {Metlitski}}\ and\ \bibinfo {author} {\bibfnamefont {S.}~\bibnamefont
  {Sachdev}},\ }\bibfield  {title} {\bibinfo {title} {Instabilities near the
  onset of spin density wave order in metals},\ }\href
  {http://stacks.iop.org/1367-2630/12/i=10/a=105007} {\bibfield  {journal}
  {\bibinfo  {journal} {New Journal of Physics}\ }\textbf {\bibinfo {volume}
  {12}},\ \bibinfo {pages} {105007} (\bibinfo {year}
  {2010}{\natexlab{c}})}\BibitemShut {NoStop}%
\bibitem [{\citenamefont {Lederer}\ \emph {et~al.}(2017)\citenamefont
  {Lederer}, \citenamefont {Schattner}, \citenamefont {Berg},\ and\
  \citenamefont {Kivelson}}]{Lederer2017}%
  \BibitemOpen
  \bibfield  {author} {\bibinfo {author} {\bibfnamefont {S.}~\bibnamefont
  {Lederer}}, \bibinfo {author} {\bibfnamefont {Y.}~\bibnamefont {Schattner}},
  \bibinfo {author} {\bibfnamefont {E.}~\bibnamefont {Berg}},\ and\ \bibinfo
  {author} {\bibfnamefont {S.~A.}\ \bibnamefont {Kivelson}},\ }\bibfield
  {title} {\bibinfo {title} {Superconductivity and non-fermi liquid behavior
  near a nematic quantum critical point},\ }\href
  {https://doi.org/10.1073/pnas.1620651114} {\bibfield  {journal} {\bibinfo
  {journal} {Proceedings of the National Academy of Sciences}\ }\textbf
  {\bibinfo {volume} {114}},\ \bibinfo {pages} {4905} (\bibinfo {year}
  {2017})}\BibitemShut {NoStop}%
\bibitem [{\citenamefont {Lee}(2018)}]{Lee2018}%
  \BibitemOpen
  \bibfield  {author} {\bibinfo {author} {\bibfnamefont {S.-S.}\ \bibnamefont
  {Lee}},\ }\bibfield  {title} {\bibinfo {title} {Recent developments in
  non-fermi liquid theory},\ }\bibfield  {booktitle} {\emph {\bibinfo
  {booktitle} {Annual Review of Condensed Matter Physics}},\ }\href
  {https://doi.org/10.1146/annurev-conmatphys-031016-025531} {\bibfield
  {journal} {\bibinfo  {journal} {Annu. Rev. Condens. Matter Phys.}\ }\textbf
  {\bibinfo {volume} {9}},\ \bibinfo {pages} {227} (\bibinfo {year}
  {2018})}\BibitemShut {NoStop}%
\bibitem [{\citenamefont {Maslov}\ and\ \citenamefont
  {Chubukov}(2010)}]{Maslov2010}%
  \BibitemOpen
  \bibfield  {author} {\bibinfo {author} {\bibfnamefont {D.~L.}\ \bibnamefont
  {Maslov}}\ and\ \bibinfo {author} {\bibfnamefont {A.~V.}\ \bibnamefont
  {Chubukov}},\ }\bibfield  {title} {\bibinfo {title} {Fermi liquid near
  pomeranchuk quantum criticality},\ }\href
  {https://doi.org/10.1103/PhysRevB.81.045110} {\bibfield  {journal} {\bibinfo
  {journal} {Phys. Rev. B}\ }\textbf {\bibinfo {volume} {81}},\ \bibinfo
  {pages} {045110} (\bibinfo {year} {2010})}\BibitemShut {NoStop}%
\bibitem [{\citenamefont {Fradkin}\ \emph {et~al.}(2010)\citenamefont
  {Fradkin}, \citenamefont {Kivelson}, \citenamefont {Lawler}, \citenamefont
  {Eisenstein},\ and\ \citenamefont {Mackenzie}}]{Fradkin2010}%
  \BibitemOpen
  \bibfield  {author} {\bibinfo {author} {\bibfnamefont {E.}~\bibnamefont
  {Fradkin}}, \bibinfo {author} {\bibfnamefont {S.~A.}\ \bibnamefont
  {Kivelson}}, \bibinfo {author} {\bibfnamefont {M.~J.}\ \bibnamefont
  {Lawler}}, \bibinfo {author} {\bibfnamefont {J.~P.}\ \bibnamefont
  {Eisenstein}},\ and\ \bibinfo {author} {\bibfnamefont {A.~P.}\ \bibnamefont
  {Mackenzie}},\ }\bibfield  {title} {\bibinfo {title} {Nematic fermi fluids in
  condensed matter physics},\ }\bibfield  {booktitle} {\emph {\bibinfo
  {booktitle} {Annual Review of Condensed Matter Physics}},\ }\href
  {https://doi.org/10.1146/annurev-conmatphys-070909-103925} {\bibfield
  {journal} {\bibinfo  {journal} {Annu. Rev. Condens. Matter Phys.}\ }\textbf
  {\bibinfo {volume} {1}},\ \bibinfo {pages} {153} (\bibinfo {year}
  {2010})}\BibitemShut {NoStop}%
\bibitem [{\citenamefont {Wang}\ \emph {et~al.}(2016)\citenamefont {Wang},
  \citenamefont {Abanov}, \citenamefont {Altshuler}, \citenamefont
  {Yuzbashyan},\ and\ \citenamefont {Chubukov}}]{Wang2016}%
  \BibitemOpen
  \bibfield  {author} {\bibinfo {author} {\bibfnamefont {Y.}~\bibnamefont
  {Wang}}, \bibinfo {author} {\bibfnamefont {A.}~\bibnamefont {Abanov}},
  \bibinfo {author} {\bibfnamefont {B.~L.}\ \bibnamefont {Altshuler}}, \bibinfo
  {author} {\bibfnamefont {E.~A.}\ \bibnamefont {Yuzbashyan}},\ and\ \bibinfo
  {author} {\bibfnamefont {A.~V.}\ \bibnamefont {Chubukov}},\ }\bibfield
  {title} {\bibinfo {title} {Superconductivity near a quantum-critical point:
  The special role of the first matsubara frequency},\ }\href
  {https://doi.org/10.1103/PhysRevLett.117.157001} {\bibfield  {journal}
  {\bibinfo  {journal} {Phys. Rev. Lett.}\ }\textbf {\bibinfo {volume} {117}},\
  \bibinfo {pages} {157001} (\bibinfo {year} {2016})}\BibitemShut {NoStop}%
\bibitem [{\citenamefont {Raghu}\ \emph {et~al.}(2015)\citenamefont {Raghu},
  \citenamefont {Torroba},\ and\ \citenamefont {Wang}}]{Raghu2015}%
  \BibitemOpen
  \bibfield  {author} {\bibinfo {author} {\bibfnamefont {S.}~\bibnamefont
  {Raghu}}, \bibinfo {author} {\bibfnamefont {G.}~\bibnamefont {Torroba}},\
  and\ \bibinfo {author} {\bibfnamefont {H.}~\bibnamefont {Wang}},\ }\bibfield
  {title} {\bibinfo {title} {Metallic quantum critical points with finite bcs
  couplings},\ }\href {https://doi.org/10.1103/PhysRevB.92.205104} {\bibfield
  {journal} {\bibinfo  {journal} {Phys. Rev. B}\ }\textbf {\bibinfo {volume}
  {92}},\ \bibinfo {pages} {205104} (\bibinfo {year} {2015})}\BibitemShut
  {NoStop}%
\bibitem [{\citenamefont {Metlitski}\ \emph {et~al.}(2015)\citenamefont
  {Metlitski}, \citenamefont {Mross}, \citenamefont {Sachdev},\ and\
  \citenamefont {Senthil}}]{Metlitski2015}%
  \BibitemOpen
  \bibfield  {author} {\bibinfo {author} {\bibfnamefont {M.~A.}\ \bibnamefont
  {Metlitski}}, \bibinfo {author} {\bibfnamefont {D.~F.}\ \bibnamefont
  {Mross}}, \bibinfo {author} {\bibfnamefont {S.}~\bibnamefont {Sachdev}},\
  and\ \bibinfo {author} {\bibfnamefont {T.}~\bibnamefont {Senthil}},\
  }\bibfield  {title} {\bibinfo {title} {Cooper pairing in non-fermi liquids},\
  }\href {https://doi.org/10.1103/PhysRevB.91.115111} {\bibfield  {journal}
  {\bibinfo  {journal} {Phys. Rev. B}\ }\textbf {\bibinfo {volume} {91}},\
  \bibinfo {pages} {115111} (\bibinfo {year} {2015})}\BibitemShut {NoStop}%
\bibitem [{\citenamefont {Lederer}\ \emph {et~al.}(2015)\citenamefont
  {Lederer}, \citenamefont {Schattner}, \citenamefont {Berg},\ and\
  \citenamefont {Kivelson}}]{Lederer2015}%
  \BibitemOpen
  \bibfield  {author} {\bibinfo {author} {\bibfnamefont {S.}~\bibnamefont
  {Lederer}}, \bibinfo {author} {\bibfnamefont {Y.}~\bibnamefont {Schattner}},
  \bibinfo {author} {\bibfnamefont {E.}~\bibnamefont {Berg}},\ and\ \bibinfo
  {author} {\bibfnamefont {S.~A.}\ \bibnamefont {Kivelson}},\ }\bibfield
  {title} {\bibinfo {title} {Enhancement of superconductivity near a nematic
  quantum critical point},\ }\href
  {https://doi.org/10.1103/PhysRevLett.114.097001} {\bibfield  {journal}
  {\bibinfo  {journal} {Phys. Rev. Lett.}\ }\textbf {\bibinfo {volume} {114}},\
  \bibinfo {pages} {097001} (\bibinfo {year} {2015})}\BibitemShut {NoStop}%
\bibitem [{\citenamefont {L\"ohneysen}\ \emph {et~al.}(2007)\citenamefont
  {L\"ohneysen}, \citenamefont {Rosch}, \citenamefont {Vojta},\ and\
  \citenamefont {W\"olfle}}]{Loehneysen2007}%
  \BibitemOpen
  \bibfield  {author} {\bibinfo {author} {\bibfnamefont {H.~V.}\ \bibnamefont
  {L\"ohneysen}}, \bibinfo {author} {\bibfnamefont {A.}~\bibnamefont {Rosch}},
  \bibinfo {author} {\bibfnamefont {M.}~\bibnamefont {Vojta}},\ and\ \bibinfo
  {author} {\bibfnamefont {P.}~\bibnamefont {W\"olfle}},\ }\bibfield  {title}
  {\bibinfo {title} {Fermi-liquid instabilities at magnetic quantum phase
  transitions},\ }\href {https://doi.org/10.1103/RevModPhys.79.1015} {\bibfield
   {journal} {\bibinfo  {journal} {Rev. Mod. Phys.}\ }\textbf {\bibinfo
  {volume} {79}},\ \bibinfo {pages} {1015} (\bibinfo {year}
  {2007})}\BibitemShut {NoStop}%
\bibitem [{\citenamefont {Monthoux}\ \emph {et~al.}(2007)\citenamefont
  {Monthoux}, \citenamefont {Pines},\ and\ \citenamefont
  {Lonzarich}}]{Monthoux2007}%
  \BibitemOpen
  \bibfield  {author} {\bibinfo {author} {\bibfnamefont {P.}~\bibnamefont
  {Monthoux}}, \bibinfo {author} {\bibfnamefont {D.}~\bibnamefont {Pines}},\
  and\ \bibinfo {author} {\bibfnamefont {G.~G.}\ \bibnamefont {Lonzarich}},\
  }\bibfield  {title} {\bibinfo {title} {Superconductivity without phonons},\
  }\href {http://dx.doi.org/10.1038/nature06480} {\bibfield  {journal}
  {\bibinfo  {journal} {Nature}\ }\textbf {\bibinfo {volume} {450}},\ \bibinfo
  {pages} {1177} (\bibinfo {year} {2007})}\BibitemShut {NoStop}%
\bibitem [{\citenamefont {Scalapino}(2012)}]{Scalapino2012}%
  \BibitemOpen
  \bibfield  {author} {\bibinfo {author} {\bibfnamefont {D.~J.}\ \bibnamefont
  {Scalapino}},\ }\bibfield  {title} {\bibinfo {title} {A common thread: The
  pairing interaction for unconventional superconductors},\ }\href
  {https://doi.org/10.1103/RevModPhys.84.1383} {\bibfield  {journal} {\bibinfo
  {journal} {Rev. Mod. Phys.}\ }\textbf {\bibinfo {volume} {84}},\ \bibinfo
  {pages} {1383} (\bibinfo {year} {2012})}\BibitemShut {NoStop}%
\bibitem [{\citenamefont {Sachdev}\ \emph {et~al.}(2012)\citenamefont
  {Sachdev}, \citenamefont {Metlitski},\ and\ \citenamefont
  {Punk}}]{Sachdev2012}%
  \BibitemOpen
  \bibfield  {author} {\bibinfo {author} {\bibfnamefont {S.}~\bibnamefont
  {Sachdev}}, \bibinfo {author} {\bibfnamefont {M.~A.}\ \bibnamefont
  {Metlitski}},\ and\ \bibinfo {author} {\bibfnamefont {M.}~\bibnamefont
  {Punk}},\ }\bibfield  {title} {\bibinfo {title} {Antiferromagnetism in
  metals: from the cuprate superconductors to the heavy fermion materials},\
  }\href {http://stacks.iop.org/0953-8984/24/i=29/a=294205} {\bibfield
  {journal} {\bibinfo  {journal} {Journal of Physics: Condensed Matter}\
  }\textbf {\bibinfo {volume} {24}},\ \bibinfo {pages} {294205} (\bibinfo
  {year} {2012})},\ \bibinfo {note} {and references therein}\BibitemShut
  {NoStop}%
\bibitem [{\citenamefont {Cyr-Choini\`ere}\ \emph {et~al.}(2018)\citenamefont
  {Cyr-Choini\`ere}, \citenamefont {Daou}, \citenamefont {Lalibert\'e},
  \citenamefont {Collignon}, \citenamefont {Badoux}, \citenamefont {LeBoeuf},
  \citenamefont {Chang}, \citenamefont {Ramshaw}, \citenamefont {Bonn},
  \citenamefont {Hardy}, \citenamefont {Liang}, \citenamefont {Yan},
  \citenamefont {Cheng}, \citenamefont {Zhou}, \citenamefont {Goodenough},
  \citenamefont {Pyon}, \citenamefont {Takayama}, \citenamefont {Takagi},
  \citenamefont {Doiron-Leyraud},\ and\ \citenamefont
  {Taillefer}}]{Cyr-Choiniere2018}%
  \BibitemOpen
  \bibfield  {author} {\bibinfo {author} {\bibfnamefont {O.}~\bibnamefont
  {Cyr-Choini\`ere}}, \bibinfo {author} {\bibfnamefont {R.}~\bibnamefont
  {Daou}}, \bibinfo {author} {\bibfnamefont {F.}~\bibnamefont {Lalibert\'e}},
  \bibinfo {author} {\bibfnamefont {C.}~\bibnamefont {Collignon}}, \bibinfo
  {author} {\bibfnamefont {S.}~\bibnamefont {Badoux}}, \bibinfo {author}
  {\bibfnamefont {D.}~\bibnamefont {LeBoeuf}}, \bibinfo {author} {\bibfnamefont
  {J.}~\bibnamefont {Chang}}, \bibinfo {author} {\bibfnamefont {B.~J.}\
  \bibnamefont {Ramshaw}}, \bibinfo {author} {\bibfnamefont {D.~A.}\
  \bibnamefont {Bonn}}, \bibinfo {author} {\bibfnamefont {W.~N.}\ \bibnamefont
  {Hardy}}, \bibinfo {author} {\bibfnamefont {R.}~\bibnamefont {Liang}},
  \bibinfo {author} {\bibfnamefont {J.-Q.}\ \bibnamefont {Yan}}, \bibinfo
  {author} {\bibfnamefont {J.-G.}\ \bibnamefont {Cheng}}, \bibinfo {author}
  {\bibfnamefont {J.-S.}\ \bibnamefont {Zhou}}, \bibinfo {author}
  {\bibfnamefont {J.~B.}\ \bibnamefont {Goodenough}}, \bibinfo {author}
  {\bibfnamefont {S.}~\bibnamefont {Pyon}}, \bibinfo {author} {\bibfnamefont
  {T.}~\bibnamefont {Takayama}}, \bibinfo {author} {\bibfnamefont
  {H.}~\bibnamefont {Takagi}}, \bibinfo {author} {\bibfnamefont
  {N.}~\bibnamefont {Doiron-Leyraud}},\ and\ \bibinfo {author} {\bibfnamefont
  {L.}~\bibnamefont {Taillefer}},\ }\bibfield  {title} {\bibinfo {title}
  {Pseudogap temperature ${T}^{*}$ of cuprate superconductors from the nernst
  effect},\ }\href {https://doi.org/10.1103/PhysRevB.97.064502} {\bibfield
  {journal} {\bibinfo  {journal} {Phys. Rev. B}\ }\textbf {\bibinfo {volume}
  {97}},\ \bibinfo {pages} {064502} (\bibinfo {year} {2018})}\BibitemShut
  {NoStop}%
\bibitem [{\citenamefont {Fernandes}\ \emph {et~al.}(2014)\citenamefont
  {Fernandes}, \citenamefont {Chubukov},\ and\ \citenamefont
  {Schmalian}}]{Fernandes2014}%
  \BibitemOpen
  \bibfield  {author} {\bibinfo {author} {\bibfnamefont {R.~M.}\ \bibnamefont
  {Fernandes}}, \bibinfo {author} {\bibfnamefont {A.~V.}\ \bibnamefont
  {Chubukov}},\ and\ \bibinfo {author} {\bibfnamefont {J.}~\bibnamefont
  {Schmalian}},\ }\bibfield  {title} {\bibinfo {title} {What drives nematic
  order in iron-based superconductors?},\ }\href
  {http://dx.doi.org/10.1038/nphys2877} {\bibfield  {journal} {\bibinfo
  {journal} {Nat Phys}\ }\textbf {\bibinfo {volume} {10}},\ \bibinfo {pages}
  {97} (\bibinfo {year} {2014})}\BibitemShut {NoStop}%
\bibitem [{\citenamefont {Wang}\ \emph {et~al.}(2015)\citenamefont {Wang},
  \citenamefont {Kivelson},\ and\ \citenamefont {Lee}}]{Wang2015}%
  \BibitemOpen
  \bibfield  {author} {\bibinfo {author} {\bibfnamefont {F.}~\bibnamefont
  {Wang}}, \bibinfo {author} {\bibfnamefont {S.~A.}\ \bibnamefont {Kivelson}},\
  and\ \bibinfo {author} {\bibfnamefont {D.-H.}\ \bibnamefont {Lee}},\
  }\bibfield  {title} {\bibinfo {title} {Nematicity and quantum paramagnetism
  in fese},\ }\href {http://dx.doi.org/10.1038/nphys3456} {\bibfield  {journal}
  {\bibinfo  {journal} {Nat Phys}\ }\textbf {\bibinfo {volume} {11}},\ \bibinfo
  {pages} {959} (\bibinfo {year} {2015})}\BibitemShut {NoStop}%
\bibitem [{\citenamefont {Wang}\ and\ \citenamefont
  {Chubukov}(2014)}]{Wang2014}%
  \BibitemOpen
  \bibfield  {author} {\bibinfo {author} {\bibfnamefont {Y.}~\bibnamefont
  {Wang}}\ and\ \bibinfo {author} {\bibfnamefont {A.}~\bibnamefont
  {Chubukov}},\ }\bibfield  {title} {\bibinfo {title} {Charge-density-wave
  order with momentum $(2q,0)$ and $(0,2q)$ within the spin-fermion model:
  Continuous and discrete symmetry breaking, preemptive composite order, and
  relation to pseudogap in hole-doped cuprates},\ }\href
  {https://doi.org/10.1103/PhysRevB.90.035149} {\bibfield  {journal} {\bibinfo
  {journal} {Phys. Rev. B}\ }\textbf {\bibinfo {volume} {90}},\ \bibinfo
  {pages} {035149} (\bibinfo {year} {2014})}\BibitemShut {NoStop}%
\bibitem [{\citenamefont {Millis}(1992)}]{Millis1992}%
  \BibitemOpen
  \bibfield  {author} {\bibinfo {author} {\bibfnamefont {A.~J.}\ \bibnamefont
  {Millis}},\ }\bibfield  {title} {\bibinfo {title} {Nearly antiferromagnetic
  fermi liquids: An analytic eliashberg approach},\ }\href
  {https://doi.org/10.1103/PhysRevB.45.13047} {\bibfield  {journal} {\bibinfo
  {journal} {Phys. Rev. B}\ }\textbf {\bibinfo {volume} {45}},\ \bibinfo
  {pages} {13047} (\bibinfo {year} {1992})}\BibitemShut {NoStop}%
\bibitem [{\citenamefont {Rech}\ \emph {et~al.}(2006)\citenamefont {Rech},
  \citenamefont {P\'epin},\ and\ \citenamefont {Chubukov}}]{Rech2006}%
  \BibitemOpen
  \bibfield  {author} {\bibinfo {author} {\bibfnamefont {J.}~\bibnamefont
  {Rech}}, \bibinfo {author} {\bibfnamefont {C.}~\bibnamefont {P\'epin}},\ and\
  \bibinfo {author} {\bibfnamefont {A.~V.}\ \bibnamefont {Chubukov}},\
  }\bibfield  {title} {\bibinfo {title} {Quantum critical behavior in itinerant
  electron systems: Eliashberg theory and instability of a ferromagnetic
  quantum critical point},\ }\href {https://doi.org/10.1103/PhysRevB.74.195126}
  {\bibfield  {journal} {\bibinfo  {journal} {Phys. Rev. B}\ }\textbf {\bibinfo
  {volume} {74}},\ \bibinfo {pages} {195126} (\bibinfo {year}
  {2006})}\BibitemShut {NoStop}%
\bibitem [{\citenamefont {{Chubukov}}\ \emph {et~al.}(2020)\citenamefont
  {{Chubukov}}, \citenamefont {{Abanov}}, \citenamefont {{Esterlis}},\ and\
  \citenamefont {A.}}]{Kivelson2020}%
  \BibitemOpen
  \bibfield  {author} {\bibinfo {author} {\bibfnamefont {A.~V.}\ \bibnamefont
  {{Chubukov}}}, \bibinfo {author} {\bibfnamefont {A.}~\bibnamefont
  {{Abanov}}}, \bibinfo {author} {\bibfnamefont {I.}~\bibnamefont
  {{Esterlis}}},\ and\ \bibinfo {author} {\bibfnamefont {K.~S.}\ \bibnamefont
  {A.}},\ }\bibfield  {title} {\bibinfo {title} {{Eliashberg theory of
  phonon-mediated superconductivity -- when it is valid and how it breaks
  down}},\ }\href@noop {} {\bibfield  {journal} {\bibinfo  {journal} {ArXiv
  e-prints}\ } (\bibinfo {year} {2020})}\BibitemShut {NoStop}%
\bibitem [{\citenamefont {Lee}(2009)}]{Lee2009}%
  \BibitemOpen
  \bibfield  {author} {\bibinfo {author} {\bibfnamefont {S.-S.}\ \bibnamefont
  {Lee}},\ }\bibfield  {title} {\bibinfo {title} {Low-energy effective theory
  of fermi surface coupled with u(1) gauge field in $2+1$ dimensions},\ }\href
  {https://doi.org/10.1103/PhysRevB.80.165102} {\bibfield  {journal} {\bibinfo
  {journal} {Phys. Rev. B}\ }\textbf {\bibinfo {volume} {80}},\ \bibinfo
  {pages} {165102} (\bibinfo {year} {2009})}\BibitemShut {NoStop}%
\bibitem [{\citenamefont {S-S.~Lee}()}]{LKCunp}%
  \BibitemOpen
  \bibfield  {author} {\bibinfo {author} {\bibfnamefont {A.~V.~C.}\
  \bibnamefont {S-S.~Lee}, \bibfnamefont {Y.~B.~Kim}},\ }\bibinfo {note}
  {unpublished}\BibitemShut {NoStop}%
\bibitem [{\citenamefont {Holder}\ and\ \citenamefont
  {Metzner}(2015{\natexlab{a}})}]{Holder2015}%
  \BibitemOpen
  \bibfield  {author} {\bibinfo {author} {\bibfnamefont {T.}~\bibnamefont
  {Holder}}\ and\ \bibinfo {author} {\bibfnamefont {W.}~\bibnamefont
  {Metzner}},\ }\bibfield  {title} {\bibinfo {title} {Anomalous dynamical
  scaling from nematic and u(1) gauge field fluctuations in two-dimensional
  metals},\ }\href {https://doi.org/10.1103/PhysRevB.92.041112} {\bibfield
  {journal} {\bibinfo  {journal} {Phys. Rev. B}\ }\textbf {\bibinfo {volume}
  {92}},\ \bibinfo {pages} {041112} (\bibinfo {year}
  {2015}{\natexlab{a}})}\BibitemShut {NoStop}%
\bibitem [{\citenamefont {Holder}\ and\ \citenamefont
  {Metzner}(2015{\natexlab{b}})}]{Holder2015a}%
  \BibitemOpen
  \bibfield  {author} {\bibinfo {author} {\bibfnamefont {T.}~\bibnamefont
  {Holder}}\ and\ \bibinfo {author} {\bibfnamefont {W.}~\bibnamefont
  {Metzner}},\ }\bibfield  {title} {\bibinfo {title} {Fermion loops and
  improved power-counting in two-dimensional critical metals with singular
  forward scattering},\ }\href {https://doi.org/10.1103/PhysRevB.92.245128}
  {\bibfield  {journal} {\bibinfo  {journal} {Phys. Rev. B}\ }\textbf {\bibinfo
  {volume} {92}},\ \bibinfo {pages} {245128} (\bibinfo {year}
  {2015}{\natexlab{b}})}\BibitemShut {NoStop}%
\bibitem [{Note1()}]{Note1}%
  \BibitemOpen
  \bibinfo {note} {Whether these logarithmic corrections give rise to the
  appearance of an anomalous fermionic residue but preserve the $\omega ^{2/3}$
  scaling for the self-energy is not known.}\BibitemShut {Stop}%
\bibitem [{Note2()}]{Note2}%
  \BibitemOpen
  \bibinfo {note} {It was argued~\cite {Lee2018} that because of these
  logarithms, the system eventually flows towards the new fixed point with the
  dynamical exponent $z=1$.}\BibitemShut {Stop}%
\bibitem [{\citenamefont {Wang}\ and\ \citenamefont
  {Chubukov}(2013)}]{Wang2013}%
  \BibitemOpen
  \bibfield  {author} {\bibinfo {author} {\bibfnamefont {Y.}~\bibnamefont
  {Wang}}\ and\ \bibinfo {author} {\bibfnamefont {A.~V.}\ \bibnamefont
  {Chubukov}},\ }\bibfield  {title} {\bibinfo {title} {Superconductivity at the
  onset of spin-density-wave order in a metal},\ }\href
  {https://doi.org/10.1103/PhysRevLett.110.127001} {\bibfield  {journal}
  {\bibinfo  {journal} {Phys. Rev. Lett.}\ }\textbf {\bibinfo {volume} {110}},\
  \bibinfo {pages} {127001} (\bibinfo {year} {2013})}\BibitemShut {NoStop}%
\bibitem [{\citenamefont {Bonesteel}\ \emph {et~al.}(1996)\citenamefont
  {Bonesteel}, \citenamefont {McDonald},\ and\ \citenamefont
  {Nayak}}]{Bonesteel1996}%
  \BibitemOpen
  \bibfield  {author} {\bibinfo {author} {\bibfnamefont {N.~E.}\ \bibnamefont
  {Bonesteel}}, \bibinfo {author} {\bibfnamefont {I.~A.}\ \bibnamefont
  {McDonald}},\ and\ \bibinfo {author} {\bibfnamefont {C.}~\bibnamefont
  {Nayak}},\ }\bibfield  {title} {\bibinfo {title} {Gauge fields and pairing in
  double-layer composite fermion metals},\ }\href
  {https://doi.org/10.1103/PhysRevLett.77.3009} {\bibfield  {journal} {\bibinfo
   {journal} {Phys. Rev. Lett.}\ }\textbf {\bibinfo {volume} {77}},\ \bibinfo
  {pages} {3009} (\bibinfo {year} {1996})}\BibitemShut {NoStop}%
\bibitem [{\citenamefont {Chubukov}\ \emph {et~al.}(1997)\citenamefont
  {Chubukov}, \citenamefont {Monthoux},\ and\ \citenamefont
  {Morr}}]{Monthoux1997}%
  \BibitemOpen
  \bibfield  {author} {\bibinfo {author} {\bibfnamefont {A.~V.}\ \bibnamefont
  {Chubukov}}, \bibinfo {author} {\bibfnamefont {P.}~\bibnamefont {Monthoux}},\
  and\ \bibinfo {author} {\bibfnamefont {D.~K.}\ \bibnamefont {Morr}},\
  }\bibfield  {title} {\bibinfo {title} {Vertex corrections in
  antiferromagnetic spin-fluctuation theories},\ }\href
  {https://doi.org/10.1103/PhysRevB.56.7789} {\bibfield  {journal} {\bibinfo
  {journal} {Phys. Rev. B}\ }\textbf {\bibinfo {volume} {56}},\ \bibinfo
  {pages} {7789} (\bibinfo {year} {1997})}\BibitemShut {NoStop}%
\bibitem [{\citenamefont {Schattner}\ \emph
  {et~al.}(2016{\natexlab{a}})\citenamefont {Schattner}, \citenamefont
  {Gerlach}, \citenamefont {Trebst},\ and\ \citenamefont
  {Berg}}]{Schattner2016a}%
  \BibitemOpen
  \bibfield  {author} {\bibinfo {author} {\bibfnamefont {Y.}~\bibnamefont
  {Schattner}}, \bibinfo {author} {\bibfnamefont {M.~H.}\ \bibnamefont
  {Gerlach}}, \bibinfo {author} {\bibfnamefont {S.}~\bibnamefont {Trebst}},\
  and\ \bibinfo {author} {\bibfnamefont {E.}~\bibnamefont {Berg}},\ }\bibfield
  {title} {\bibinfo {title} {Competing orders in a nearly antiferromagnetic
  metal},\ }\href {https://doi.org/10.1103/PhysRevLett.117.097002} {\bibfield
  {journal} {\bibinfo  {journal} {Phys. Rev. Lett.}\ }\textbf {\bibinfo
  {volume} {117}},\ \bibinfo {pages} {097002} (\bibinfo {year}
  {2016}{\natexlab{a}})}\BibitemShut {NoStop}%
\bibitem [{\citenamefont {Gerlach}\ \emph {et~al.}(2017)\citenamefont
  {Gerlach}, \citenamefont {Schattner}, \citenamefont {Berg},\ and\
  \citenamefont {Trebst}}]{Gerlach2017}%
  \BibitemOpen
  \bibfield  {author} {\bibinfo {author} {\bibfnamefont {M.~H.}\ \bibnamefont
  {Gerlach}}, \bibinfo {author} {\bibfnamefont {Y.}~\bibnamefont {Schattner}},
  \bibinfo {author} {\bibfnamefont {E.}~\bibnamefont {Berg}},\ and\ \bibinfo
  {author} {\bibfnamefont {S.}~\bibnamefont {Trebst}},\ }\bibfield  {title}
  {\bibinfo {title} {Quantum critical properties of a metallic
  spin-density-wave transition},\ }\href
  {https://doi.org/10.1103/PhysRevB.95.035124} {\bibfield  {journal} {\bibinfo
  {journal} {Phys. Rev. B}\ }\textbf {\bibinfo {volume} {95}},\ \bibinfo
  {pages} {035124} (\bibinfo {year} {2017})}\BibitemShut {NoStop}%
\bibitem [{\citenamefont {Schattner}\ \emph
  {et~al.}(2016{\natexlab{b}})\citenamefont {Schattner}, \citenamefont
  {Lederer}, \citenamefont {Kivelson},\ and\ \citenamefont
  {Berg}}]{Schattner2016}%
  \BibitemOpen
  \bibfield  {author} {\bibinfo {author} {\bibfnamefont {Y.}~\bibnamefont
  {Schattner}}, \bibinfo {author} {\bibfnamefont {S.}~\bibnamefont {Lederer}},
  \bibinfo {author} {\bibfnamefont {S.~A.}\ \bibnamefont {Kivelson}},\ and\
  \bibinfo {author} {\bibfnamefont {E.}~\bibnamefont {Berg}},\ }\bibfield
  {title} {\bibinfo {title} {Ising nematic quantum critical point in a metal: A
  monte carlo study},\ }\href {https://doi.org/10.1103/PhysRevX.6.031028}
  {\bibfield  {journal} {\bibinfo  {journal} {Phys. Rev. X}\ }\textbf {\bibinfo
  {volume} {6}},\ \bibinfo {pages} {031028} (\bibinfo {year}
  {2016}{\natexlab{b}})}\BibitemShut {NoStop}%
\bibitem [{\citenamefont {Wang}\ \emph {et~al.}(2017)\citenamefont {Wang},
  \citenamefont {Schattner}, \citenamefont {Berg},\ and\ \citenamefont
  {Fernandes}}]{Wang2017}%
  \BibitemOpen
  \bibfield  {author} {\bibinfo {author} {\bibfnamefont {X.}~\bibnamefont
  {Wang}}, \bibinfo {author} {\bibfnamefont {Y.}~\bibnamefont {Schattner}},
  \bibinfo {author} {\bibfnamefont {E.}~\bibnamefont {Berg}},\ and\ \bibinfo
  {author} {\bibfnamefont {R.~M.}\ \bibnamefont {Fernandes}},\ }\bibfield
  {title} {\bibinfo {title} {Superconductivity mediated by quantum critical
  antiferromagnetic fluctuations: The rise and fall of hot spots},\ }\href
  {https://doi.org/10.1103/PhysRevB.95.174520} {\bibfield  {journal} {\bibinfo
  {journal} {Phys. Rev. B}\ }\textbf {\bibinfo {volume} {95}},\ \bibinfo
  {pages} {174520} (\bibinfo {year} {2017})}\BibitemShut {NoStop}%
\bibitem [{\citenamefont {Liu}\ \emph {et~al.}(2018)\citenamefont {Liu},
  \citenamefont {Xu}, \citenamefont {Qi}, \citenamefont {Sun},\ and\
  \citenamefont {Meng}}]{Liu2018}%
  \BibitemOpen
  \bibfield  {author} {\bibinfo {author} {\bibfnamefont {Z.~H.}\ \bibnamefont
  {Liu}}, \bibinfo {author} {\bibfnamefont {X.~Y.}\ \bibnamefont {Xu}},
  \bibinfo {author} {\bibfnamefont {Y.}~\bibnamefont {Qi}}, \bibinfo {author}
  {\bibfnamefont {K.}~\bibnamefont {Sun}},\ and\ \bibinfo {author}
  {\bibfnamefont {Z.~Y.}\ \bibnamefont {Meng}},\ }\bibfield  {title} {\bibinfo
  {title} {Itinerant quantum critical point with frustration and a non-fermi
  liquid},\ }\href {https://doi.org/10.1103/PhysRevB.98.045116} {\bibfield
  {journal} {\bibinfo  {journal} {Phys. Rev. B}\ }\textbf {\bibinfo {volume}
  {98}},\ \bibinfo {pages} {045116} (\bibinfo {year} {2018})}\BibitemShut
  {NoStop}%
\bibitem [{\citenamefont {Xu}\ \emph {et~al.}(2017{\natexlab{a}})\citenamefont
  {Xu}, \citenamefont {Qi}, \citenamefont {Liu}, \citenamefont {Fu},\ and\
  \citenamefont {Meng}}]{Xu2017}%
  \BibitemOpen
  \bibfield  {author} {\bibinfo {author} {\bibfnamefont {X.~Y.}\ \bibnamefont
  {Xu}}, \bibinfo {author} {\bibfnamefont {Y.}~\bibnamefont {Qi}}, \bibinfo
  {author} {\bibfnamefont {J.}~\bibnamefont {Liu}}, \bibinfo {author}
  {\bibfnamefont {L.}~\bibnamefont {Fu}},\ and\ \bibinfo {author}
  {\bibfnamefont {Z.~Y.}\ \bibnamefont {Meng}},\ }\bibfield  {title} {\bibinfo
  {title} {Self-learning quantum monte carlo method in interacting fermion
  systems},\ }\href {https://doi.org/10.1103/PhysRevB.96.041119} {\bibfield
  {journal} {\bibinfo  {journal} {Phys. Rev. B}\ }\textbf {\bibinfo {volume}
  {96}},\ \bibinfo {pages} {041119} (\bibinfo {year}
  {2017}{\natexlab{a}})}\BibitemShut {NoStop}%
\bibitem [{\citenamefont {Xu}\ \emph {et~al.}(2017{\natexlab{b}})\citenamefont
  {Xu}, \citenamefont {Sun}, \citenamefont {Schattner}, \citenamefont {Berg},\
  and\ \citenamefont {Meng}}]{Xu2017a}%
  \BibitemOpen
  \bibfield  {author} {\bibinfo {author} {\bibfnamefont {X.~Y.}\ \bibnamefont
  {Xu}}, \bibinfo {author} {\bibfnamefont {K.}~\bibnamefont {Sun}}, \bibinfo
  {author} {\bibfnamefont {Y.}~\bibnamefont {Schattner}}, \bibinfo {author}
  {\bibfnamefont {E.}~\bibnamefont {Berg}},\ and\ \bibinfo {author}
  {\bibfnamefont {Z.~Y.}\ \bibnamefont {Meng}},\ }\bibfield  {title} {\bibinfo
  {title} {Non-fermi liquid at ($2+1$)$\mathrm{D}$ ferromagnetic quantum
  critical point},\ }\href {https://doi.org/10.1103/PhysRevX.7.031058}
  {\bibfield  {journal} {\bibinfo  {journal} {Phys. Rev. X}\ }\textbf {\bibinfo
  {volume} {7}},\ \bibinfo {pages} {031058} (\bibinfo {year}
  {2017}{\natexlab{b}})}\BibitemShut {NoStop}%
\bibitem [{\citenamefont {Xu}\ \emph {et~al.}(2019)\citenamefont {Xu},
  \citenamefont {Liu}, \citenamefont {Pan}, \citenamefont {Qi}, \citenamefont
  {Sun},\ and\ \citenamefont {Meng}}]{Xu2019}%
  \BibitemOpen
  \bibfield  {author} {\bibinfo {author} {\bibfnamefont {X.~Y.}\ \bibnamefont
  {Xu}}, \bibinfo {author} {\bibfnamefont {Z.~H.}\ \bibnamefont {Liu}},
  \bibinfo {author} {\bibfnamefont {G.}~\bibnamefont {Pan}}, \bibinfo {author}
  {\bibfnamefont {Y.}~\bibnamefont {Qi}}, \bibinfo {author} {\bibfnamefont
  {K.}~\bibnamefont {Sun}},\ and\ \bibinfo {author} {\bibfnamefont {Z.~Y.}\
  \bibnamefont {Meng}},\ }\bibfield  {title} {\bibinfo {title} {Revealing
  fermionic quantum criticality from new monte carlo techniques},\ }\href
  {https://doi.org/10.1088/1361-648x/ab3295} {\bibfield  {journal} {\bibinfo
  {journal} {Journal of Physics: Condensed Matter}\ }\textbf {\bibinfo {volume}
  {31}},\ \bibinfo {pages} {463001} (\bibinfo {year} {2019})}\BibitemShut
  {NoStop}%
\bibitem [{\citenamefont {Berg}\ \emph {et~al.}(2019)\citenamefont {Berg},
  \citenamefont {Lederer}, \citenamefont {Schattner},\ and\ \citenamefont
  {Trebst}}]{Berg2019}%
  \BibitemOpen
  \bibfield  {author} {\bibinfo {author} {\bibfnamefont {E.}~\bibnamefont
  {Berg}}, \bibinfo {author} {\bibfnamefont {S.}~\bibnamefont {Lederer}},
  \bibinfo {author} {\bibfnamefont {Y.}~\bibnamefont {Schattner}},\ and\
  \bibinfo {author} {\bibfnamefont {S.}~\bibnamefont {Trebst}},\ }\bibfield
  {title} {\bibinfo {title} {Monte carlo studies of quantum critical metals},\
  }\href@noop {} {\bibfield  {journal} {\bibinfo  {journal} {Annual Review of
  Condensed Matter Physics}\ }\textbf {\bibinfo {volume} {10}},\ \bibinfo
  {pages} {null} (\bibinfo {year} {2019})}\BibitemShut {NoStop}%
\bibitem [{\citenamefont {Yamase}\ and\ \citenamefont
  {Metzner}(2012)}]{Yamase2012}%
  \BibitemOpen
  \bibfield  {author} {\bibinfo {author} {\bibfnamefont {H.}~\bibnamefont
  {Yamase}}\ and\ \bibinfo {author} {\bibfnamefont {W.}~\bibnamefont
  {Metzner}},\ }\bibfield  {title} {\bibinfo {title} {Fermi-surface truncation
  from thermal nematic fluctuations},\ }\href
  {https://doi.org/10.1103/PhysRevLett.108.186405} {\bibfield  {journal}
  {\bibinfo  {journal} {Phys. Rev. Lett.}\ }\textbf {\bibinfo {volume} {108}},\
  \bibinfo {pages} {186405} (\bibinfo {year} {2012})}\BibitemShut {NoStop}%
\bibitem [{\citenamefont {Punk}(2016)}]{Punk2016}%
  \BibitemOpen
  \bibfield  {author} {\bibinfo {author} {\bibfnamefont {M.}~\bibnamefont
  {Punk}},\ }\bibfield  {title} {\bibinfo {title} {Finite-temperature scaling
  close to ising-nematic quantum critical points in two-dimensional metals},\
  }\href {https://doi.org/10.1103/PhysRevB.94.195113} {\bibfield  {journal}
  {\bibinfo  {journal} {Phys. Rev. B}\ }\textbf {\bibinfo {volume} {94}},\
  \bibinfo {pages} {195113} (\bibinfo {year} {2016})}\BibitemShut {NoStop}%
\bibitem [{\citenamefont {Abrikosov}\ \emph {et~al.}(1975)\citenamefont
  {Abrikosov}, \citenamefont {Gorkov},\ and\ \citenamefont
  {Dzyaloshinski}}]{Abrikosov1975}%
  \BibitemOpen
  \bibfield  {author} {\bibinfo {author} {\bibfnamefont {A.}~\bibnamefont
  {Abrikosov}}, \bibinfo {author} {\bibfnamefont {L.}~\bibnamefont {Gorkov}},\
  and\ \bibinfo {author} {\bibfnamefont {I.}~\bibnamefont {Dzyaloshinski}},\
  }\href {https://books.google.com/books?id=E\_9NtwNY7UcC} {\emph {\bibinfo
  {title} {Methods of Quantum Field Theory in Statistical Physics}}},\ Dover
  Books on Physics Series\ (\bibinfo  {publisher} {Dover Publications},\
  \bibinfo {year} {1975})\BibitemShut {NoStop}%
\bibitem [{\citenamefont {Chubukov}(2005)}]{Chubukov2005}%
  \BibitemOpen
  \bibfield  {author} {\bibinfo {author} {\bibfnamefont {A.~V.}\ \bibnamefont
  {Chubukov}},\ }\bibfield  {title} {\bibinfo {title} {Ward identities for
  strongly coupled eliashberg theories},\ }\href
  {https://doi.org/10.1103/PhysRevB.72.085113} {\bibfield  {journal} {\bibinfo
  {journal} {Phys. Rev. B}\ }\textbf {\bibinfo {volume} {72}},\ \bibinfo
  {pages} {085113} (\bibinfo {year} {2005})}\BibitemShut {NoStop}%
\bibitem [{Note3()}]{Note3}%
  \BibitemOpen
  \bibinfo {note} {In actuality, it is only exact for $\Omega _m/q \to 0$ or
  $\Omega _m/q \to \infty $, but because of the constraint imposed at these two
  limits, the corrections for finite $\Omega _m/q$ are at most of order
  one.}\BibitemShut {Stop}%
\bibitem [{\citenamefont {Tuck}(1967)}]{tuck_simple_1967}%
  \BibitemOpen
  \bibfield  {author} {\bibinfo {author} {\bibfnamefont {E.~O.}\ \bibnamefont
  {Tuck}},\ }\bibfield  {title} {\bibinfo {title} {A {Simple}
  "{Filon}-{Trapezoidal}" {Rule}},\ }\href {https://doi.org/10.2307/2004168}
  {\bibfield  {journal} {\bibinfo  {journal} {Mathematics of Computation}\
  }\textbf {\bibinfo {volume} {21}},\ \bibinfo {pages} {239} (\bibinfo {year}
  {1967})}\BibitemShut {NoStop}%
\bibitem [{Note4()}]{Note4}%
  \BibitemOpen
  \bibinfo {note} {For $T < 0.1$ the QMC data shows that the nematic coherence
  length is already cut off by superconductivity, so we did not use that
  data}\BibitemShut {NoStop}%
\bibitem [{Note5()}]{Note5}%
  \BibitemOpen
  \bibinfo {note} {Note that the asymptotic expression in Eq. \protect \textup
  {\hbox {\mathsurround \z@ \protect \normalfont (\ignorespaces \ref
  {eq:Pi-INM-T}\unskip \@@italiccorr )}} for the MET applies when $v_Fq \ll
  \omega _T$, which is not the case for the QMC data. We performed the angular
  integration in that equation numerically.}\BibitemShut {Stop}%
\bibitem [{\citenamefont {Chubukov}\ \emph {et~al.}(2019)\citenamefont
  {Chubukov}, \citenamefont {Abanov}, \citenamefont {Wang},\ and\ \citenamefont
  {Wu}}]{chubukov2019interplay}%
  \BibitemOpen
  \bibfield  {author} {\bibinfo {author} {\bibfnamefont {A.~V.}\ \bibnamefont
  {Chubukov}}, \bibinfo {author} {\bibfnamefont {A.}~\bibnamefont {Abanov}},
  \bibinfo {author} {\bibfnamefont {Y.}~\bibnamefont {Wang}},\ and\ \bibinfo
  {author} {\bibfnamefont {Y.-M.}\ \bibnamefont {Wu}},\ }\href@noop {}
  {\bibinfo {title} {The interplay between superconductivity and non-fermi
  liquid at a quantum-critical point in a metal}} (\bibinfo {year} {2019}),\
  \Eprint {https://arxiv.org/abs/1912.01797} {arXiv:1912.01797
  [cond-mat.supr-con]} \BibitemShut {NoStop}%
\bibitem [{\citenamefont {Chubukov}\ and\ \citenamefont
  {Maslov}(2012)}]{Chubukov2012}%
  \BibitemOpen
  \bibfield  {author} {\bibinfo {author} {\bibfnamefont {A.~V.}\ \bibnamefont
  {Chubukov}}\ and\ \bibinfo {author} {\bibfnamefont {D.~L.}\ \bibnamefont
  {Maslov}},\ }\bibfield  {title} {\bibinfo {title} {First-matsubara-frequency
  rule in a fermi liquid. i. fermionic self-energy},\ }\href
  {https://doi.org/10.1103/PhysRevB.86.155136} {\bibfield  {journal} {\bibinfo
  {journal} {Phys. Rev. B}\ }\textbf {\bibinfo {volume} {86}},\ \bibinfo
  {pages} {155136} (\bibinfo {year} {2012})}\BibitemShut {NoStop}%
\bibitem [{\citenamefont {Chubukov}\ and\ \citenamefont
  {Maslov}(2017)}]{Chubukov2017}%
  \BibitemOpen
  \bibfield  {author} {\bibinfo {author} {\bibfnamefont {A.~V.}\ \bibnamefont
  {Chubukov}}\ and\ \bibinfo {author} {\bibfnamefont {D.~L.}\ \bibnamefont
  {Maslov}},\ }\bibfield  {title} {\bibinfo {title} {Optical conductivity of a
  two-dimensional metal near a quantum critical point: The status of the
  extended drude formula},\ }\href {https://doi.org/10.1103/PhysRevB.96.205136}
  {\bibfield  {journal} {\bibinfo  {journal} {Phys. Rev. B}\ }\textbf {\bibinfo
  {volume} {96}},\ \bibinfo {pages} {205136} (\bibinfo {year}
  {2017})}\BibitemShut {NoStop}%
\end{thebibliography}%

\clearpage
\onecolumngrid
\appendix

\section*{Appendix: Quantum self energy at the first Matsubara frequency}

In this Appendix we explicitly compute the value of the quantum part of the self energy at the first Matsubara frequency $\Sg_Q(\pi T)$. Our purpose is to justify the somewhat puzzling result, shown in Figs. \ref{fig:sfm-an} and \ref{fig:INM-se-panels}, that $\Sg_Q(\pi T)$ is negative. We will show that $\Sg_Q(\pi T)$ in MET is always negative, and that this is a consequence of the so-called ``first Matsubara frequency rule'' - namely that $\Sg_Q(\pi T) = 0$ in ET \cite{Chubukov2012,Chubukov2017}.

We assume that we are at the QCP ($M=0$) and write down an expression for  $\Sg_Q(\pi T)$ directly from Eq. \eqref{eq:sig-low},
\begin{align}
  \label{eq:sig-low-appen}
  \Sigma_Q(\pi T) &\approx  
                    \gb T \sum_{n\neq 0} \int_0^\infty \frac{|\p| dp}{\tp}\frac{\sigma(\W_n+\pi T)}{\sqrt{(\W_n+\pi T + \Sigma(\W_n+\pi T))^2+ v_F^2 |\p|^2}}
                    \frac{1}{|\p|^2+\Pi(|\p|,\W_n)}, \nn \\
                  & = \gb T
                    \sum_{n\geq 1} \int_0^\infty \frac{|\p| dp}{\tp}  \frac{
                    1}{|\p|^2+\Pi(|\p|,\W_n)} \left(\frac{1}{\sqrt{(\W_n+\pi T + \Sigma(\W_n+\pi T))^2+ v_F^2 |\p|^2}} -\right.\nn\\
  &\qquad\qquad\qquad\qquad\qquad\qquad\qquad\qquad\qquad\qquad\qquad\qquad\left.\frac{1}{\sqrt{(\W_n-\pi T + \Sigma(\W_n-\pi T))^2+ v_F^2 |\p|^2}}\right),
\end{align} 
For simplicity we set $\vf(\theta) = \vf$, $\nu_F (\theta) = \nu_F$, $f(\theta) =1$ and $N_b = 1$.

It is immediately evident that if we neglect the dynamical part 
in the square-root of the denominator of Eq. \eqref{eq:sig-low-appen} as is done in ET (see Sec. \ref{sec:t=0-theory}) we obtain $\Sg_Q(\pi T) = 0$, because contributions from positive and negative frequencies exactly cancel out. This is the first Matsubara frequency rule. The correction coming from MET is therefore coming from the small asymmetry between positive and negative frequencies that arise from the non-factorization of momentum integration, see the discussion in the Introduction and in Sec.  \ref{sec:t=0-theory}. To proceed we assume and then verify that (i) the integration and summation in \eqref{eq:sig-low-appen} are dominated by large $v_Fp, \W_n \gg \pi T$, and that (ii) we may neglect the self-energy terms.
Expanding the square-roots we obtain,
\begin{flalign}
  \label{eq:sig-1}
  \Sigma_Q(\pi T) \approx -\frac{\gb T 
  }{2\pi}
  \sum_{n \geq 1} \int_0^\infty \frac{|\p| dp}{|\p|^2+\Pi(|\p|,\W_n)}\frac{2\pi T\W_n}{(\W_n^2+(\pi T)^2 +  v_F^2 |\p|^2)^{3/2}}
\end{flalign}

We first solve for the SFM.
Here $\vf^2\Pi^\sfm(|\p|,\W_n|) = \w_{b}\W_n$ where $\w_{b} = \frac{\gb N \nu_F\vf}{2\pi Q_{hs}}$ (by order of magnitude, $\omega_b \sim {\bar g}$).
Rescaling the $p$ integral, we find
\begin{equation}
  \label{eq:sig-2}
  \Sigma_Q^\sfm(\pi T) \approx -\frac{\gb T}{2\pi}
  \sum_{n \geq 1} \int_0^\infty \frac{x dx}{x^2+\frac{\w_b\W_n}{\W_n^2+(\pi T)^2}}\frac{2\pi T\W_n}{(\W_n^2+(\pi T)^2)^{3/2}(x^2+1)^{3/2}}.
\end{equation}
By inspection, one may verify that the integral is dominated by $x \sim 1, \W_n \sim \w_b$. Since $\w_b$ is a large parameter in the theory, this justifies our previous assumptions, and also allows us to safely replace the summation with an integral.
We then obtain
\begin{align}
  \label{eq:sig-3}
  \Sigma_Q^\sfm(\pi T) &\approx -\frac{\gb T }{2\pi}
                         \int_{\pi T}^\infty \frac{d\W}{\W^2}\int_0^\infty \frac{x dx}{\left(x^2+\frac{\w_b}{\W}\right)(x^2+1)^{3/2}} \nn\\
                       &\approx -\frac{\gb T }{2\pi\w_b} \log\frac{\w_b}{\pi T}
\end{align}
We see that  $\Sigma_Q^\sfm(\pi T)$ is {\it negative} and of order ${\bar g} T/\omega_b \sim T$. 
For other Matsubara frequencies, $\omega_n = O(T)$, and $\Sigma_Q^\sfm(\w_n)$ is positive and of order $(T \omega_{\sfm})^{1/2} \sim (T {\bar g})^{1/2}\sim T(\gb/T)^{1/2}$, i.e., is much larger than $\Sg_Q^\sfm(\pi T)$.
  
For the INM, we write $\vf^2\Pi^{\inm}(|\p|,\W_n) = \w_{b'}^2\frac{\W_n}{\vf|\p|}$, where $\w_{b'}^2 = \gb \nu_F \vf^2/\pi$
(by order of magnitude $\w_{b'} \sim ({\bar g} E_F)^{1/2}$).
Repeating the same computational steps we find,
\begin{flalign}
  \label{eq:sig-4}
  \Sigma_Q^\inm(\pi T) \approx -\frac{\gb T}{2\pi}
  \sum_{n \geq 1} \int_0^\infty \frac{x^2 dx}{x^3+\frac{\w_{b'}^2
      \W_n}{(\W_n^2+(\pi T)^2)^{3/2}}}\frac{2\pi T\W_n}{(\W_n^2+(\pi T)^2)^{3/2}(x^2+1)^{3/2}}.
\end{flalign}
Again, we find $x\sim 1, \W_n \sim \w_{b'}$ justifying our assumptions. Replacing the summation by an integral we obtain,
\begin{align}
  \label{eq:sig-5}
  \Sg_Q^\inm(\pi T) &\approx -\frac{\gb T}{2\pi}
               \int_{\pi T}^\infty \frac{d\W}{\W^2} \int_0^\infty \frac{x^2 dx}{\left(x^3+\frac{\w_{b'}^2}{\W^2}\right)(x^2+1)^{3/2}} \nn \\
  & \approx -\frac{\gb T
  }{4\sqrt{\pi}\w_{b'}}\Gamma^2(3/4),
\end{align}
and $\Gamma(3/4) = 1.23$.
We see that $\Sg_Q^\inm(\pi T)$ is again negative. For the INM it is of order $T {\bar g}/\w_{b'} \sim T \left({\bar g}/E_F\right)^{1/2}$.  For other Matsubara frequencies $\omega_n = O(T)$, 
$\Sg_Q^\inm(\W_n)$ is positive and is of order $(T^2 {\bar g}^2/E_F)^{1/3} \sim \left(T \left({\bar g}/E_F\right)^{1/2}\right)  \left({\bar g} E_F/T^2\right)^{1/6}$, i.e., is again much larger than $\Sg_Q^\inm(\pi T)$.
\end{document}